\documentclass[a4paper,11pt]{article}

\pdfoutput=1
\usepackage{macros}

\preprint{}

\title{3d $\Ncal=4$ mirror symmetry with 1-form symmetry}

 \author{Satoshi Nawata${}^1$, Marcus Sperling${}^2$, Hao Ellery Wang${}^3$, Zhenghao Zhong${}^{4,5}$}

\affiliation[1]{Department of Physics and Center for Field Theory and Particle Physics, Fudan University, \\
220, Handan Road, 200433 Shanghai, China}
\affiliation[2]{Shing-Tung Yau Center, Southeast University \\
 Xuanwu District, Nanjing, Jiangsu, 210096, China}

\affiliation[3]{Yau Mathematical Sciences Center, Tsinghua University\\
 Haidian District, Beijing, 100084, China}

\affiliation[4]{Theoretical Physics Group, The Blackett Laboratory, Imperial College London \\
Prince Consort Road,
London, SW7 2AZ, UK}

\affiliation[5]{Mathematical Institute, University of Oxford,\\
Andrew Wiles Building, Woodstock Road, Oxford, OX2 6GG, UK}

\emailAdd{snawata@gmail.com}
\emailAdd{msperling@seu.edu.cn}
\emailAdd{yukawahaow@gmail.com}
\emailAdd{zhenghao.zhong@maths.ox.ac.uk}

\abstract{
The study of 3d mirror symmetry has greatly enhanced our understanding of various aspects of 3d $\mathcal{N}=4$ theories. In this paper, starting with known mirror pairs of 3d $\mathcal{N}=4$ quiver gauge theories and gauging discrete subgroups of the flavour or topological symmetry, we construct new mirror pairs with non-trivial 1-form symmetry. By providing explicit quiver descriptions of these theories, we thoroughly specify their symmetries (0-form, 1-form, and 2-group) and the mirror maps between them.}

\begin{document}
\maketitle

\section{Introduction}
Supersymmetric theories with 8 supercharges in space-time dimension 3 exhibit a rich set of intriguing features; One of the most prominent is 3d mirror symmetry \cite{Intriligator:1996ex}. 
Given a 3d theory that has a mirror dual theory, 3d mirror symmetry exchanges Coulomb branch and Higgs branch. In particular, this also implies the exchange of flavour symmetries $G_f$ (Higgs branch isometries) and the topological symmetries $G_t$ (Coulomb branch isometries).

The notion of symmetries has been generalised to include novel types beyond the standard symmetries of local operators \cite{Gaiotto:2014kfa}. Among others, these include higher-form symmetries. Specifically for 3d theories, discrete 1-form symmetries can be generated by gauging discrete 0-form symmetries. 
The structure of generalised symmetries in 3d supersymmetric theories has been the focus of recent research, including \cite{Benini:2017dus,Eckhard:2019jgg,Bergman:2020ifi,Apruzzi:2021mlh,Beratto:2021xmn,Bhardwaj:2022yxj,Bhardwaj:2022dyt,Bartsch:2022mpm,Bhardwaj:2022scy,Mekareeya:2022spm,vanBeest:2022fss,Bartsch:2022ytj} and references therein\footnote{See also \cite{Argyres:2022mnu,Cordova:2022ruw} for recent review articles.
}.
Given the vast catalogue of 3d mirror pairs with trivial 1-form symmetry, one might wonder what mirror symmetry implies for 3d theories with 1-form symmetry.

In this paper, we start with a known mirror pair $(\Tcal,\Tcal^\vee)$ of 3d $\Ncal=4$ theories that admit UV quiver descriptions, and gauge a discrete $\Gamma^{[0]}$ subgroup of the 0-form symmetry to generate new theories with $\Gamma^{[1]}$ 1-form symmetry. Depending on whether $\Gamma \equiv\Gamma^{[0]}$ is a subgroup of the flavour or topological symmetry, the resulting mirror pair $(\Tcal\slash \Gamma,(\Tcal\slash \Gamma)^\vee)$ changes. For $\Gamma \equiv \Gamma^f \subset G_f$, the field theory description of $\Tcal\slash \Gamma^f$ is straightforward, but for $\Gamma \equiv \Gamma^t \subset G_t^\vee$, the description of $\Tcal^\vee \slash \Gamma^t$ is less transparent. In this paper, we consider $\Gamma=\Z_q$ suitably embedded into a Cartan $\urm(1)$ which enables us to derive explicit quiver descriptions for these cases and allows to specify the global form of the 0-form symmetries of $(\Tcal\slash \Gamma,(\Tcal\slash \Gamma)^\vee)$.
It is known that the resulting 0-form and the newly introduced discrete 1-form symmetry may not just be a direct product, but can form an extension, called 2-group symmetry \cite{Sharpe:2015mja,Tachikawa:2017gyf,Cordova:2018cvg,Benini:2018reh,Hsin:2020nts,Bhardwaj:2022dyt}. We comment on such extensions throughout this work.

The remainder of this paper is organised as follows: in Section \ref{sec:gauging_0-form}, we consider known mirror pairs and gauge discrete $0$-form symmetries to generate mirror pairs with non-trivial $1$-form symmetry. We first study abelian theories, followed by non-abelian $T[\surm(N)]$ and $T_{\rho}^{\sigma}[\surm(N)]$ theories with non-abelian product gauge groups $\prod_i \urm(N_i)$.  This class of examples has the benefit that all $0$-form symmetries are manifest in the UV description.
Thereafter, $\sorm(k)$ and $\sprm(k)$ gauge groups are considered by studying $T[\sorm(2N)]$ theories, $\sprm(k)$ SQCD, and linear orthosymplectic quivers. While the flavour $0$-form symmetry is manifest in this set of examples, the topological symmetry is at most accessible by discrete $\Z_2$ subgroups, which turns out to be sufficient for the intents and purposes here.
Lastly, we consider mixed types: i.e.\ $D$ and $C$-type Dynkin quivers composed of unitary gauge groups and their mirror $\sprm(k)$ and $\orm(2k)$ SQCD theories, respectively.
The advantage of this class of mirror pairs is that the flavour symmetry of the SQCD theories and the topological symmetry of the unitary Dynkin quivers are fully manifest.
Before closing, some magnetic quiver examples are considered.
Conclusions are provided in Section \ref{sec:conclusion}. 
Several appendices complement the main text and provide computational details.

\paragraph{\emph{Note added.}}
\textit{During the course of this project, we were informed of a related work done by Bhardwaj, Bullimore, Ferrari, and Sch\"afer-Nameki \cite{Bhardwaj:2023zix}.  We are grateful to them for coordinating the submission of our papers.}


\section{Gauging discrete 0-form symmetries}
\label{sec:gauging_0-form}
In this section, mirror theories with non-trivial 1-form symmetry are constructed. Gauging discrete subgroups of the 0-form symmetry, which results in 1-from symmetries and a potential 2-group structure, has, for example, been considered in \cite{Tachikawa:2017gyf,Cordova:2018cvg,Benini:2018reh,Hsin:2020nts,Bhardwaj:2022dyt}.

The principle is simple: start from a known mirror pair $(\Tcal,\Tcal^\vee)$ and gauge discrete 0-form symmetries $\Gamma$ (finite, cyclic) such that $\Gamma\equiv \Gamma^t \subset G_t(\Tcal)$ and $\Gamma \equiv \Gamma^f \subset G_f(\Tcal^\vee)$. This ensures that the resulting theories $(\Tcal\slash \Gamma^t,\Tcal^\vee\slash \Gamma^f)$ are mirror pairs with 1-form symmetry $\Gamma$. The aims here are (i) to provide explicit quiver descriptions for $(\Tcal\slash \Gamma^t,\Tcal^\vee\slash \Gamma^f)$ and (ii) to detail the resulting symmetries (0-form, 1-form, and 2-group).

\subsection{Abelian theories}
As a first example, consider 3d $\Ncal=4$ SQED with $N$ hypermultiplets of charge $1$ and its abelian mirror quiver theory \cite{Intriligator:1996ex}, see Figure \ref{fig:SQED}. The global 0-form symmetries are well-known: for SQED one finds $\urm(1)_t \times \psurm(N)_f$, while the abelian mirror quiver enjoys a $\urm(1)_f\times \psurm(N)_t$ symmetry.

\subsubsection{SQED with higher charge}
\label{sec:SQED_higher_charges}
Suppose that one gauges a discrete $\Z_q$ subgroup of the abelian $\urm(1)$ factor of the global 0-form symmetry. The resulting theories are straightforwardly derived. Gauging a $\Z_q\subset \urm(1)_t$ for SQED with $N$ charge $1$ hypermultiplets leads to SQED with $N$ charge $q$ hypermultiplets, see also \cite{Mekareeya:2022spm}. Similarly, gauging a $\Z_q\subset \urm(1)_f$ of the abelian mirror quiver leads to an abelian quiver with a $\Z_q$ 1-form symmetry. The two theories obtained are then mirror to each other, see Figure \ref{fig:SQED}. The quiver notation is summarised in Table \ref{tab:notation} of Appendix \ref{app:notation}. 

\begin{figure}[t]
    \centering
 \includegraphics[page=1]{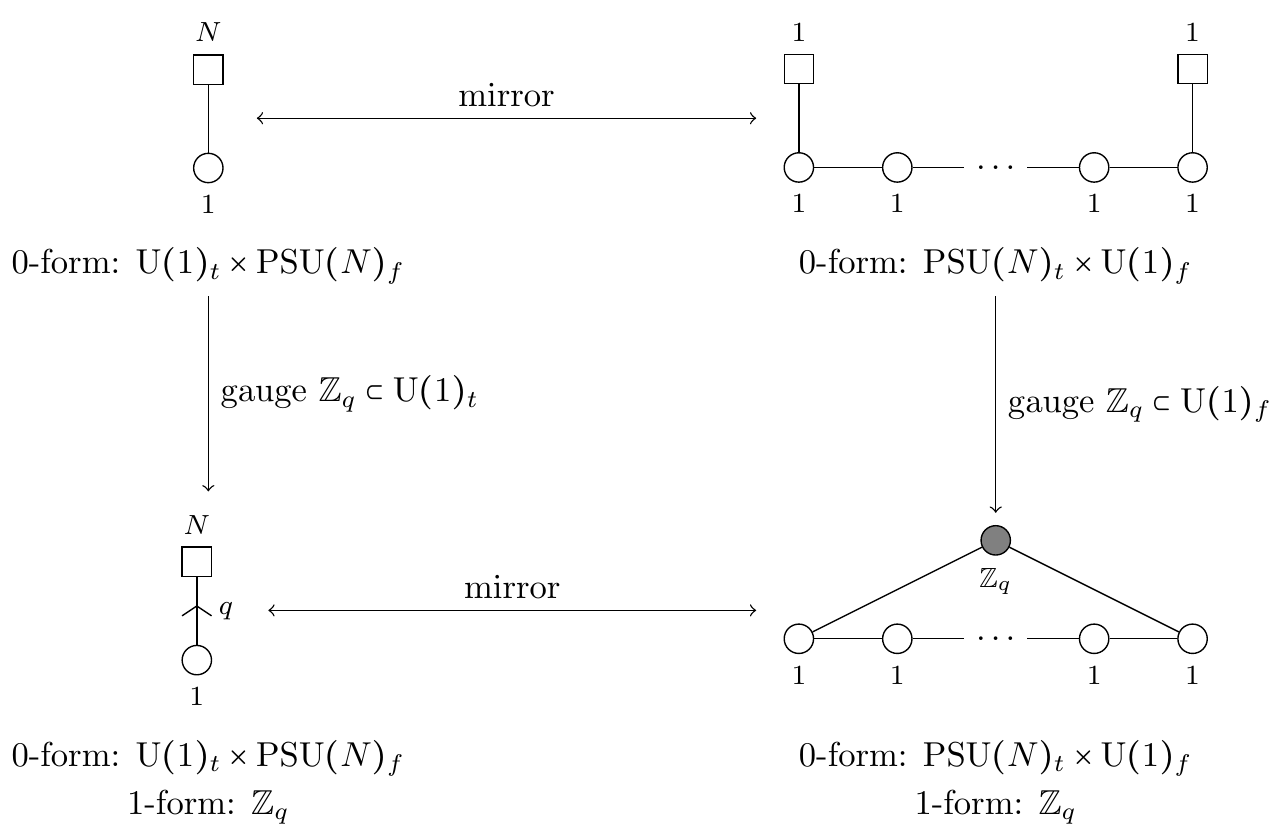}
        \caption{Gauging of discrete 0-form symmetries in SQED and its mirror. For SQED with charge 1 hypermultiplets, a $\Z_q^t$ gauging results in SQED with charge $q$ hypermultiplets. These are indicated by an arrow with the label $q$. For the abelian mirror quiver, the $\Z_q^f$ gauging is realised by acting on the flavours.  The fundamental flavours that are charged under the discrete $\Z_q$ are connected to a grey node. See Appendix \ref{app:notation} for conventions. }
    \label{fig:SQED}
\end{figure}

\paragraph{Consistency checks.}
The proposed mirror symmetry can be verified by Hilbert series techniques \cite{Benvenuti:2006qr,Feng:2007ur,Gray:2008yu,Cremonesi:2013lqa}. The Higgs branch Hilbert series is insensitive to the gauging of the $\Z_q$ inside the topological symmetry of the SQED theory; similarly, the Coulomb branch of the mirror does not perceive changes upon gauging a discrete subgroup of the flavour symmetry. See Appendix \ref{app:Hilbert_series} for conventions.

Performing the discrete gauging for the SQED theory reduces to a $\Z_q$ Molien{-}Weyl sum over the Coulomb branch Hilbert series
\begin{align}
    \HS_{\mathrm{SQED}_N^{q=1}}^{\Coulomb} (w|t)&= \frac{1}{1-t} \sum_{m\in \Z} w^m t^{\frac{1}{2}N |m|} \notag \\
    &= \PE[t + (w +w^{-1}) t^{\frac{N}{2}}  - t^{N}]   \\
    \HS_{\mathrm{SQED}_N^{q}}^{\Coulomb} (z|t) &= 
    \frac{1}{q}\sum_{p=0}^{q-1} \HS_{\mathrm{SQED}_N^{q'=1}}^{\Coulomb} (w|t)\big|_{w=z^{\frac{1}{q}}(\zeta_q)^p }
    \qquad \zeta_q = e^{\frac{2\pi \im }{q}} \in \Z_q \notag \\
    &=  \frac{1}{1-t} \sum_{m\in \Z} z^m t^{\frac{1}{2}N |q\cdot m|} \notag \\
    &=\PE[t + (z +z^{-1}) t^{\frac{1}{2}q N}  - t^{qN}]   \,.
    \end{align}
Likewise, one performs the $\Z_q$ Molien{-}Weyl sum on the Higgs branch Hilbert series of the mirror theory
    \begin{align}
    \HS_{\mathrm{mirror}}^{\Higgs} (y|t) &=
    \prod_{a=1}^{N{-}1} \oint \frac{\diff x_a}{2\pi i x_a}
    \PE\left[ \sum_{b=1}^{N{-}2} \left(\frac{x_b}{x_{b+1}} + \frac{x_{b+1}}{x_{b}} \right)t^{\frac{1}{2}}  -(N{-}1)t   \right]
    \notag  \\
     &\qquad \cdot
     \PE\left[ \left( 
     \frac{y^{\frac{1}{2}}}{x_{1}} 
     +  \frac{x_{1}}{y^{\frac{1}{2}}}  
     +\frac{y^{-\frac{1}{2}}}{x_{N{-}1}}
     + \frac{x_{N{-}1}}{y^{-\frac{1}{2}}}
     \right)t^{\frac{1}{2}} \right]
     \notag \\
     &= \PE[t + (y +y^{-1}) t^{\frac{N}{2}}  - t^{N}] \\
      \HS_{\mathrm{mirror}/\Z_q}^{\Higgs} (z|t)&=
      \frac{1}{q}\sum_{p=0}^{q-1}
      \HS_{\mathrm{mirror}}^{\Higgs} (y|t) \bigg|_{y=z^{\frac{1}{q}} (\zeta_q)^p} \notag \\
     &=\PE[t + (z +z^{-1}) t^{\frac{1}{2}q N}  - t^{qN}]   \,.
\end{align}
In summary, both results confirm the expectation and provide the explicit parameter map.
As a remark, the superconformal index is equally well suited to probe such dualities; see for instance \cite{Mekareeya:2022spm} for SQED with charge $q=2$ hypermultiplets. Since either Higgs or Coulomb branch operators are unaffected by gauging a $\Z_q^{t/f}$, the Hilbert series is a more convenient tool. 

\paragraph{Symmetries.}
Using the techniques of \cite{Apruzzi:2021mlh}, one can inspect the interplay between the discrete 1-form symmetry $\Z_q \subset \urm(1)_t$ and the global 0-form symmetry $\psurm(N)_f$ for the SQED theory. The centre symmetry $Z_F=\Z_N$ of $\surmL(N)_f$ is generated by $\alpha_F=\zeta_N$, while the $\urm(1)$ gauge group supports a $Z_G=\Z_{N\cdot q}$ centre generated by $\alpha_{G}=\zeta_{N\cdot q}$. The diagonal $\alpha_D=(\alpha_G,\alpha_F)$ generates a $\Ecal=\Z_{N\cdot q} \subset Z_G \times Z_F$. The 1-form symmetry $\Gamma^{[1]} =\Z_q$ is generated by $\alpha_D^N = (\alpha_{G}^N,\alpha_{F}^N) = (\alpha_{G}^N,1)$, which acts trivially on the matter content. The short exact sequence
\begin{align}
\label{eq:SES_SQED}
    0 \to \Gamma^{[1]}  =  \Z_q \to\Ecal= \Z_{q\cdot N} \to Z_F =\Z_N \to 0
\end{align}
splits whenever $\gcd(q,N)=1$, i.e.\ $q$ and $N$ are co-prime.  In other words, $\gcd(q,N)>1$ is a necessary condition for the existence of an extension to a 2-group structure. A sufficient condition is to have the non-trivial Postnikov class in $H^3(B\psurm(N);\Z_q)$ \cite{Hsin:2020nts}, which is the image of the obstruction class for lifting a $\psurm(N)$-bundle to an $\surm(N)$-bundle, under the Bockstein map $H^2(B\psurm(N);\Z_N)\to H^3(B\psurm(N);\Z_q)$. In fact, the Postnikov class is non-trivial if and only if $\gcd(q,N)>1$.\footnote{We would like to thank A. Milivojević for providing the proof at \href{https://mathoverflow.net/questions/444507/non-triviality-of-a-postnikov-class-in-h3-leftb-operatornamepsun-math}{mathoverflow}.} Therefore, the short exact sequence \eqref{eq:SES_SQED} represents a non-trivial 2-group extension if and only if $\gcd(q,N)>1$. See also \cite{Bhardwaj:2022dyt,Mekareeya:2022spm} for a recent discussion of SQED with 2 flavours of charge 2.

\paragraph{Comments on lines.}
As explained in \cite{Bhardwaj:2021wif,Lee:2021crt,Bhardwaj:2022dyt}, 1-form symmetries and 2-group structures can be understood via equivalence classes of line defects\footnote{In brief, lines $L_{1,2}$ are equivalent if there exists a local operator $\Ocal$ at the junction between them. The set of equivalence classes $\{L\}\slash \sim $ forms the Pontryagin dual $\widehat{\Gamma}^{[1]}$ of the 1-form symmetry $\Gamma^{[1]}$. Refining the equivalence relation by keeping track of $0$-form symmetry representations $R$ leads to the following equivalence relation: $(L_1,R_1) \sim (L_2,R_2)$ iff there exists a local operator transforming as $R_1 \otimes R_2^\ast$ (or $R_1^\ast \otimes R_2$) at the junction of the lines. The equivalence classes give rise to $\widehat{\Ecal}$ (Pontryagin dual of $\Ecal$), which encodes the interplay between the centres of  gauge symmetry and 0-form symmetry. These groups fit into the short exact sequence $0\to \widehat{Z}\to \widehat{\Ecal} \to \widehat{\Gamma}^{[1]}\to0$, which is the Pontryagin dual of the sequences discussed in the text, e.g.\ \eqref{eq:SES_SQED}, \eqref{eq:SES_SQCD_1}, \eqref{eq:SES_SQCD_2}. Whenever these short exact sequences split, the 2-group is necessarily trivial. For non-split sequences, the Postnikov class controls whether the 2-group is trivial or not.}. Here, we illustrate how the higher-form symmetry is also realised on the line defects.

Consider SQED with $N$ hypermultiplets of charge $q$. A Wilson line of charge $h$ with $h\in \{1,2,\ldots,q-1\}$ cannot end on a local operator because local operators are either constructed as polynomials in the fundamental hypermultiplets of charge $q$ or are monopole operators, which are gauge singlets for 3d $\Ncal=4$ theories. Thus, the 1-form symmetry $\Gamma^{[1]}$ (or its Pontryagin dual) is generated by the $(q-1)$ Wilson lines that cannot end.
Refining with respect to the flavour symmetry shows that a Wilson line of charge $q$ is equivalent to a flavour Wilson line transforming as $[1,0,\ldots,0]_{A_{N-1}}$. This however is not an allowed representation of $G_f =\psurm(N)$, and signals the existence of a 2-group structure. In fact, the $N$-th power of such a Wilson line is well-defined under $G_f$, because the $N$-th tensor product of $[1,0,\ldots,0]_{A_{N-1}}$ contains a singlet. Such lines generate the group $\widehat{\Ecal}=\Z_{q\cdot N}$.

Turning to the abelian mirror quiver, one can straightforwardly see that the fundamental Wilson lines, i.e.\ those of unit charge under a single $\urm(1)$ gauge group factor, can end on a local operator constructed out of the hypermultiplets. Therefore, one needs to turn to the vortex lines to understand the 1-form symmetry. It is known \cite{Dimofte:2019zzj} that the junctions between vortex lines are significantly more challenging than those between Wilson lines. It would be interesting to systematically address this in explicit examples.

\subsubsection{SQED with discrete gauge factor}
\label{sec:SQED_discrete_topol}
Next, revert the logic: gauge a $\Z_q$ subgroup of the $\psurm(N)_f$ symmetry of SQED. Conversely, on the mirror side, one gauges a $\Z_q$ subgroup of the $\psurm(N)_t$ topological symmetry of the abelian quiver theory.

\begin{figure}[t]
    \centering
 \includegraphics[page=2]{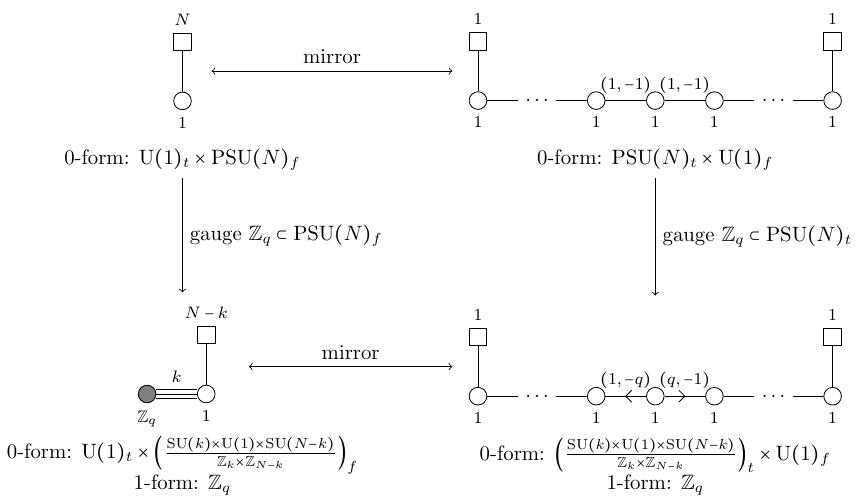}
        \caption{Gauging of discrete 0-form symmetries in SQED and its mirror. The centre symmetries act with charges $(-q\bmod k,q\bmod (N{-}k))$ on the $\urm(1)$ factor. For SQED, the $\Z_q$ acts on $k$ of the $N$ hypermultiplets, which is indicated by $k$ edges connected to a grey node. The remaining hypermultiplets are uncharged under the discrete group. One can use a global $\urm(1)$ rotation to move the $\Z_q$-action onto the other set of hypermultiplets as well, which renders the entire setup symmetric, cf.\ Appendix~\ref{app:mirror-map_SQED_after}.
        For the abelian linear quiver, gauging along the Cartan $\urm(1)_t$ at the $k$-th gauge nodes leads to hypermultiplets with charge  $q$ under the $k$-th $\urm(1)$, while still of unit charge under the adjacent gauge factors. This is indicated by an arrow with label, cf.\ Appendix~\ref{app:notation}. }
    \label{fig:SQED_2}
\end{figure}

For the abelian quiver theory, discrete gauging along a Cartan $\urm(1)_t$ of the topological symmetry alters the linear quiver theory by modifying the charges of the bifundamental hypermultiplets attached to a single gauge node. This follows from analogous arguments as for SQED with charge $q$ hypermultiplets or the arguments used in Appendices \ref{app:discrete_gauge_TSUN} -- \ref{app:discrete_gauging_index}.
For the SQED theory, gauging of a discrete flavour 0-form symmetry affects some of the fundamental flavours. To see this, one uses the original mirror map \eqref{eq:mirror_map_min_A-type_orbit} between the parameters to identify which flavour fugacities are affected by gauging along a Cartan $\urm(1)_t$ factor in the abelian mirror. As a result, the flavours of the SQED split into two sets: one charged under $\Z_q$ and the other is trivial. This is shown in Figure \ref{fig:SQED_2}.

\paragraph{Global symmetry: abelian mirror point of view.}
The global symmetry is affected as follows: suppose that one gauges a $\Z_q \subset \urm(1)_k \subset \psurm(N)_t$ subgroup of the topological Cartan $\urm(1)$ at the $k$-th node of the abelian quiver 
\begin{align}
\label{eq:abelian_quiver}
       \raisebox{-.5\height}{
    \includegraphics[page=1]{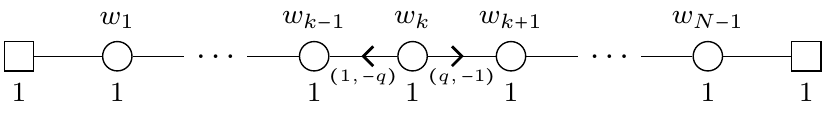}
    }  \;.
\end{align}
The 0-form symmetry algebra after gauging is $\surmL(k)\oplus \urmL(1)\oplus \surmL(N{-}k)$. As exemplified in Appendix \ref{app:abelian_quiver}, the 0-form symmetry group is 
\begin{align}
\label{eq:sym_group_abelian_quiver}
   G_t\eqref{eq:abelian_quiver}= \frac{ \surm(k) \times \urm(1)_Q \times \surm(N{-}k) }{\Z_k \times \Z_{N{-}k}} 
\end{align}
where the centre symmetry $\Z_\ell \subset \surm(\ell)$ acts on the fundamental representation $[1,0,\ldots,0]_{A_{\ell-1}}$ with charge $+1$ under $\Z_\ell$, for $\ell \in \{ k,N{-}k \}$, see also Table \ref{tab:congruency_class_algebras}. Moreover, the $\Z_k \times \Z_{N{-}k}$ act with charge $\left(-q\mod k, q\mod (N{-}k)\right)$ on the $\urm(1)_Q$ variable. Roughly $Q \sim w_k$, see \eqref{eq:abelian_quiver_fugacities} for details.

The global structure \eqref{eq:sym_group_abelian_quiver} can also be inferred directly from the set of balanced nodes in \eqref{eq:abelian_quiver}. The unbalanced gauge node $\urm(1)_k$ is connected to two balanced sets of gauge nodes, forming the $A_{k-1}$ and $A_{N-k-1}$ Dynkin diagram. Generalising the arguments of \cite{Gaiotto:2008ak}, there are monopole operators transforming as $[0,0,\ldots,q]_{A_{k-1}}\times (1)_Q$ (and its conjugate) and $[q,0,\ldots,0]\times (1)_Q$ (and conjugate). This follows because the $\urm(1)_k$ node is attached to the $k-1$-th node of the $A_{k-1}$ Dynkin diagram and the $1$-st node of $A_{N-k-1}$ Dynkin diagram. Compared to the standard case of unit charge bifundamental hypermultiplets, the increased charge $q$ modifies the appearing $A_\ell$ representations accordingly. The existence of these monopole operators in the Coulomb branch chiral ring leads to the isometry \eqref{eq:sym_group_abelian_quiver}.

\paragraph{Global symmetry: SQED point of view.}
To illuminate this result, it is instructive to also consider the SQED side:
\begin{align}
\label{eq:abelian_quiver2}
       \raisebox{-.5\height}{
    \includegraphics[page=2]{figures/figures_1-form_abelian_quivers.pdf}
    }  \;,
\end{align}
where the two distinct sets of fundamentals are denoted as $X$ and $\tilde{X}$ (by convention, both have charge $-1$ under the $\urm(1)$ gauge group). 
Here, as a different notation from elsewhere, we use the arrowed lines to symbolize 3d $\Ncal=2$ chiral multiplets (inflow into the gauge node) and anti-chiral multiplets (outflow from the gauge node).
Computationally, gauging a discrete $\Z_q^f$ is realised via the following flavour fugacities, see Appendix \ref{app:mirror_map_SQED}
\begin{subequations}
\label{eq:HB_SQED_para}
\begin{alignat}{3}
&X_a:  \qquad & a&=1,\ldots, k: \qquad &
 y_a &=\zeta_q \cdot Q_1  \cdot  \begin{cases}
    x_1\,, & a=1 \\
    \frac{x_a}{x_{a-1}} \,,& 1<a<k \\
    \frac{1}{x_{k{-}1}}\,, & a=k
    \end{cases}
    \\
 &\tilde{X}_d:  \qquad &    d&=k+1,\ldots,N:  \qquad &
   y_d &=  Q_2  \cdot \begin{cases}
    u_1 \,,& d=k+1 \\
    \frac{u_{d-k}}{u_{d-k{-}1}}\,, & k+1<d<N{-}k \\
    \frac{1}{u_{N{-}k{-}1}}\,, & d=N{-}k
    \end{cases}
\end{alignat}
with $\zeta_q \in \Z_q^f$. The $x_a$ and $u_d$ are weight space fugacities for $\surmL(k)$ and $\surmL(N{-}k)$, respectively.

The first observation is that if $k | q$ then the $\Z_k$ centre symmetry of $\surm(k)$ is gauged, such that a global $\psurm(k)_f$ factor arises. Similarly, if $(N{-}k)|q$ the $\Z_{N{-}k}$ centre of $\surm(N{-}k)$ is gauged, leading to a $\psurm(N{-}k)_f$ factor. For the general case, one fixes the two so far arbitrary $\urm(1)_{Q_{1,2}}$ symmetries\footnote{The definition of $Q$ is a \emph{choice}. Here, it is chosen such that the operator $\Ocal$ in \eqref{eq:charges_Ocal_SQED}, as Higgs branch operator with lowest R-charge that is charged under the $\urm(1)$, has the unit charge. }:
\begin{align}
\begin{cases}
   Q_1^k \cdot Q_2^{N{-}k} &\stackrel{!}{=}1 \\
   \left(\frac{Q_1}{Q_2}\right)^q &\stackrel{!}{=} Q
\end{cases} 
\quad \Rightarrow \quad 
\begin{cases}
Q_1 &= Q^{\frac{N{-}k}{q\cdot N}} \\
Q_2 &= Q^{\frac{-k}{q\cdot n}} 
\end{cases}
\end{align}
\end{subequations}
which agrees with \eqref{eq:mirror_map_SQED_after_gauging_rotated} of Appendix \ref{app:mirror_map_SQED}.
Next, consider a gauge invariant operator $\Ocal$ built from the  fields $\{X_a\}_{a=1}^k$ transforming as  $(\zeta_q \cdot Q_1 \cdot [1,0,\ldots,0]_{A_{k{-}1}},-1)$ under flavour-gauge transformations and fields $\{\tilde{X}_d^\dagger\}_{d=k+1}^{N}$ transforming as $( Q_2^{-1}\cdot [0,\ldots,0,1]_{A_{N{-}k{-}1}},+1)$. Thus,  $X_a \tilde{X}_d^\dagger$ is  $\urm(1)$ gauge invariant. For $\Z_q$ invariance, one also requires $q$-copies of $X_a$ in the form of $\mathrm{Sym}^q (X_a)$, which leads to the $q$-th symmetric representation $\mathrm{Sym}^q [1,0,\ldots,0]$ of $\surm(k)_f$. As a consequence, one also requires $q$ copies of $\{\tilde{X}_d^\dagger\}$ in the form $\mathrm{Sym}^q (\tilde{X}_d)$, which leads to the $q$-th conjugate symmetric representation $\mathrm{Sym}^q [0,\ldots,0,1]$ of $\surm(N{-}k)_f$. Such a gauge invariant operator has charges
\begin{align}
    \Ocal&= \mathrm{Sym}^q_{a_1,\ldots,a_q} (X_{a_1}) \cdot \mathrm{Sym}^q_{d_1,\ldots,d_q} (\tilde{X}_{d_i}) \notag \\
    &\qquad \leftrightarrow
    \mathrm{Sym}^q [1,0,\ldots,0]_{\surm(k)}\otimes
    \mathrm{Sym}^q [0,\ldots,0,1]_{\surm(N{-}k)} \otimes \underbrace{\left( \frac{Q_1}{Q_2}\right)^q }_{=Q}
    \label{eq:charges_Ocal_SQED}
\end{align}
The operator $\Ocal$ has $\Z_k \times \Z_{N{-}k}$ centre charges $(q \bmod k,-q \bmod (N{-}k))$. Hence, the $\Z_k\times \Z_{N{-}k}$ transformations can be compensated by a global $\urm(1)_Q$ rotation if $Q$ has charges $(-q \bmod k, q \bmod (N{-}k))$ under the centre symmetries. This confirms \eqref{eq:sym_group_abelian_quiver} as flavour symmetry $G_f$.
The operator $\Ocal$ can be detected in the Hilbert series at $R$-charge $q\cdot 2 \cdot \frac{1}{2}=q$.

\paragraph{Comments on lines.}
Returning to the quiver \eqref{eq:abelian_quiver}, consider a Wilson lines $W_a$ of charge $1$ under the $a$-th $\urm(1)$ gauge group factor. For each  $a\neq k$, $W_a$ can end on a local operator composed of concatenated bifundamental hypermultiplets. For $a=k$, $W_k$ cannot end since the bifundamentals connected to the $k$-th gauge node are of charge $q$. Further, monopole operators cannot screen gauge Wilson lines, because monopole operators are gauge singlets for 3d $\Ncal=4$ theories. Thus, the lines $(W_k)^h$ with $h\in \{0,1,\ldots,q-1\}$ cannot end and generate the abelian group $\widehat{\Gamma}^{[1]}=\Z_q$.

\subsection{An illustrative example}
\label{sec:example}
One of the main messages of this paper is that gauging discrete $\Z_q$ subgroups of the topological symmetry for quiver gauge theories $\Tcal$ with unitary gauge nodes can result in theories $\Tcal \slash \Z_q^t$ which admit a simple quiver description. To illustrate this fact, consider $\urm(k)$ SQCD with $N\geq 2k$ fundamental flavours
\begin{align}
    \raisebox{-.5\height}{
    \includegraphics[page=1]{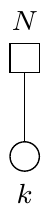}
    }  
\end{align}
with the well-known $0$-form symmetries: $G_f =\psurm(N)$, $G_t=\urm(1)_t$ for $N>2k$ and $G_t=\sorm(3)$ for $N=2k$. 

Next, express the gauge group as $\urm(k)\cong \frac{\urm(1)\times \surm(k)}{\Z_k}$ where $\Z_k$ acts as centre on $\surm(k)$ and via $\Z_k$ charge $(k{-}1)$ on the $\urm(1)$ factor. Rewriting $\urm(k)$ magnetic fluxes $\bm{m} \in \Z^k$ into $\urm(1)\times \surm(k)$ fluxes $(h,\bm{l})$ requires the co-character lattice to be $\Gamma=\bigcup_{i=0}^{k{-}1} (\Z + \frac{i}{k})^k$. Effectively, the SQCD theory can be understood as $\surm(k)\times \urm(1)$ gauge theory with $N$ copies of bifundamentals and an ``unusual'' magnetic lattice $\Gamma$. One can introduce a (topological) fugacity $z$  that keeps track of the components of $\Gamma$. If $w$ denotes the topological fugacity of $\Tcal$, one employs $w \to z w^{\frac{1}{k}}$. 
Next, gauge a discrete $\Z_q$ subgroup of the topological symmetry by performing a discrete Molien{-}Weyl sum over $z$. It is convenient to choose either $q|k$ or $k|q$. One can show rigorously (e.g.\ using the superconformal index or the Coulomb branch Hilbert series, see Appendix \ref{app:discrete_gauging}) the following:

\subsubsection{\texorpdfstring{Gauge a subgroup with $\mathbf{q|k}$}{Gauge a subgroup with q|k} }
If $q|k$ then only the subgroup $\Z_q \subset \Z_k$ is gauged. The theory becomes
\begin{align}
\left[
    \raisebox{-.5\height}{
    \includegraphics[page=2]{figures/figures_1-form_example.pdf}
    } \right] \slash \Z_{\frac{k}{q}}
    \qquad
    \text{ with magnetic lattice }
    \bigcup_{i=0}^{\frac{k}{q}-1} \left(\Z + i\cdot \frac{q}{k}\right)^k \,,
\end{align}
where the quotient $\Z_{\frac{k}{q}}$ signals that this discrete group is not gauged, in the sense of \cite{Bourget:2020xdz}. 
The resulting theory has a $\urm(1)_t \times \psurm(N)_f$ 0-form symmetry and a $\Z_q$ 1-form symmetry. The potential interplay can be analysed via the action of the centre symmetries: defining $\alpha_G=((\zeta_k)^{\frac{k}{q}},\zeta_{q \cdot N})\in \Z_k \times \Z_{q\cdot N}$ (because only a $\Z_q \subset \Z_k$ is gauged) and $\alpha_F=\zeta_N \in \Z_N$, the diagonal combination $\alpha_D=(\alpha_G,\alpha_F)$ generates a $\Ecal=\Z_{q\cdot N}$ group. The element $N\cdot \alpha_D =
((\zeta_k)^{\frac{k}{q}\cdot N},\zeta_{q \cdot N}^N, 1)$ generates a $\Gamma^{(1)}=\Z_q$ subgroup that acts trivial on the matter fields. By definition, this establishes the 1-form symmetry. The short exact sequence 
\begin{align}
\label{eq:SES_SQCD_1}
0\to \Gamma^{(1)} = \Z_q \to\Ecal = \Z_{N\cdot q} \to \Z_N \to0 
\end{align}
splits if $\gcd(q,N)=1$. Whenever $\gcd(q,N)>1$, this short exact sequence exhibits a non-trivial 2-group extension of $\Gamma^{(1)}$ and $\psurm(N)_f$ as discussed below \eqref{eq:SES_SQED}.

\paragraph{Symmetries via lines.}
One can again illustrate this higher-form symmetry by using line defects and their equivalence classes. 
A gauge Wilson line $W$ in the representation $[0,\ldots,0]_{A_{k-1}} \times (-1)$ cannot end on any local operator; Neither polynomials of the hypermultiplets nor monopole operators, because of a mismatch in gauge charges. However, $W^q$ can end on the determinant operator $\Ocal \sim \det (X)$, obtained by contracting hypermultiplets $X$ with the invariant $\epsilon$ tensor of $\surm(k)$. This operator has charges $[0,\ldots,0]_{A_{k-1}}\times k$. Since $q|k$, $\Ocal$ has the same centre charges as $W^q$, such that $W^q$ can end on it. Therefore, the lines $W^a$ with $a\in \{1,2,\ldots, q-1\}$ cannot end on any local operator and generate the abelian group $\widehat{\Gamma}^{[1]}=\Z_q$. Taking flavour charges into account, $W^q$ is equivalent to a flavour Wilson line transforming as $\wedge^k [0,\ldots,0,1]_{A_{N-1}}$, which follows from the flavour charges of $\Ocal$. This is not a representation of $\psurm(N)_f$, but taking $N$-th tensor $(W^q)^{\otimes N}$ is equivalent to a singlet of the flavour symmetry. Thus, these lines generate the group $\widehat{\Ecal}= \Z_{N\cdot q}$ and the 
1-form symmetry potentially forms a 2-group with the flavour symmetry (depending on the $\gcd(N,q)$).

\subsubsection{\texorpdfstring{Gauge a discrete group $\Z_q$ with $\mathbf{k|q}$}{Gauge a discrete group Zq with k|q}}
If $k|q$ then the $\surm(k)$ centre $\Z_k$ is a subgroup of $\Z_q$ and fully gauged. The theory becomes
\begin{align}
    \raisebox{-.5\height}{
    \includegraphics[page=3]{figures/figures_1-form_example.pdf}
    } 
    \qquad 
    \text{with magnetic lattice } \Z^k \,,
\end{align}
The difference is now that the $N$ hypermultiplets transform as $\surm(k)$ fundamental with charge $\frac{q}{k}\in \N$ under the $\urm(1)$. This is indicated by the arrow, cf.\ Table \ref{tab:notation}.

In terms of symmetries, the theory $\Tcal \slash \Z_q^t$ has a $\urm(1)_t$ topological symmetry, $\psurm(N)_f$ flavour symmetry, a $\Z_q$ 1-form symmetry. Moreover, inspecting the gauge-flavour centre symmetries shows: $\alpha_G = (\zeta_k,\zeta_{ N\cdot q})\in \Z_k \times \Z_{N \cdot q}$ and $\alpha_{F}=\zeta_{N} \in \Z_{N}$. The diagonal generator $\alpha_d =(\alpha_G,\alpha_F)$ spans a $\Ecal=\Z_{ N\cdot q}$, and the element $N \cdot\alpha_d 
=  (\zeta_k^N,\zeta_{ N\cdot q}^N,1 )$ generates a $\Gamma^{(1)}=\Z_{k\cdot q}$ subgroup, using that $k|q$. This subgroup acts trivial on the matter fields; thus, defining the 1-form symmetry $\Gamma^{(1)}$. The short exact sequence 
\begin{align}
\label{eq:SES_SQCD_2}
0\to \Gamma^{(1)} = \Z_q \to\Ecal = \Z_{N\cdot q} \to \Z_N \to0
\end{align}
splits if $\gcd(q,N)=1$. In all other cases, there exists a non-trivial extension giving rise to a 2-group structure between $\Gamma^{(1)}$ and $G_f=\psurm(N)$.

\paragraph{Symmetries via lines.}
Again, let us illustrate these structures with line defects. The gauge Wilson line $W$ transforming as $[0,\ldots,0]_{A_{k-1}} \times (-1)$ under $\surm(k)\times \urm(1)$ cannot end on a local operator, which either has to be a polynomial in the hypermultiplet $X$ transforming as $[1,0,\ldots,0]_{A_{k-1}} \times (-\tfrac{q}{k})$ or has to be a monopole operator, which is gauge singlets. In contrast, the Wilson line $W^q$ can end on the local operator constructed as the determinant: i.e.\ the $\surm(k)$ gauge group is equipped with the invariant $\epsilon_{i_1,\ldots,i_k}$ tensor. Contracting $k$ hypermultiplets yields an operator $\Ocal \sim \det (X)$ which transforms as $[0,\ldots,0]_{A_{k-1}} \times (-q)$. Hence, the set of Wilson lines $W^a$ with $a\in\{1,2,\ldots,q-1\}$ cannot end and generate the abelian group $\widehat{\Gamma}^{[1]}=\Z_q$.
If one also keeps track of the flavour symmetry representations, one finds that  $\Ocal$ transforms as $\wedge^k [0,\ldots,0,1]_{A_{N-1}}$ which is not a representation of $\psurm(N)_f$. Hence, this gauge Wilson line is equivalent to a flavour Wilson line and the centres of gauge and flavour symmetry intertwine to give rise to a 2-group structure.

The following sections apply the analogous argument to other quiver gauge theories. The relevant questions are: (i) What is the resulting theory? (ii) What are its symmetries? (iii) What is the mirror dual theory?
\subsection{\texorpdfstring{$T[\surm(N)]$ theories}{TSUn theories}}
\label{sec:TSUN}
\begin{figure}[t]
    \centering
 \includegraphics[page=4]{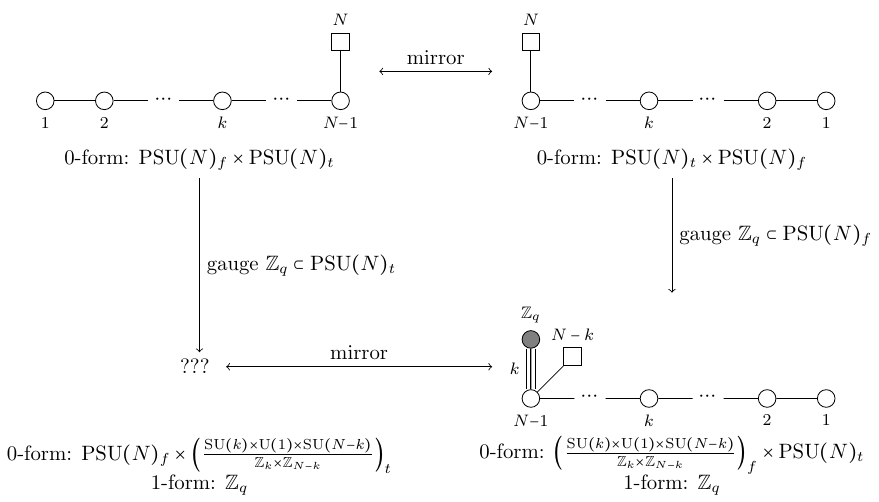}
        \caption{Gauging of discrete 0-form symmetries in $T[\surm(N)]$ theories. The centre symmetries act with charges $(-q\bmod k,q \bmod (N{-}k))$ on the $\urm(1)$ factor. The quiver description for $T[\surm(N)]\slash \Z_q^t$, here denoted by $???$, is provided in \eqref{eq:TSUN_topol_gauged}. The quiver for $T[\surm(N)]^\vee\slash \Z_q^f$ shows again a split of the $N$ fundamental flavours into two sets: $k$ of them are charged under $\Z_q^f$, which is indicated by an edge of multiplicity $k$  to the grey node; the remaining $N{-}k$ flavours are uncharged.}
    \label{fig:TSUn}
\end{figure}

Moving on to quiver theories with non-abelian gauge factors, consider the self-mirror $T[\surm(N)]$ theories \cite{Gaiotto:2008ak}, see Figure \ref{fig:TSUn}. The global 0-form symmetry group is given by $\psurm(N)_t \times \psurm(N)_f$.
In the same spirit as above, one can gauge a discrete $\Z_q$ 0-form symmetry inside, say, the topological symmetry. The mirror of the resulting theory is then obtained by gauging a $\Z_q$ 0-form symmetry inside the flavour symmetry. The question is how the $\Z_q$ is embedded inside the flavour symmetry, given that the $\Z_q$ is embedded into a Cartan $\urm(1)$ of  the topological symmetry of the mirror. To answer this, one utilises the mirror map \eqref{eq:mirror_map_TSUN}.

In more detail, let us consider gauging a $\Z_q \subset \psurm(N)_t$ of a $T[\surm(N)]$ theory; one inquires about the nature of the resulting theory $T[\surm(N)]\slash \Z_q^t$. Analogous to Section \ref{sec:example}, see also Appendix \ref{app:discrete_gauge_TSUN}, for a specific $\Z_q$ embedded in the $k$-th topological Cartan $\urm(1)$ factor, the resulting theories $T[\surm(N)]\slash \Z_q^t$ are in fact related to versions of $T[\surm(N)]$ encountered in \cite{Bourget:2020xdz}. These quiver theories differ from $T[\surm(N)]$ as follows: the $k$-th node is replaced by $\urm(k)\to\surm(k)$, and the flavour node becomes a $\urm(1)$ gauge nodes with an $N$ copies of bifundamental hypermultiplets between $\urm(N{-}1)$ and the ``new'' $\urm(1)$ gauge node. 
Restricting to the case that either $q|k$ or $k|q$, the resulting theory is given by
\begin{subequations}
\label{eq:TSUN_topol_gauged}
\begin{alignat}{3}
&q|k &\;\text{ with }  d&=\frac{k}{q}: \qquad&
  &\left[\raisebox{-.5\height}{
    \includegraphics[page=1]{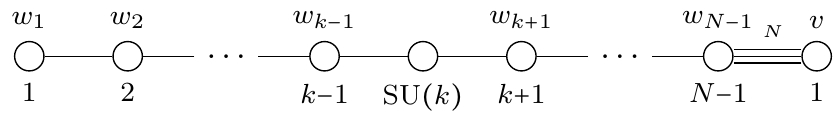}
    } \right]\slash \Z_d \notag
\\
& & & & &\text{magnetic lattice: }\bigcup_{i=0}^{d-1} \left(\Gamma +\frac{i}{d} \right)  \\
&k|q&\; \text{ with } q&= a\cdot k: \qquad &
  &\raisebox{-.5\height}{
    \includegraphics[page=2]{figures/figures_1-form_TSUN.pdf}
    } \\
    & & & & &\text{magnetic lattice: }\Gamma  \notag 
\end{alignat}
\end{subequations}
wherein $\Gamma$ denotes the standard integer lattice one assigns to the quiver based on \cite{Cremonesi:2013lqa}. The shifts by $\frac{i}{d}$ are to be understood as in \cite{Bourget:2020xdz}. As a comment, restricting to $k|q$ or $q|k$ ensures that the theory after discrete gauging has a simple quiver description. If this constraint is relaxed, there may not be a simple quiver, but the gauging is perfectly well-defined on the level of Hilbert series and index.

The mirror theory $T[\surm(N)]^\vee \slash \Z_q^f$ is obtained from $T[\surm(N)]^\vee =T[\surm(N)]$ by gauging a $\Z_q \subset \psurm(N)_f$. The mirror map \eqref{eq:mirror_map_TSUN} dictates that this is realised by splitting the $N$ fundamental flavours into two sets of $k$ and $N{-}k$ flavours, and gauging the $\Z_q$ symmetry in the overall $\urm(1)$ flavour symmetry of one of the two sets.
For concreteness, consider gauging the $\Z_q$ on the set of $k$ fundamental flavours:
\begin{align}
    \raisebox{-.5\height}{
    \includegraphics[page=3]{figures/figures_1-form_TSUN.pdf}
    }  \label{eq:TSUN_flavour_gauged}
\end{align}
and Appendix \ref{app:TSUN_calc} provides exemplary Hilbert series computations that confirm the mirror symmetry between the theories with non-trivial 1-form symmetry.

The mirror map between the parameters of $T[\surm(N)]\slash \Z_q^t$ in \eqref{eq:TSUN_topol_gauged} and $T[\surm(N)]^\vee\slash \Z_q^f$ in \eqref{eq:TSUN_flavour_gauged} can be derived exactly. For concreteness, consider the case $q=k$, then the map between the parameters in \eqref{eq:TSUN_topol_gauged} and \eqref{eq:TSUN_flavour_gauged} is established via
\begin{align}
\label{eq:mirror_paras_TSUN}
 \begin{cases}
w_i&= \frac{y_i}{y_{i+1}},\;  i\neq k \;, \cr
v&= y_{N}^N \,.
\end{cases}
\end{align}
Further details on this map are provided in Appendix \ref{app:mirror_map_TSUN}.

\paragraph{Global symmetry.}
Building on the understanding of the 0-form symmetry group \eqref{eq:sym_group_abelian_quiver} for (balanced) abelian quivers, one can utilise a similar logic for the balanced $T[\surm(N)]$ theories. Consider the quiver \eqref{eq:TSUN_topol_gauged} the topological symmetry algebra is $\surmL(k)\oplus \urmL(1)\oplus \surmL(N{-}k)$. The global form is then given by
\begin{align}
 \label{eq:global_sym_TSUN}
    G_t &= \frac{\surm(k) \times \urm(1)_Q \times \surm(N{-}k)}{\Z_k \times \Z_{N{-}k}}  \\
    &\text{with $Q$ has $\Z_k\times \Z_{N{-}k}$ charges $\left( -q \bmod k, \ q \bmod (N{-}k)\right)$}  \notag 
\end{align}
where the centre symmetries $\Z_\ell$ act in the standard way on $\surm(\ell)$.  Note that for $k|q$ there is a $\psurm(k)$ factor in the global symmetry.
The examples in the next paragraph, as well as the explicit character decomposition in Appendix \ref{app:TSUN_calc}, confirm this structure. 


This structure \eqref{eq:global_sym_TSUN} is also apparent from the Higgs branch isometry of the mirror \eqref{eq:TSUN_flavour_gauged}, i.e.\ denote the two distinct sets of fundamentals by
\begin{align}
    \raisebox{-.5\height}{
    \includegraphics[page=5]{figures/figures_1-form_TSUN.pdf}
    }  \label{eq:TSUN_flavour_HB} 
    \;.
\end{align}
Here, as a distinctive notation from elsewhere, we use arrowed lines to represent the arrowed lines to symbolize 3d $\Ncal=2$ chiral multiplets (inflow into the gauge node) and anti-chiral multiplets (outflow from the gauge node).
Analogously to \eqref{eq:HB_SQED_para}, one can perform the $\Z_q^f$ gauging by assigning (c.f.\ Appendix \ref{app:mirror_map_TSUN}) 
\begin{subequations}
\label{eq:HB_fugacities_TSUN}
\begin{alignat}{3}
&X_a: \qquad  & a&=1,\ldots, k: \qquad &
 y_a &= \zeta_q \cdot Q_1  \cdot  \begin{cases}
    x_1\,, & a=1 \\
    \frac{x_a}{x_{a+1}} \,,& 1<a<k \\
    \frac{1}{x_{k{-}1}}\,, & a=k
    \end{cases}
    \\
  &\tilde{X}_d: \qquad  &   d&=k+1,\ldots, N:  \qquad &
   y_d &=  
   Q_2  \cdot \begin{cases}
    u_1 \,,& d=k+1 \\
    \frac{u_{d-k}}{u_{d-k{-}1}}\,, & k+1<d<N{-}k \\
    \frac{1}{u_{N{-}k{-}1}}\,, & d=N{-}k
    \end{cases}
\end{alignat}
with $\zeta_q \in \Z_q^f$. The $x_i$ and $u_j$ are weight space fugacities of $\surmL(k)$ and $\surmL(N{-}k)$, respectively. The two appearing $\urm(1)$ fugacities $Q_{1,2}$ effectively reduce to a single $\urm(1)_Q$; for instance by imposing $\prod_i y_i=1$, i.e.\
\begin{align}
    \begin{cases}
    Q_1^k \cdot Q_2^{N{-}k} &\stackrel{!}{=}1 \\
    \left(\frac{Q_1}{Q_2} \right)^q &\stackrel{!}{=}Q
    \end{cases}
    \qquad \Rightarrow \qquad 
    \begin{cases}
    Q_1 = Q^{\frac{N{-}k}{N\cdot q}} \\
    Q_2 = Q^{-\frac{k}{N \cdot q}}
    \end{cases}
\end{align}
\end{subequations}
which agrees\footnote{Again, the definition of $Q$ is a \emph{choice}. It is motivated by assigning the unit charge to the Higgs branch operator $\Ocal$ in \eqref{eq:HB_GIO_TSUN}, which is the operator with the lowest R-charge that is charged under the $\urm(1)_Q$.} with \eqref{eq:mirror_map_TSUN_after_gauging_rotated} of Appendix \ref{app:mirror_map_TSUN}.
Note also that for $q|k$ the $(\zeta_q)^{\frac{k}{q}}  \in \Z_q^f$ acts as the $ \Z_k\subset \surm(k)_{x_i}$ centre symmetry; thus the global factor is $\psurm(k)_{x_i}$ in this case. The $\urm(1)_Q$ may transform non-trivially under the $\Z_{\ell}\subset \surm(\ell)$ centre symmetries, depending on the charge of $Q$. To determine the charge, one again considers a specific gauge-invariant operator $\Ocal$ build out of the two sets of fundamentals: $X$ transforms as $(\zeta_q  \cdot Q_1 \cdot [1,0,\ldots,0]_{A_{k{-}1}}, \overline{\boldsymbol{N{-}1}})$ and $\tilde{X}^\dagger$ transforms as $( Q_2^{-1} \cdot [0,\ldots,0,1]_{A_{N{-}k{-}1}}, \boldsymbol{N{-}1})$. $\urm(N{-}1)$ gauge invariance imposes $\Tr(X \tilde{X}^\dagger)$, wherein the trace is taken over the gauge indices. $\Z_q$ gauge invariance requires $\Ocal = \mathrm{Sym}^q \Tr(X \tilde{X}^\dagger)$, where the symmetrisation acts on the flavour indices. The resulting operator transforms as 
\begin{align}
\label{eq:HB_GIO_TSUN}
    \Ocal: \quad 
    \mathrm{Sym}^q [1,0,\ldots,0]_k 
    \otimes 
    \mathrm{Sym}^q [0,\ldots,0,1]_{N{-}k} 
    \otimes 
    \underbrace{\left( \frac{Q_1}{Q_2} \right)^q}_{=Q}
\end{align}
such that the $\Z_k\times \Z_{N{-}k}$ centre charges are $(q \bmod k,-q \bmod (N{-}k))$. These can be compensated by a global $\urm(1)_Q$ rotation provided the centre charges of $Q$ are $(-q \bmod k, q \bmod (N{-}k)) $. This confirms \eqref{eq:global_sym_TSUN} as flavour symmetry for the quiver \eqref{eq:TSUN_flavour_gauged}. 

As a remark, the operator $\Ocal$ can be detected in the Hilbert series as the first non-trivial term in $Q$. The $R$-charge of $\Ocal$ is simply $q \times 2\cdot \frac{1}{2} = q$.
The appendix \ref{app:TSUN_calc} provides examples that illustrate this point.

By analogous arguments as in Section \ref{sec:example}, one can verify that theories \eqref{eq:TSUN_topol_gauged} indeed have the expected $\Gamma^{(1)}=\Z_q$ 1-form symmetry. One finds that the centre generators of the combined gauge-flavour symmetry span a $\Ecal = \Z_{q \cdot N}$ group, such that there exists a non-trivial 2-group extension between $\Gamma^{(1)}$ and $G_f=\psurm(N)$ whenever $\gcd(q,N)>1$. Similarly, the same conclusion is reached by inspecting the screening of Wilson lines.

\paragraph{Example 1.}
For an illustrative purpose, let us consider $N=4$.
Gauging a specific $\Z_2$ 0-form symmetry leads to a mirror pair:
\begin{align}
    \raisebox{-.5\height}{
    \includegraphics[page=6]{figures/figures_1-form_TSUN.pdf}
    } 
    \xleftrightarrow{\quad \text{mirror}\quad }
    \raisebox{-.5\height}{
    \includegraphics[page=7]{figures/figures_1-form_TSUN.pdf}
    }  \,.
\end{align}
The Hilbert series in \eqref{eq:HS_TSU4_SU2} confirms that the Coulomb branch symmetry algebra for the left quiver (and the Higgs branch isometry algebra of the right quiver) is $\gfrak = \surmL(2)\oplus \surmL(2)\oplus \urmL(1)$. Moreover, the appearing $\surm(2)$ representations are all of integer spin; thus, suggesting the global form $G = \sorm(3)\times \sorm(3) \times \urm(1)$.

Choosing to gauge a specific discrete $\Z_3$ subgroup of the 0-form symmetry results in the pair:
\begin{align}
    \raisebox{-.5\height}{
    \includegraphics[page=8]{figures/figures_1-form_TSUN.pdf}
    } 
    \xleftrightarrow{\quad \text{mirror}\quad }
    \raisebox{-.5\height}{
    \includegraphics[page=9]{figures/figures_1-form_TSUN.pdf}
    } \,.
\end{align}
The explicit Hilbert series in \eqref{eq:HS_TSU4_SU3} shows that the Coulomb branch symmetry algebra of the left quiver (and the Higgs branch isometry algebra of the right theory) is $\gfrak = \surmL(3)\oplus \urmL(1)$. Moreover, all appearing characters are neutral under the $\Z_3$ centre symmetry of $\surm(3)$; hence, the global form is $G = \psurm(3)\times \urm(1)$.
\paragraph{Example 2.}
Considering discrete symmetries of the type $\Z_q$ with $q=a \cdot k $ allows us to uncover equivalent descriptions. Consider $T[\surm(5)]$ and gauge a $\Z_6$ 0-form symmetry. Among the choices considered here, gauging a $\Z_6 \subset \psurm(5)_t$ is realised by turning the $\urm(3)$ gauge node into $\surm(3)$ together with charge $2$ for the ``new'' $\urm(1)$ node
\begin{align}
    \raisebox{-.5\height}{
    \includegraphics[page=10]{figures/figures_1-form_TSUN.pdf}
    } 
    \xleftrightarrow{\quad \text{mirror}\quad }
    \raisebox{-.5\height}{
    \includegraphics[page=11]{figures/figures_1-form_TSUN.pdf}
    } \,, \label{eq:TSU5_SU3_charge_2}
\end{align}
or by turing the $\urm(2)$ node into $\surm(2)$ together with charge $3$ for the ``new'' $\urm(1)$ gauge factor
\begin{align}
    \raisebox{-.5\height}{
    \includegraphics[page=12]{figures/figures_1-form_TSUN.pdf}
    } 
    \xleftrightarrow{\quad \text{mirror}\quad }
    \raisebox{-.5\height}{
    \includegraphics[page=13]{figures/figures_1-form_TSUN.pdf}
    } \,.\label{eq:TSU5_SU2_charge_3}
\end{align}
Without the additional charges the theories are clearly distinct, for instance by the 0-form and 1-form symmetries, see \eqref{eq:global_sym_TSUN} and Figure \ref{fig:TSUn}.  However, with the modification, both become equivalent as, for example, the monopole formula in \eqref{eq:HS_TSU5_with_charges} confirms.

Equivalently, on the mirror, one gauges a $\Z_6$ subgroup of the flavour 0-form symmetry, but one time acting on three fundamental hypermultiplets and one time on two. For \eqref{eq:TSU5_SU3_charge_2} and \eqref{eq:TSU5_SU2_charge_3}, this is realised by
\begin{align}
    \{y_i\}_{i=1}^5 \to 
    \begin{cases}
    \{y_1,\ y_2,\  \zeta_6\ y_3 ,\ \zeta_6\ y_4,\ \zeta_6\ y_5\} , &\text{ for } \eqref{eq:TSU5_SU3_charge_2}\\
    &\text{or}\\
     \{\zeta_6\ y_1,\ \zeta_6\ y_2,\   y_3 ,\  y_4,\  y_5\}
     , &\text{ for } \eqref{eq:TSU5_SU2_charge_3}
    \end{cases} 
\end{align}
with $\zeta_6 \in \Z_6^f$. But in both cases, the $\Z_2 \times \Z_3$ centre symmetries are gauged by the discrete gauging of the $\Z_6^f$ 0-form symmetry.  Hence, the global symmetry is simply $\psurm(2) \times \urm(1)_Q \times \psurm(3)$ for both.

The observed equivalence can now be understood as follows: on the level of the Coulomb branch quivers, the global symmetry algebra arises from the split $\surmL(5)\to \surmL(3)\oplus \surmL(2)\oplus \urmL(1)$. Without the higher charge, the $\urm(1)$ factor transform non-trivially under the centre symmetry, see \eqref{eq:global_sym_TSUN}. In fact, it transforms differently in both cases. However, the higher charge is just tuned such that the $\urm(1)$ becomes independent of the discrete centre symmetries. This then also implies that the operators charged under the $\urm(1)$ factor coincide in both theories. 
On the Higgs branch side, the equivalence is a simple consequence of a global $\urm(1)$ rotation that takes the $\Z_6$ action from 2 fundamental flavours to the other 3 fundamental flavours. See also \eqref{eq:mirror_map_TSUN_after_gauging}--\eqref{eq:mirror_map_TSUN_after_gauging_rotated}.
\paragraph{Comment.}
The considerations so far implicitly assume that the gauge node $\urm(k)$ at which the discrete subgroup of the Cartan $\urm(1)_t$ of the topological symmetry is gauged has $k>1$, see for instance Appendix \ref{app:discrete_gauge_TSUN} and \ref{app:discrete_gauging_index}. Gauging the Cartan $\urm(1)_t$ of the $\urm(1)$ gauge node of $T[\surm(N)]$ is in spirit similar to Section \ref{sec:SQED_higher_charges}. Concretely, after gauging $\Z_q^t$ at the $\urm(1)$ node, the bifundamental between $\urm(1)$ and $\urm(2)$ is modified to have charge $q$ under the $\urm(1)$. Thus, the mirror pair becomes
\begin{align}
    \raisebox{-.5\height}{
    \includegraphics[page=14]{figures/figures_1-form_TSUN.pdf}
    } 
    \quad \xleftrightarrow{\text{ mirror }}
    \quad
    \raisebox{-.5\height}{
    \includegraphics[page=15]{figures/figures_1-form_TSUN.pdf}
    } 
    \label{eq:TSUN_U1_higher_charge}
\end{align}
and the global symmetry becomes
\begin{align}
    G_t(\text{left quiver \eqref{eq:TSUN_U1_higher_charge}}) =
    G_f(\text{right quiver \eqref{eq:TSUN_U1_higher_charge}})
    = \frac{\urm(1)_Q \times \surm(N{-}1)}{\Z_{N{-}1}}
\end{align}
where $Q$ has $\Z_{N{-}1}$ charge $q \bmod (N{-}1)$.


\subsection{\texorpdfstring{$T_{\bm{\rho}}^{\bm{\sigma}}[\surm(N)]$ theories}{T-rho-sigma theories}}
\label{sec:T_sigma_rho}
The class of linear quiver gauge theories with unitary gauge groups and fundamental or bifundamental hypermultiplets is given by the 
$T^\sigma_\rho [\surm(N)]$ theories \cite{Gaiotto:2008ak}, where $\rho, \sigma$ are two partitions of $N$. For $\sigma = \rho  = (1,\ldots,1)\equiv (1^N)$, the corresponding theory is simply $T^{(1^N)}_{(1^N)} [\surm(N)] =T[\surm(N)]$ and the partition data can be dropped. Mirror symmetry exchanges the partitions $\sigma$ and $\rho$, i.e.\ $T^\sigma_\rho [\surm(N)]^\vee =T^\rho_\sigma [\surm(N)]$. 

Analogous to the cases considered so far, the gauging of a discrete 0-form symmetry (either inside the topological symmetry group $G_t$ or the flavour symmetry group $G_f$) leads to a theory with non-trivial 1-form symmetry. Again, consider the two options in turn. While the process of gauging a discrete subgroup of $G_t$ is by now understood (see Section \ref{sec:TSUN} and Appendix \ref{app:discrete_gauging}), determining the action of the discrete group on the flavour symmetry of the mirror theory becomes more challenging when $G_f$ is a generic product group.  Thus, special attention is paid to determining the mirror theory of $T^\sigma_\rho [\surm(N)]\slash \Z_q^t$.

\paragraph{Gauging a $\Z_{N_k}\subset G_t$.}
A $T^\sigma_\rho [\surm(N)]$ is a linear quiver theory with gauge/flavour groups specified by a sequence of integers $\{N_i\}$ and $\{M_i\}$, respectively. The partitions determine the integers as detailed in \cite{Gaiotto:2008ak} and the quiver becomes
\begin{align}
   T^\sigma_\rho [\surm(N)]:\qquad    \raisebox{-.5\height}{
    \includegraphics[page=6]{figures/figures_1-form_T_rho_sigma.pdf}
    } \,.
    \label{eq:quiver_T-rho_sigma}
\end{align}
For concreteness, take the node $N_k$, with $N_k>1$, and gauge a $\Z_{N_k} \subset \urm(1)_k\subset G_t$ inside the Cartan factor of the topological symmetry associated to the $k$-th node. By the same arguments as in Appendices \ref{app:discrete_gauge_TSUN} and \ref{app:discrete_gauging_index}, one straightforwardly derives the resulting theory
\begin{align}
   T^\sigma_\rho [\surm(N)]\slash \Z_{N_k}^t:\qquad    \raisebox{-.5\height}{
    \includegraphics[page=7]{figures/figures_1-form_T_rho_sigma.pdf}
    } 
\end{align}
which has a non-trivial $\Z_{N_k}$ 1-form symmetry.
Now, one constructs the mirror theory.

\paragraph{Gauging a $\Z_{N_k} \subset G_f^\vee$.}
The mirror quiver gauge theory of \eqref{eq:quiver_T-rho_sigma} is given by
\begin{align}
   T^\sigma_\rho [\surm(N)]^\vee= T^\rho_\sigma [\surm(N)] :\qquad    \raisebox{-.5\height}{
    \includegraphics[page=8]{figures/figures_1-form_T_rho_sigma.pdf}
    } \,.
    \label{eq:mirror_T-rho_sigma}
\end{align}
and the integers $\{N_i^\vee\}$ and $\{M_i^\vee\}$ are determined by the partition data $\rho$, $\sigma$.

To determine which $\Z_{N_k}\subset G_f^\vee$ subgroup needs to be gauged, one has two options: one could derive the mirror map of parameters for the specific pair $(T_\rho^\sigma[\surm(N)], T^\rho_\sigma[\surm(N)])$ and compute which flavours are charged under $\Z_{N_k}^f$. In principle, this is straightforward but likely to be tedious.
Alternatively, one can employ the following train of thought: The mirror theory $T^\rho_\sigma [\surm(N)]$ can be rewritten in an unframed form
    \begin{align}
     \raisebox{-.5\height}{
    \includegraphics[page=8]{figures/figures_1-form_T_rho_sigma.pdf}
    }
    \cong
     \raisebox{-.5\height}{
    \includegraphics[page=9]{figures/figures_1-form_T_rho_sigma.pdf}
    } /// \urm(1)_{\diag}
    \end{align}
    where no explicit flavour group appears. For such a theory, it is implied that an overall $\urm(1)_{\text{diag}}$ subgroup decouples so the two quiver diagrams express the same theory. 

 The next step is to turn the unitary gauge group $\urm(N_k)$ in \eqref{eq:quiver_T-rho_sigma} into a special unitary gauge group $\surm(N_k)$. This theory still has a trivial 1-form symmetry due to the flavour groups. However, the 3d mirror theory can be found by using the algorithm in \cite{Bourget:2021jwo}.
 Schematically, one finds 
    \begin{align}
    \label{eq:T_rho_sigma_with_SU}
         \raisebox{-.5\height}{
    \includegraphics[page=10]{figures/figures_1-form_T_rho_sigma.pdf}
    }
    \xleftrightarrow{\text{mirror}}
         \raisebox{-.5\height}{
    \includegraphics[page=11]{figures/figures_1-form_T_rho_sigma.pdf}
    }/// \urm(1)_{\diag}
    \end{align}
Turning $\urm(N_k)$ into $\surm(N_k)$ means the 3d mirror has an additional $\urm(1)$ gauge group. The additional $\urm(1)$ gauge group in the unframed quiver is the result of gauging the flavour symmetry. Now, there are two $\urm(1)$ gauge groups connected to the rest of the linear quiver. The number of bonds $M_i^\vee$ attached to each $\urm(1)$ depends precisely on the choice of $\surm(N_k)$, i.e.\ which $k$. The splitting can also occur where the same gauge node, for example, $\urm(N_2^\vee)$ is connected to one $\urm(1)$ gauge group with an edge of multiplicity $M_2^\vee-x$ and to the other $\urm(1)$ with an edge of multiplicity $x$; see for instance \emph{Example 2} below.

The final step is to gauge the diagonal $\urm(1)$ flavour symmetry in the left quiver of \eqref{eq:T_rho_sigma_with_SU} to obtain $T_{\rho}^\sigma [\surm(N)] \slash \Z_{N_k}^t$. However, simply introducing a new $\urm(1)$ gauge node leads to the ambiguity of the global form of the product gauge group, which can either be $\Gcal=\urm(1) \times \surm(N_k)\times \prod_{i\neq k} \urm(N_i)$ or $\Gcal$ removed by a subgroup of its centre. For $\Gcal \slash \Z_{N_k}$, with $\Z_{N_k}$ embedded into $\surm(N_k)$ as centre and into the diagonal $\urm(1) \subset \urm(N_i)$ of each of the other gauge group factors, one obtains back the original theory $T_\rho^\sigma[\surm(N)]$, see \cite{Bourget:2020xdz} or Appendix \ref{app:discrete_gauging}. But this theory exhibits a $\Z_{N_k}^t$ topological symmetry, see also Section \ref{sec:comment_magQuiv}.
In order to generate the desired $T_{\rho}^\sigma [\surm(N)] \slash \Z_{N_k}^t$ theory with gauge group $\Gcal$, one needs to gauge the discrete $\Z_{N_k}^t$ symmetry, which effectively reduces the magnetic lattice to the standard integer lattice.
For the 3d mirror, this means that one first gauges a $\urm(1)_t$ topological symmetry, which effectively removes a $\urm(1)$ gauge degree of freedom. But one also needs to gauge a $\Z_{N_k}^f$ in a subsequent step. This $\Z_{N_k}^f$  can be thought of as embedded in the $\urm(1)$ that one has to be removed. 
Hence, the intermediate step is given by 
\begin{align}
T_{\rho}^\sigma [\surm(N)] \slash \Z_{N_k}^t
     \xleftrightarrow{\text{mirror}}
         \raisebox{-.5\height}{
    \includegraphics[page=11]{figures/figures_1-form_T_rho_sigma.pdf}
    }/// \left(\frac{\substack{\urm(1)_{\diag}  \times \urm(1)} }{ \Z_{N_k}^f} \right)
\end{align}
From the unframed quiver on the right, one has to ungauge a $\urm(1)_{\text{diag}}\times \urm(1)$ and also keep a $\Z_{N_k}^f$ gauged. The natural choice is to ungauge the two $\urm(1)$ gauge groups on top; thus, turning them into flavour groups up to a choice of $\Z_{N_k}^f$. The last step is to choose in which of the two $\urm(1)$s one embeds the $\Z_{N_k}^f$.  This is because, as with the $T[\surm(N)]$ theories, one knows the only difference between $T^\sigma_\rho [\surm(N)]^\vee$ and $\left(T^\sigma_\rho [\surm(N)] \slash \Z_{N_k}^t \right)^\vee$ should be the splitting of the flavour groups along with a discrete quotient. Schematically, one finds
\begin{align}
    &\raisebox{-.5\height}{
    \includegraphics[page=11]{figures/figures_1-form_T_rho_sigma.pdf}
    }///
    \left(\frac{\substack{\urm(1)_{\diag}  \times \urm(1)} }{ \Z_{N_k}^f} \right) \\
    &\quad\cong
    \raisebox{-.5\height}{
    \includegraphics[page=12]{figures/figures_1-form_T_rho_sigma.pdf}
    }\notag 
    %
   \quad  \cong \quad 
    \raisebox{-.5\height}{
    \includegraphics[page=13]{figures/figures_1-form_T_rho_sigma.pdf}
    } \notag 
\end{align}
and the two framed mirrors show that the discrete quotient can be applied diagonally on either one of the two sets of flavour hypermultiplets. This is also clear from Sections \ref{sec:SQED_discrete_topol} and \ref{sec:TSUN}, and Appendices \ref{app:mirror_map_SQED} and \ref{app:mirror_map_TSUN}, as an overall $\urm(1)$ rotation can be used to shuffle the discrete $\Z_{N_k}$ charges from one set of fundamental flavours to another.

\begin{figure}[ht]
    \centering
    \includegraphics[page=4,scale=1]{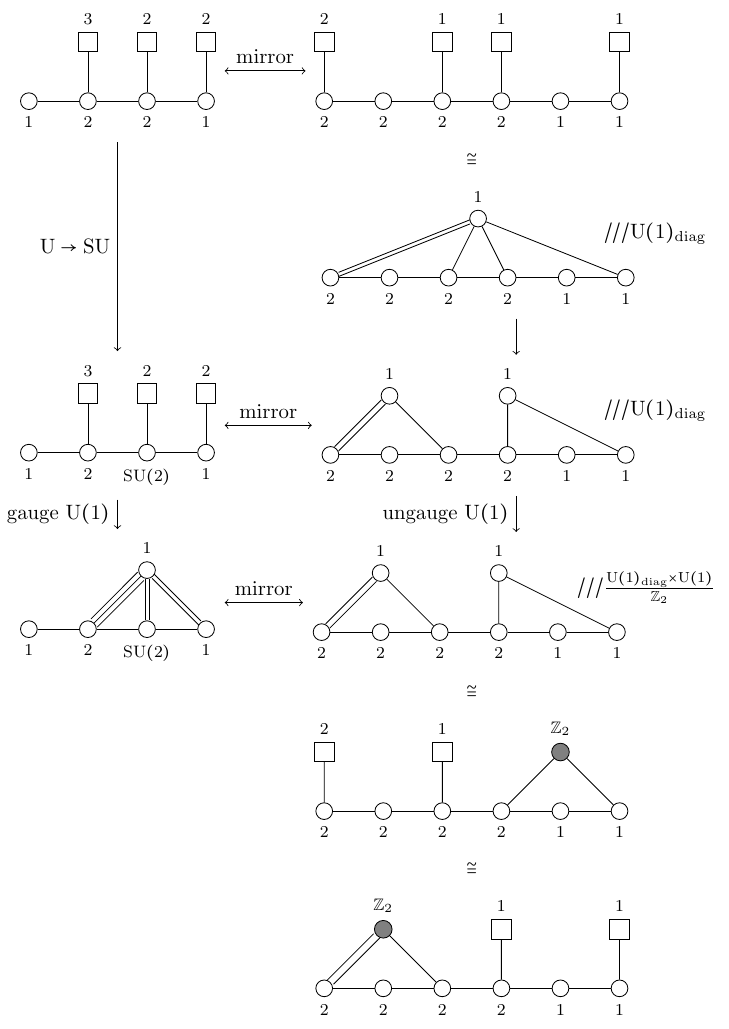}
    \caption{Starting from the mirror pair $T^{\sigma}_{\rho}[\surm(15)]$ and $T^{\rho}_{\sigma}[\surm(15)]$ with $\sigma=(3^3,2^2,1^2)$ and $\rho=(6,4,3,1^2)$, one can gauge a discrete $\Z_2$ 0-form symmetry to create a new mirror pair with $\Z_2$ 1-form symmetry. See also Appendix \ref{app:mirror_map_T-rho_sigma} for the choice of $\Z_2^f$ gauging.}
    \label{fig:T-rho-sigma-ex1}
\end{figure}
\paragraph{Example 1.}
One can apply the above procedure to $T^{\sigma}_{\rho}[\surm(15)]$ where $\sigma=(3^3,2^2,1^2)$ and $\rho=(6,4,3,1^2)$ for the example as in Figure \ref{fig:T-rho-sigma-ex1}.
The global form of the 0-form symmetry is expected to be 
\begin{align}
  G_t&(\text{bottom left quiver of Figure  \ref{fig:T-rho-sigma-ex1}})  = 
  G_f(\text{bottom right quiver(s) of Figure  \ref{fig:T-rho-sigma-ex1}})  \notag  \\ &\quad =
  \frac{\surm(2)\times\urm(1)_1}{\Z_2} \times \urm(1)_2 \times \urm(1)_3 \cong \urm(2) \times \urm(1)_2 \times \urm(1)_3
\end{align}
and one can explicitly verify this structure as demonstrated in \eqref{eq:HS_U1-U2-SU2-U1}. Alternatively, the Coulomb branch quiver indicates this isometry group as follows: only the leftmost $\urm(1)$ is balanced, leading to a topological $\surmL(2)^t$ because there are monopole operators of $\urm(1)$ magnetic flux $\pm1$ at $R$-charge 1 (see also \cite{Gaiotto:2008ak}). The remaining $\urm(2)$ and $\urm(1)$ gauge nodes provide one $\urm(1)_{i=1,2,3}^t$ topological symmetry factor each. Let the one associated with $\urm(2)$ be denoted by $\urm(1)_1^t$. Since this node is connected to the balanced node, arguments similar to \cite{Gaiotto:2008ak} show the existence of a chiral ring operator that transforms as a spinor under $\surmL(2)^t$ and has charge $\pm1$ under $\urm(1)_1^t$. Therefore, the $\Z_2$ centre action can be absorbed into $\urm(1)_1^t$, resulting in a $\urm(2)^t$ topological symmetry factor.

\begin{figure}[ht]
    \centering
    \includegraphics[page=5,scale=1]{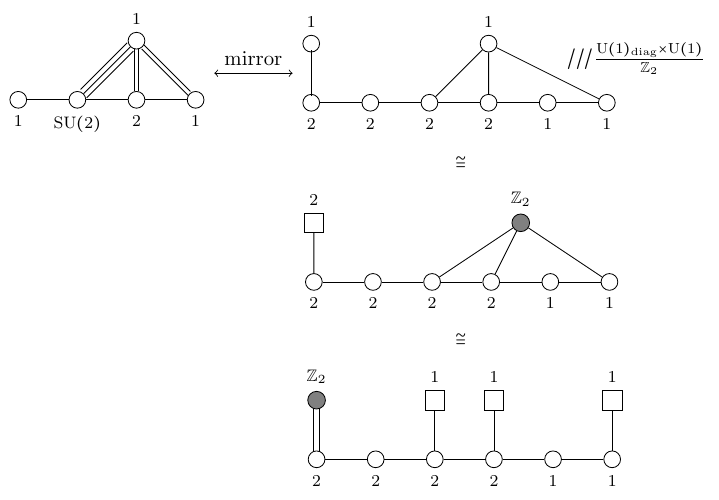}
    \caption{Again starting from the mirror pair $T^{\sigma}_{\rho}[\surm(15)]$ and $T^{\rho}_{\sigma}[\surm(15)]$ with $\sigma=(3^3,2^2,1^2)$ and $\rho=(6,4,3,1^2)$, one can gauge a different discrete $\Z_2$ 0-form symmetry to generate a another mirror pair with $\Z_2$ 1-form symmetry. See Appendix \ref{app:mirror_map_T-rho_sigma} for the choice of $\Z_2^f$ in the mirror.}
    \label{fig:T-rho-sigma-ex2}
\end{figure}

One can also choose the other $\surm(2)$ in $T^{\sigma}_{\rho}[\surm(15)]$ which gives the mirror pairs displayed in Figure \ref{fig:T-rho-sigma-ex2}.
Comparing Figure \ref{fig:T-rho-sigma-ex1} and \ref{fig:T-rho-sigma-ex2}, one observes that the global form of the 0-form symmetry in Figure \ref{fig:T-rho-sigma-ex2} is simply
\begin{align}
    \psurm(2)\times \prod_{i=1}^3 \urm(1)_i \,,
\end{align}
which is supported by the explicit calculations in \eqref{eq:HS_U1-SU2-U2-U1}. This conclusion can also be drawn by examining the Coulomb branch quiver.  Since the balanced $\urm(1)$ gauge node is not directly connected to any of the $\urm(2)$ or $\urm(1)$ gauge groups, there is no expectation on a chiral ring operator that transforms non-trivially under the $\Z_2$ centre of the $\surm(2)^t$ topological symmetry.

\paragraph{Example 2.}
Consider the mirror pair $T_{\rho}^{\sigma}[\surm(9)]$ with $\rho=(3,2^3)$ and $\sigma=(3^2,1^3)$
\begin{align}
  \raisebox{-.5\height}{
    \includegraphics[page=18]{figures/figures_1-form_T_rho_sigma.pdf}
    }
\qquad \longleftrightarrow\qquad
    \raisebox{-.5\height}{
    \includegraphics[page=19]{figures/figures_1-form_T_rho_sigma.pdf}
    }
    \label{eq:T-sigma-rho-ex3}
\end{align}
whose symmetry algebra is $\surmL(3)\oplus \urmL(1)$, as apparent from the balanced set of nodes. The global form is evaluated to be 
\begin{align}
G_t ( \text{LHS \eqref{eq:T-sigma-rho-ex3}}) = 
G_f ( \text{RHS \eqref{eq:T-sigma-rho-ex3}}) =
\frac{\surm(3)\times \urm(1)}{\Z_3}
\cong \urm(3)
\end{align}
because the $\urm(1)$ has charge $-1$ under the $\Z_3$ centre symmetry. See \eqref{eq:HS_T-sigma_rho_ex3} for details. Alternatively, the left-hand-side quiver in \eqref{eq:T-sigma-rho-ex3} allows us to derive this by using the balanced set of nodes. Since the  unbalanced gauge nodes connect to the $A_2$ Dynkin diagram (formed by the balanced nodes) on its first node, there exists a chiral ring operator transforming as $[1,0]\times (+1)$ (plus conjugate) under the topological $\surm(3)^t\times \urm(1)^t$. Thus, the $\Z_3$ centre can be compensated by suitable embedding into the $\urm(1)^t$ factor.

To create a new mirror pair, we can gauge a $\Z_2$ symmetry on both sides of the dual theories. For example, we can gauge the topological $\Z_2^t$ symmetry on $w_3$. The mirror map, as shown in \eqref{eq:mirror_map_T-sigma-rho_ex3}, indicates that gauging the $\Z_2^f$ symmetry leads to the following mirror pair
\begin{align}
 \raisebox{-.5\height}{
    \includegraphics[page=20]{figures/figures_1-form_T_rho_sigma.pdf}
    }
\quad \longleftrightarrow\quad 
    \raisebox{-.5\height}{
    \includegraphics[page=21]{figures/figures_1-form_T_rho_sigma.pdf}
    }
    \label{eq:T-sigma-rho-ex3_gauged}
\end{align}
whose symmetry algebra is $\surmL(2)\oplus \urmL(1)\oplus \urmL(1)$. The Hilbert series \eqref{eq:HS_T-sigma_rho_ex3_gauged} then suggests a symmetry group of 
\begin{align}
G_t ( \text{LHS \eqref{eq:T-sigma-rho-ex3_gauged}}) = 
G_f ( \text{RHS \eqref{eq:T-sigma-rho-ex3_gauged}}) =
\frac{\surm(2)\times \urm(1) \times \urm(1)}{\Z_2}
\cong \urm(2) \times \urm(1)
\end{align}
because the centre $\Z_2$ acts trivial on one $\urm(1)$ factor and with charge $-1$ on the other.
This can also be read off from the Coulomb branch quiver. As there is a $\urm(2)$ node connected to the balanced $\urm(2)$ node, there exists a chiral ring operator transforming as $[1]_{A_1} \times (\pm1)$ under the associated $\surm(2)^t\times \urm(1)^t$ topological symmetry factors. Therefore, the $\Z_2$ centre symmetry then gives rise to a $\urm(2)^t$ isometry factor. The other topological Cartan $\urm(1)^t$ is uncharged under the $\Z_2$ centre, as the gauge nodes are not connected to each other.

\paragraph{Gauging $\Z_q^t$ on a $\urm(1)$ node.}
Analogous to Section \ref{sec:TSUN}, one can also gauge discrete subgroups of the topological symmetry associated to a $\urm(1)$ gauge node. From the examples considered, it is clear what the theory after gauge the $\Z_q^t$ is: the same quiver as before, but all hypermultiplets connected to the specific $\urm(1)$ gauge node have now charge $q$. The question is then, what the corresponding mirror theory is. This can be determined by utilising the mirror map between the fugacities, as demonstrated in Appendix \ref{app:mirror_map}. 
\subsection{\texorpdfstring{$T[\sorm(2N)]$ theories}{TSO2N theories}}
\label{sec:TSO2N}
\begin{figure}[ht]
    \centering
 \includegraphics[page=1]{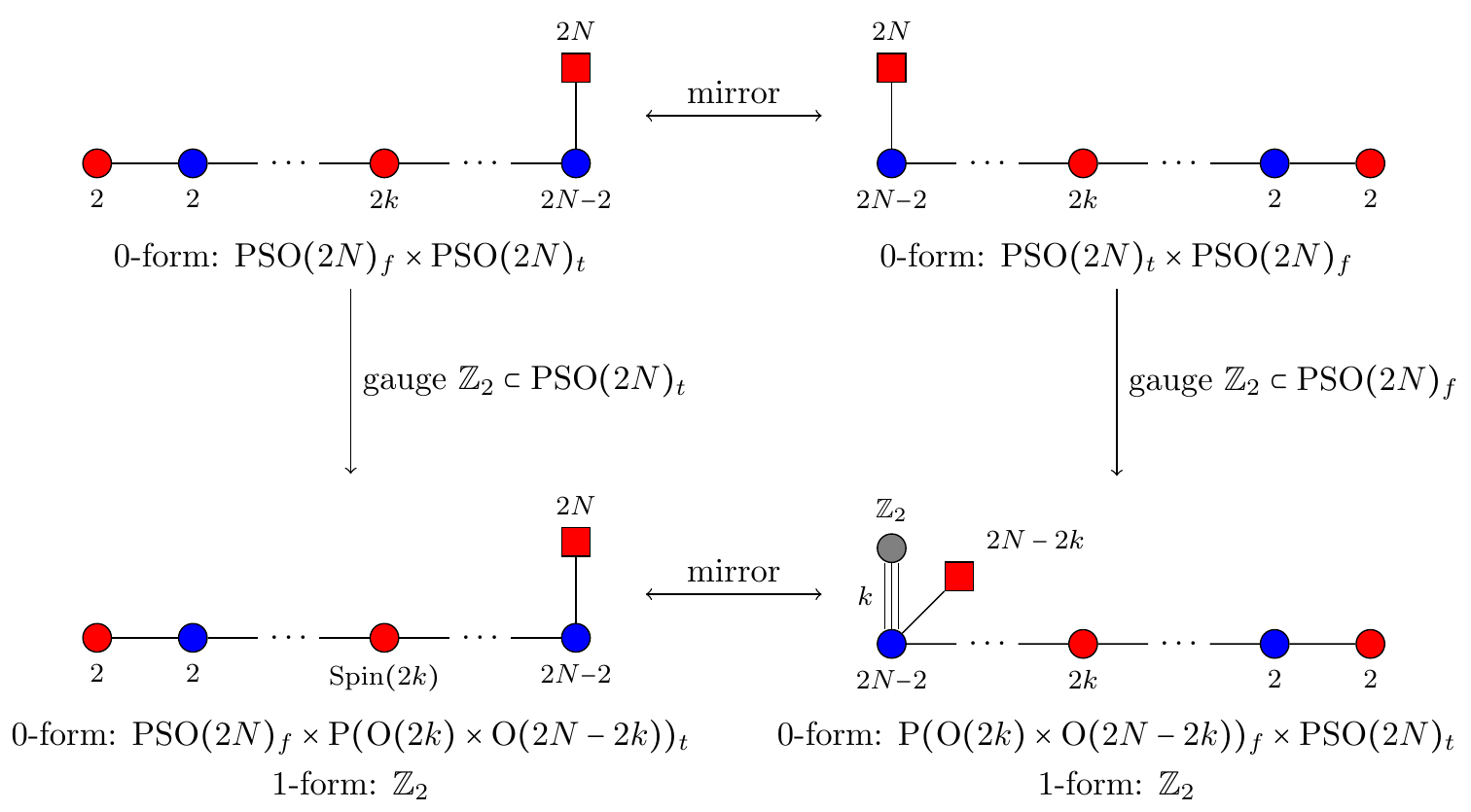}
        \caption{Gauging of discrete 0-form symmetries in $T[\sorm(2N)]$ theories. Gauging the $\Z_2^t$ for an $\sorm(2k)$ gauge leads to a $\spin(2k)$ gauge group. Gauging the mirror dual $\Z_2^f$ is realised by splitting the fundamental flavours into two sets: one set is uncharged and the other set is charged under $\Z_2^f$, indicated by an edge with multiplicity $k$ connected to a grey node.}
    \label{fig:TSO2N}
\end{figure}

In a similar vein to $T[\surm(N)]$, one can consider the self-mirror theory $T[\sorm(2N)]$ \cite{Gaiotto:2008ak}, see Figure \ref{fig:TSO2N}. For quiver theories composed of alternating $\sorm(n)$ and $\sprm(m)$ gauge nodes, only the $\Z_2$ factors of the $\sorm(n)$ gauge nodes are the discrete parts of the topological symmetry visible in the UV description. If we gauge any of these, we get a $T[\sorm(2N)]$-type quiver with a single replacement $\sorm(2k)\to\spin(2k)$\footnote{This follows as gauging the $\Z_2$ topological symmetry of an $\sorm(2n)$ gauge group leads to an $\spin(2k)$ gauge group. Conversely \cite{Kapustin:2014gua,Hsin:2020nts}, gauging the $\Z_2$ 1-form symmetry in $\spin(2k)$ recovers the $\sorm(2k)$ theory.}. The corresponding mirror theory is obtained from $T[\sorm(2N)]$ by gauging a suitable $\Z_2$ inside the flavour symmetry. This leads to a splitting of the flavour node as indicated in Figure \ref{fig:TSO2N}. Appendix \ref{app:TSO2N} provides examples and consistency checks for $T[\sorm(6)]$ and $T[\sorm(8)]$.

Considering the theory $T[\sorm(2N)]\slash \Z_2^t$ obtained from gauging $\Z_2^t$, the quiver description allows us to use the techniques of \cite{Apruzzi:2021mlh} to verify the 1-form symmetry and its interplay with the flavour 0-form symmetry. One finds the discrete groups summarised in Table \ref{tab:TSO2N_2-group} which constitute the short exact sequence
\begin{align}
    0 \to \Gamma^{[1]} \to \mathcal{E} \to Z \to 0 \,.
\end{align}
As expected, the flavour 0-form symmetry is always $\psorm(2N)$ since the flavour centre $\mathcal{Z}$ is maximal. Moreover, only for $T[\sorm(4N)]\slash \Z_2^t$ with a $\spin(4l+2)$  gauge node does the 1-form symmetry and the flavour 0-form symmetry form a non-trivial extension hinting to a 2-group symmetry. 
\begin{table}[t]
\centering
\begin{tabular}{l|ccc}
\toprule
theory     & $\Gamma^{[1]}$  & $\mathcal{E}$ & $Z$ \\ \midrule
$T[\sorm(4N)]\slash \Z_2^t$ with $\spin(4l)$ &  $\Z_2$   &  $\Z_2 \times \Z_2 \times \Z_2$  & $\Z_2\times \Z_2$  \\
$T[\sorm(4N)]\slash \Z_2^t$ with $\spin(4l+2)$     & $\Z_2$   &  $\Z_2 \times \Z_4 $  & $\Z_2\times \Z_2$  \\
$T[\sorm(4N+2)]\slash \Z_2^t$ with $\spin(4l)$     &  $\Z_2$   &  $\Z_2 \times \Z_4$  & $\Z_4$ \\
$T[\sorm(4N+2)]\slash \Z_2^t$ with $\spin(4l+2)$     &  $\Z_2$   &  $\Z_2 \times \Z_4$  & $\Z_4$  \\ \bottomrule
\end{tabular}
\caption{Interplay of 1-form symmetry and the flavour centre for $T[\sorm(2N)]\slash \Z_2^t$ theories.}
\label{tab:TSO2N_2-group}
\end{table}

Following \cite{Hsin:2020nts,Lee:2021crt}, it is straightforward to illustrate the 1-form symmetry and 2-group structure via line operators. For the $\spin(2l)$ gauge group, a Wilson line $W_s$ in the spinor representation cannot end on a local operator, because all half-hypermultiplets transform in the vector representation. For $l$ even, the tensor product of the spinor with itself contains a singlet; therefore, $W^2_s$ is equivalent to the identity line without the need for any local operator. The lines that cannot end generate the (Pontryagin dual of the) 1-form symmetry and there is no 2-group structure. For $l$ odd, the tensor product of the spinor with itself contains the vector.  Now $W_s^2$ is equivalent to a flavour Wilson line because it can end on a local operator build from the half-hypermultiplets. However, the vector representation is not an allowed representation of $\psorm(2N)$, which means that the $\Z_2$ 1-form symmetry forms a 2-group with the flavour symmetry. This is consistent with Table \ref{tab:TSO2N_2-group}.

On the other hand, in the theory $T[\sorm(2N)]/\Z_2^f$ obtained by gauging the $\Z_2^f$ symmetry, there are two distinct sets of flavour hypermultiplets, each forming half-hypermultiplets $H$ and $h$ in the vector-vector representation of $\sorm(2N{-}2k) \times \sprm(N{-}1)$ and $\sorm(2k) \times \sprm(N{-}1)$, respectively, i.e.\
\begin{align}
    \raisebox{-.5\height}{
    \includegraphics[page=2]{figures/figures_1-form_TSO2N.pdf}
    }  \label{eq:TSO2N_flavour_HB} 
    \;.
\end{align}
The only difference is that $h$ is also charged under $\Z_2$. As in Sections \ref{sec:SQED_discrete_topol} and \ref{sec:TSUN},  to study the global form of the flavour symmetry of this theory,  one can consider gauge-invariant operators. Using the invariant $\sprm(N{-}1)$ anti-symmetric tensor $J$, the standard mesons-type invariants are $HJH$ and $hJh$, both of which then transform in the adjoint representation $[0,1,\ldots,0]_D $ of $\sormL(2N{-}2k)$ and $\sormL(2k)$, respectively. Likewise, one can consider $ h J H$, which is $\sprm(N{-}1)$ gauge invariant, but not $\Z_2$ invariant due to the $\Z_2$ charge of $h$. Hence, $\Ocal = \mathrm{Sym}^2 (h JH)$ is indeed a gauge invariant operator transforming as $[2,0,\ldots,0]_{D_k} \otimes [2,0,\ldots,0]_{D_{N{-}k}}$. All of these gauge-invariant Higgs branch operators have trivial charges under the $\sormL(2k)$ or $\sormL(2N{-}2k)$ centre symmetries. This suggests that the global form of the flavour symmetry is $\psorm(2k)\times \psorm(2N{-}2k)$.

\subsection{\texorpdfstring{$\sprm(k)$ SQCD and its orthosymplectic mirror}{Sp(k) SQCD and its orthosymplectic mirror}}
The lessons learnt can be readily applied to other orthosymplectic quivers, such as $\sprm(k)$ SQCD with $N$ fundamental hypermultiplets and its orthosymplectic mirror quiver \cite{Feng:2000eq}.
Focusing on $N\geq 2k+1$, the SQCD theory admits a manifest flavour symmetry, while there is no topological symmetry for $N>2k+1$ and a $\urm(1)_t$ symmetry for $N=2k+1$. Thus, it is quite natural to consider gauging discrete subgroups of the flavour 0-form symmetry.
Conversely, the mirror orthosymplectic quiver does not have a continuous flavour symmetry for $N>2k_1$ (i.e.\ no mass parameter) and an $\sorm(2)_f$ symmetry for $N=2k+1$ (i.e.\ one mass parameter). While the topological symmetry is not manifest in the UV description, certain remnants are: each $\sorm(l)$ gauge group admits a manifest $\Z_2^t$ symmetry.

\begin{figure}[ht]
    \centering
    \hspace*{-1.5cm}
 \includegraphics[page=5]{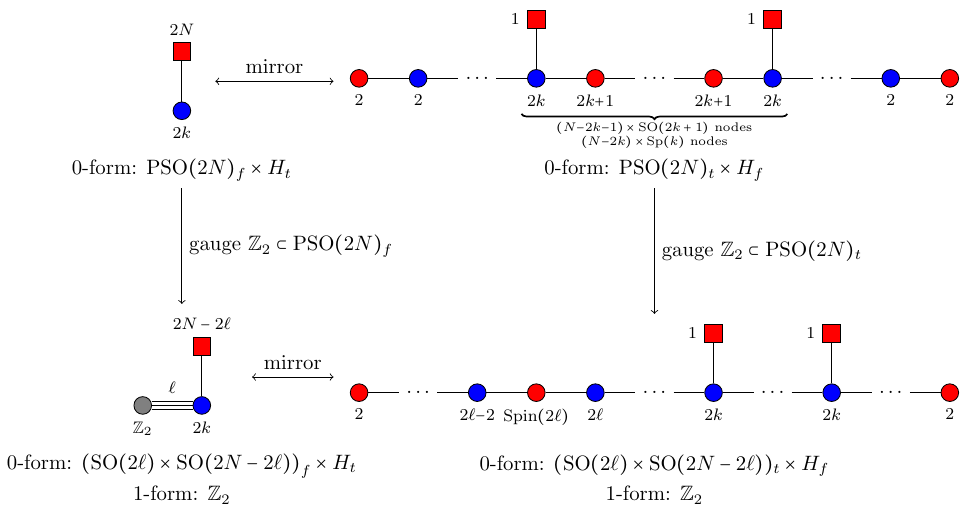}
        \caption{Gauging of discrete 0-form symmetries in $\sprm(k)$ SQCD with $N$ fundamentals and its linear orthosymplectic mirror quiver. Here, the isometry group $H_{t/f}$ is trivial for $N>2k+1$ and $\urm(1)$ for $N=2k+1$. Again, the split into two sets of flavours can made symmetric, as a global rotation can shift the $\Z_2^f$ action onto either set.}
    \label{fig:Sp_SQCD}
\end{figure}

Therefore, one can gauge a $\Z_2$ topological symmetry of a specific $\sorm(\ell)$ gauge node and inquire about the implications. It is straightforward to observe that this gauging modifies the particular gauge group $\sorm(\ell)\to \spin(\ell)$, see for instance \cite{Cremonesi:2013lqa,Hanany:2016ezz}. On the mirror side, one gauges a $\Z_2 \subset \sorm(2N)$ flavour symmetry, which then leads to a split of the flavour symmetry. This is summarised in Figure \ref{fig:Sp_SQCD}. Exemplary cases with explicit calculations are provided in Appendix \ref{app:Sp_SQCD}.

The interplay of the discrete $\Z_2$ 1-form symmetry with the continuous $0$-form symmetry is simple here. Consider the linear orthosymplectic mirror quiver.
For $N>2k+1$, there is no continuous $0$-form flavour symmetry that could mix with the 1-form symmetry $\Z_2$. For $N=2k+1$, there exists an enhance $\urm(1)$ 0-form symmetry, but the 1-form and 0-form symmetry are simply a product of each other.

\subsection{\texorpdfstring{$\sprm(k)$ SQCD and its unitary $D$-type mirror quiver}{Spk SQCD and its unitary D-type mirror quiver}}\label{sec:D-type}
It is well-known that $\sprm(k)$ SQCD with $N$ fundamental flavours admits a second mirror description \cite{Hanany:1999sj}, based on a $D_N$-type Dynkin quiver:
\begin{align}
    \raisebox{-.5\height}{
    \includegraphics[page=1]{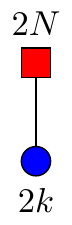}
    } 
    \quad \longleftrightarrow \quad
     \raisebox{-.5\height}{
    \includegraphics[page=2]{figures/figures_1-form_Sp_SQCD_uni.pdf}
    } 
    \label{eq:Sp_SQCD_unitary_mirror} 
    \;.
\end{align}
This mirror pair has the advantage that the $\psorm(2N)$ global symmetry is manifest as Higgs branch isometry in the SQCD theory and as Coulomb branch isometry in the $D$-type Dynkin quiver. It is, hence, natural to study gaugings of discrete $\Z_q$ symmetries in this manifest 0-form symmetry.

Starting with the Dynkin quiver, there are two distinct choices:
Firstly, gauging a $\Z_l$ on a $\urm(l)$ node which satisfies $2<l<2k$, one obtains\footnote{The cases $l=1,2$ are addressed separately below.} 
\begin{align}
  \raisebox{-.5\height}{
    \includegraphics[page=3]{figures/figures_1-form_Sp_SQCD_uni.pdf}
    } 
    \label{eq:D-type_SU_before_flavour}
\end{align}
which has a $\Z_l$ 1-form symmetry and the Coulomb branch isometry algebra is $\surmL(l)\oplus \urmL(1)_Q \oplus \sormL(N-l)$. For the global form, one can study the action of the centre symmetries of the non-abelian factors. One finds
\begin{align}
G_t \eqref{eq:D-type_SU_before_flavour} = \psurm(l) \times \frac{\urm(1)_Q \times \spin(2N-2l)}{ \Zcal_{D_{N-l}}}
\label{eq:sym_D-type_SU_before_flavour}
\end{align}
where the $\Zcal_{D_{N-l}}$ charges of $Q$ are given by the 
charges of the congruence class of the $j$-th fundamental representation $[0,\ldots,0,1,0,\ldots,0]_D$ with $j=2k-l$, see Appendix \ref{app:congruency} for details. For  explicit examples including Hilbert series computations see Appendix \ref{app:Sp_SQCD_D-type_mirror}.

Alternatively, gauging a $\Z_l$ on a $\urm(l)$ node which satisfies $l\geq 2k$, one obtains 
\begin{align}
  \raisebox{-.5\height}{
    \includegraphics[page=4]{figures/figures_1-form_Sp_SQCD_uni.pdf}
    } 
    \label{eq:D-type_SU_behind_flavour}
\end{align}
and the Coulomb branch isometry algebra is the same as in \eqref{eq:D-type_SU_before_flavour}. However, the ``extra'' $\urm(1)$ node is now attached to the balanced $A$-type Dynkin diagram such that the global form is given by
\begin{align}
G_t \eqref{eq:D-type_SU_behind_flavour} =  \frac{\psurm(l) \times \urm(1)_Q}{ \Z_l}
 \times \psorm(2N-2l)
 \label{eq:sym_D-type_SU_behind_flavour}
\end{align}
where $Q$  carries $\Z_l$ charge $2k \bmod l$, i.e.\ the 
charges of the congruence class of the $2k$-th fundamental representation, see Appendix \ref{app:congruency}. Explicit examples for this discrete gauging are given in Appendix \ref{app:Sp_SQCD_D-type_mirror}.

\paragraph{Global form via the mirror.}
Analogous to the discussion in Sections \ref{sec:SQED_discrete_topol}, \ref{sec:TSUN}, and \ref{sec:TSO2N}, one can confirm this global symmetry via the Higgs branch of the mirror theory.
The starting point is the mirror map \eqref{eq:mirror_map_D-type_Dynkin} between the flavour fugacities of $\sprm(k)$ SQCD and its unitary $D$-type Dynkin quiver, see Appendix \ref{app:mirror_map_SQED}. This allows us to identify which flavour fugacities are involved in the discrete $\Z_l^f$ gauging on the SQCD side. 
\begin{compactitem}
    \item \ul{$l<2k$:} The familiar argument then proceeds by splitting the fundamental flavours into two distinct groups: the first $l$ fundamental flavours are grouped as $X$, transforming as $\zeta_l Q^{-\frac{1}{l}} [1,0,\ldots,0]_{A_l}$, and the remaining $N-l$ fundamental flavours, transforming as $[1,0,\ldots,0]_{D_{N-l}}$. Building a gauge invariant Higgs branch operator proceeds in two steps: firstly, using the $\sprm(k)$ invariant tensors $J$ on constructs operators of the form $X J \tilde{X}$, which transform as $\zeta_l$ under the discrete symmetry. Secondly, $\Z_l$ invariance is achieved via $\Ocal = \mathrm{Sym}^l (X J \tilde{X})$, which transforms as $Q^{-1} [l,0,\ldots,0]_{A_{l-1}} \otimes [l,0,\ldots,0]_{D_{N-l}}$. The $\Z_l$ centre surely acts trivial on  $[l,0,\ldots,0]_{A_{l-1}}$, while the centre charges of $[l,0,\ldots,0]_{D_{N-l}}$ are $(0,l\mod 2)$ if $N-l$ is even or $(2l \bmod 4)$ if $N-l$ is odd. Thus, the non-trivial transformations under the centres can be compensated if $Q$ transforms as follows:
    \begin{subequations}
    \begin{alignat}{3}
N-l &= \text{even} & 
\qquad 
    \Z_l \times \Z_2 \times \Z_2 &\text{ charges of $Q$: } &\quad  &(0,0, l \bmod 2)
       \\
         N-l &= \text{odd} &
         \qquad 
    \Z_l \times \Z_4  &\text{ charges of $Q$: }
    & \quad &(0, 2l \bmod 4)
    \end{alignat}
    \end{subequations}
    which confirms \eqref{eq:sym_D-type_SU_before_flavour}. To see this, recall from Appendix \ref{app:congruency} that the congruence class of the $j$-th fundamental representation of $D_{N-l}$ with $j=2k-l$ is $(0,l\bmod 2)$ for $N-l$ even and $2l \bmod 4$ for $N-l$ odd.
    %
    \item \ul{$l>2k$:} The argument is slightly modified: the first set $X$ of flavours transforms as $\zeta_{2k} Q^{-\frac{1}{2k}} [1,0,\ldots,0]_{A_{l-1}}$, while the second set $\tilde{X}$ transforms as $[1,0,\ldots,0]_{D_{N-l}}$. The Higgs branch operator $\Ocal = \mathrm{Sym}^{2k} (X J \tilde{X})$ transforms as $Q^{-1} [2k,0,\ldots,0]_{A_{l-1}} \otimes [2k,0,\ldots,0]_{D_{N-l}}$, which has trivial $D$-type centre charges. To see this, for $N-l$ even, the $\Z_2\times \Z_2$ charges are $(0, 2k \bmod 2) =(0,0)$; while for $N-l$ odd, the $\Z_4$ charge is $2 \cdot 2k \bmod 4 =0$. Thus, to compensate potential irreps that are  non-trivial under $\Z_{2k}$, one requires that $Q$ has the following charges:
       \begin{subequations}
    \begin{alignat}{3}
N-l &= \text{even} & 
\qquad 
    \Z_l \times \Z_2 \times \Z_2 &\text{ charges of $Q$: } &\quad  &(2k\bmod l,0, 0)
       \\
         N-l &= \text{odd} &
         \qquad 
    \Z_l \times \Z_4  &\text{ charges of $Q$: }
    & \quad &(2k \bmod l, 0)
    \end{alignat}
    \end{subequations}
    which then confirms \eqref{eq:sym_D-type_SU_behind_flavour}.
\end{compactitem}

 \paragraph{Two special cases.}
In the $l=2$ case of \eqref{eq:sym_D-type_SU_before_flavour}, a symmetry enhancement is observed in the explicit computations \eqref{eq:HS_D7_node2} and \eqref{eq:HS_D8_node2}. These show that there is not only the expected $\surmL(2)^t$, but the topological Cartan symmetry of the ``new'' $\urm(1)$ gauge node is also enhanced to a non-abelian $\surmL(2)^t$. These two $\surmL(2)^t$ symmetries can both be interpreted as $\psorm(4)^t \cong \psorm(3)^t \times \psorm(3)^t$.

As in previous sections, one can also gauge a discrete $\Z_q^t$ along the topological fugacity $w_1$ associated to the first $\urm(1)$ gauge node. The $D$-type Dynkin quiver is modified in the by now familiar way: the bifundamental of $\urm(1)\times \urm(2)$ turns into a hypermultiplet that transforms as fundamental under $\urm(2)$ but is of $\urm(1)$ charge $q$. In the mirror theory, the $\Z_q^f$ acts on a single fundamental flavour, as dictated by the mirror map \eqref{eq:mirror_map_D-type_Dynkin}. In summary, the mirror pair with $\Z_q$ 1-form symmetry is
\begin{align}
    \raisebox{-.5\height}{
    \includegraphics[page=5]{figures/figures_1-form_Sp_SQCD_uni.pdf}
    } 
    \quad \longleftrightarrow \quad
     \raisebox{-.5\height}{
    \includegraphics[page=6]{figures/figures_1-form_Sp_SQCD_uni.pdf}
    } 
    \;.
\end{align}
and the global Higgs / Coulomb branch isometry is
\begin{align}
    G= \frac{\urm(1)_Q \times \spin(2N-2)}{ \Zcal_{D_{N-1}}} \;,\quad 
    \text{$\Zcal_{D_{N-1}}$ charges of $Q$ }
    \begin{cases}
    (0,q\bmod 2)\,, & N=\text{even} \\
        2q\bmod 4\,, & N=\text{odd} 
    \end{cases}
\end{align}
where $Q$ is the topological fugacity of the left-most $\urm(1)$ gauge node.
\subsection{Examples of non-simply laced unitary quivers and their mirrors}
The last class of quiver theories considered here are non-simply laced unitary quivers\footnote{See, for example, \cite{Hanany:2001iy,Bourget:2021siw} for the appearance of such quiver theories via branes and ON planes.}, whose monopole formula has been proposed in \cite{Cremonesi:2014xha}. Consider the following example
\begin{align}
  \raisebox{-.5\height}{
    \includegraphics[page=1]{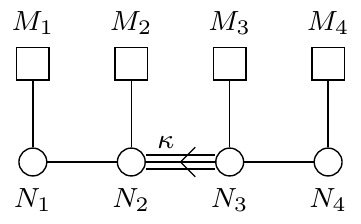}
    } 
    \label{eq:non-simply_laced}
\end{align}
with nodes $\urm(N_{1,2})$ on the ``short'' side and $\urm(N_{3,4})$ on the ``long'' side; wherein the naming is borrowed from Dynkin diagrams. The multiplicity of the non-simply laced edge is denoted by $\kappa$.  
Even though these quiver theories are non-Lagrangian (hence superconformal index and Higgs branch Hilbert series are not computable by the standard methods), we can still study their Coulomb branch using Hilbert series techniques. This allows us to investigate the effects of gauging a discrete $\Z_q^t$ symmetry.

\paragraph{Gauging at the long side.}
To begin with, attempt to gauge a discrete $\Z_{N_3}^t$ topological symmetry associated to the $\urm(N_3)$ node at the long side, with $N_3 >1$. As a first step, one rewrites \eqref{eq:non-simply_laced} by expressing $\urm(N_3)\cong (\surm(N_3)\times \urm(1))\slash \Z_{N_3}$. By analogous arguments as in Appendix \ref{app:discrete_gauging}, one arrives at
\begin{align}
     \label{eq:non-simply_laced_RHS_full}
 & \left[ \raisebox{-.5\height}{
    \includegraphics[page=3]{figures/figures_1-form_non-simply_laced.pdf}
    }  \right] \slash \Z_{N_3} 
     \qquad \text{with }
    (\bm{m}_1,\bm{m}_2,\bm{l}, \bm{m}_4,h) \in \widetilde{\Gamma}\\
    \widetilde{\Gamma} &\coloneqq
    \bigcup_{a=0}^{N_3-1} 
    \left( \Z + \tfrac{\kappa \cdot a}{N_3} \right)^{N_1}
    \times 
    \left( \Z + \tfrac{\kappa \cdot a}{N_3} \right)^{N_2}
    \times 
    \left( \Z + \tfrac{a}{N_3} \right)^{N_3-1}
    \times 
    \left( \Z + \tfrac{a}{N_3} \right)^{N_4}
    \times 
    \left( \Z + \tfrac{a}{N_3} \right) \notag 
\end{align}
and the Coulomb branch moduli space is the same as that of \eqref{eq:non-simply_laced}. The green edges transform in the fundamental representation of $\urm(N_{1,2})$ and with charge $\kappa$ under the $\urm(1)$.
Next, the $\Z_{N_3}$ symmetry is gauged.
One obtains the following quiver description: 
\begin{align}
  &\raisebox{-.5\height}{
    \includegraphics[page=3]{figures/figures_1-form_non-simply_laced.pdf}
    } 
     \qquad \text{with }
    (\bm{m}_1,\bm{m}_2,\bm{l}, \bm{m}_4,h) \in \Gamma 
    \label{eq:non-simply_laced_RHS} \\
      \Gamma &\coloneqq
    \bigcup_{a=0}^{N_3-1} 
   \Z^{N_1}
    \times 
     \Z^{N_2}
    \times 
     \Z^{N_3-1}
    \times 
     \Z^{N_4}
    \times 
     \Z  \notag 
\end{align}
where $\Gamma$ is again short-hand for the integer magnetic lattice. This theory exhibits a $\Z_{N_3}$ 1-form symmetry, by construction.

\paragraph{Gauging at the short side.}
Now, consider gauging a $\Z_{N_2}^t$ on the topological fugacity associated to the $\urm(N_2)$ gauge node, with $N_2>1$. Again, the first step is to simply rewrite $\urm(N_2) \cong (\surm(N_2)\times \urm(1))\slash \Z_{N_2}$. By adopting the arguments of Appendix \ref{app:discrete_gauging}, one finds
\begin{align}
 \label{eq:non-simply_laced_LHS_full}
&\left[\raisebox{-.5\height}{
    \includegraphics[page=2]{figures/figures_1-form_non-simply_laced.pdf}
    }  \right] \slash \Z_{N_2}
    \qquad \text{with }
    (\bm{m}_1,\bm{l},\bm{m}_3, \bm{m}_4,h)
    \in   \widehat{\Gamma} 
     \\
    \widehat{\Gamma} &=
    \bigcup_{a=0}^{N_2 \cdot \kappa-1} 
    \left(\Z + \tfrac{a}{N_2} \right)^{N_1}
    \times
    \left(\Z + \tfrac{a}{N_2} \right)^{N_2-1}
    \times
    \left(\Z + \tfrac{a}{N_2 \cdot \kappa} \right)^{N_3}
    \times
     \left(\Z + \tfrac{a}{N_2 \cdot \kappa} \right)^{N_4}
      \times
    \left(\Z + \tfrac{a}{N_2 \cdot \kappa} \right)
     \notag 
\end{align}
whose Coulomb branch coincides with that of \eqref{eq:non-simply_laced}.
Moreover, the edges highlighted in green transform in the fundamental representation of $\urm(N_{1,2})$ and with charge $\kappa$ under the $\urm(1)$ node.  As a next step, gauging the $\Z_{N_2}$ results in the following theory: 
\begin{align}
  &\raisebox{-.5\height}{
    \includegraphics[page=2]{figures/figures_1-form_non-simply_laced.pdf}
    } 
    \qquad \text{with }
    (\bm{m}_1,\bm{l},\bm{m}_3, \bm{m}_4,h)
    \in   \Gamma
    \label{eq:non-simply_laced_LHS}
    \\
      \Gamma &=
    \bigcup_{a=0}^{ \kappa-1} 
    \left(\Z + a \right)^{N_1}
    \times
    \left(\Z + a \right)^{N_2-1}
    \times
    \left(\Z + \tfrac{a}{ \kappa} \right)^{N_3}
    \times
     \left(\Z + \tfrac{a}{ \kappa} \right)^{N_4}
        \times
    \left(\Z + \tfrac{a}{\kappa} \right)
     \notag
\end{align}
where $\Gamma$ is a short-hand notation for several shifted copies of the standard integer lattice of the magnetic charges. 

\paragraph{Comment.}
One could also gauge a discrete $\Z_q^t$ along the topological Cartan $\urm(1)$ of a $\urm(1)$ gauge node. In this case, the connected hypermultiplets are modified to have charge $q$ under the $\urm(1)$, but no other changes to the quiver occur.

\subsubsection{$C$-type quivers}
A representative example is the mirror pair of $\orm(2k)$ SQCD with $N$ hypermultiplets in the vector representation and its $C$-type Dynkin mirror quiver 
\begin{align}
 \raisebox{-.5\height}{
    \includegraphics[page=4]{figures/figures_1-form_non-simply_laced.pdf}
    } 
    \qquad \longleftrightarrow \qquad
     \raisebox{-.5\height}{
    \includegraphics[page=5]{figures/figures_1-form_non-simply_laced.pdf}
    } \label{eq:O(2n)_SQCD}
\end{align}
which can be realised by a systems of D3-D5-NS5 branes with O5 and ON planes, respectively.
The logic is the same as before: Choose a $\Z_q^t$ in the $C$-type Dynkin quiver, by selecting a gauge node and its associated topological fugacity. Using the mirror map \eqref{eq:mirror_map_C-type_Dynkin} for \eqref{eq:O(2n)_SQCD} one identifies how the $\Z_q^f$ acts on the vectors.
For concreteness, consider examples for $k=1$ and $N=4$: 

\paragraph{Example: gauging on the long side.}
Gauging a $\Z_2^t$ on the fourth node yields the mirror pair (i.e.\ using \eqref{eq:mirror_map_C-type_Dynkin} with discrete variable on $w_4$)
\begin{align}
 \raisebox{-.5\height}{
    \includegraphics[page=6]{figures/figures_1-form_non-simply_laced.pdf}
    } 
    \qquad \longleftrightarrow \qquad 
     \raisebox{-.5\height}{
    \includegraphics[page=7]{figures/figures_1-form_non-simply_laced.pdf}
    } 
    \label{eq:C-type_ex1}
\end{align}
where the `new' $\urm(1)$ node is connected with a hypermultiplet of charge $2$. The global symmetry algebra is $\surmL(4) \oplus \urmL(1)$, as read off from the balanced set of nodes. Explicit Hilbert series \eqref{eq:HS_C4_quiver_SU_4th} show that the global form is
\begin{align}
G_f(\text{LHS }\eqref{eq:C-type_ex1}) 
= G_t(\text{RHS }\eqref{eq:C-type_ex1}) = \psurm(4) \times \urm(1)_Q
\end{align}
because the $\Z_4$ centre acts trivial on all appearing representations.

\paragraph{Example: gauging on the short side.}
Gauging a $\Z_2^t$ on the third gauge node results in the new mirror pair (i.e.\ using \eqref{eq:mirror_map_C-type_Dynkin} with discrete variable on $w_3$)
\begin{align}
 \raisebox{-.5\height}{
    \includegraphics[page=8]{figures/figures_1-form_non-simply_laced.pdf}
    } 
    \qquad \longleftrightarrow \qquad 
     \raisebox{-.5\height}{
    \includegraphics[page=9]{figures/figures_1-form_non-simply_laced.pdf}
    } 
    \label{eq:C-type_ex2}
\end{align}
the global symmetry algebra is $\surmL(3)\oplus \urmL(1)\oplus \sprmL(1)$, as suggested by the balanced nodes in the unitary quiver. Recalling the maximal subalgebra $\surmL(3)\oplus \urmL(1)\subset \sprmL(3)$, an analysis of the Hilbert series then suggests that the global form is 
\begin{align}
G_f(\text{LHS }\eqref{eq:C-type_ex2}) 
= G_t(\text{RHS }\eqref{eq:C-type_ex2}) =  \psprm(3) \times \psprm(1) \,.
\end{align}
See \eqref{eq:HS_C4_quiver_SU_3rd} for explicit computations.

\paragraph{Example: gauging on the short side.}
Gauging a $\Z_2^t$ on the second gauge node results in the new mirror pair (i.e.\ using \eqref{eq:mirror_map_C-type_Dynkin} with discrete variable on $w_2$)
\begin{align}
 \raisebox{-.5\height}{
    \includegraphics[page=10]{figures/figures_1-form_non-simply_laced.pdf}
    } 
    \qquad \longleftrightarrow \qquad 
     \raisebox{-.5\height}{
    \includegraphics[page=11]{figures/figures_1-form_non-simply_laced.pdf}
    } 
    \label{eq:C-type_ex3}
\end{align}
and the balanced set of nodes suggests the symmetry algebra $\surmL(2) \oplus \urmL(1) \oplus \sprmL(2)$. A Hilbert series computation 
\eqref{eq:HS_C4_quiver_SU_2nd} then indicates the following symmetry group
\begin{align}
G_f(\text{LHS }\eqref{eq:C-type_ex3}) 
= G_t(\text{RHS }\eqref{eq:C-type_ex3}) =  \psprm(2) \times \psprm(2) \,.
\end{align}
This suggests that the $\surmL(2)\oplus \urmL(1)$ realise a maximal subalgebra in one $\sprmL(2)$ factor.

\subsubsection{\texorpdfstring{$B$-type quivers}{B-type quivers}}
Alternatively, we could consider an $\sprm(k)$ gauge theory with $\sorm(2n+1)$ flavour symmetry. However, to prevent a parity anomaly, we would need to include a suitable Chern-Simons term. The Higgs branch, which is not affected by Chern-Simons levels, is known to be the closure of a $B$-type nilpotent orbit. Therefore, a natural mirror theory would be a $B$-type Dynkin quiver, for which analogous arguments apply as above.

\subsubsection{\texorpdfstring{A comment on $F_4$ Coulomb branch quivers}{A comment on F4 Coulomb branch quivers}}
The reasoning can be also applied to other non-simply laced Coulomb branch quivers, even if there may not exist a known mirror. Such an example is the $F_4$ Coulomb branch quiver of \cite{Hanany:2017ooe}. 
Table \ref{tab:min_F4_gauging_topol} summarises the resulting theories after a suitable $\Z_n^t$ is gauged, following the prescriptions \eqref{eq:non-simply_laced_LHS} and \eqref{eq:non-simply_laced_RHS}.

Here, a few remarks in comparison to the ``ungauging scheme'' of \cite{Hanany:2020jzl} are in order. The ungauging scheme involves removing a $\urm(1)$ factor from a selected $\urm(n)$ gauge group, which in the context of the monopole formula means setting one of the magnetic charges to zero.  For simply-laced quivers, this procedure leads to the same consequence as replacing a $\urm(n)$ gauge group with an $\surm(n)$ and quotienting out a diagonal $\mathbb{Z}_n$. 

However, the ungauging scheme becomes problematic when applied to a node on the short side of non-simply laced quivers. If the short node is a $\urm(1)$ gauge group, then the ungauging simply converts it into a flavour group. In \cite{Hanany:2020jzl}, the ungauging of the short $\urm(1)$ node in the $F_4$ quiver leads to a Coulomb branch that is the next-to-next-to minimal nilpotent orbit closure of $\sormL(9)$. In contrast, if the short node is non-abelian, such as the $\urm(2)$ node in the $F_4$ quiver, the resulting moduli space cannot be identified with any known space and the procedure has been argued to be ``invalid'' in \cite{Hanany:2020jzl}.

On the other hand, by replacing the short $\urm(2)$ node with an $\surm(2)$ and following the prescriptions in \eqref{eq:non-simply_laced_LHS} and \eqref{eq:non-simply_laced_RHS}, one is able to obtain consistent results,
as shown in the fourth row of Table \ref{tab:min_F4_gauging_topol}. The resulting Coulomb branch is the next-to-next-to minimal nilpotent orbit closure of $\mathfrak{so}(9)$ as well.\footnote{In general, for non-simply laced quivers, the prescriptions \eqref{eq:non-simply_laced_LHS} do not always provide the same Coulomb branch for all the short nodes.}
It is to be noted, that if one uses the prescriptions \eqref{eq:non-simply_laced_LHS_full} and \eqref{eq:non-simply_laced_RHS_full}, then one recovers the original minimal nilpotent orbit closure of $F_4$.

\begin{table}[t]
\ra{1.5}
    \centering
    \begin{tabular}{c|c|l}
    \toprule
       quiver  & symmetry   &  Coulomb branch Hilbert series \\ \midrule
     \raisebox{-.5\height}{
    \includegraphics[page=18,scale=0.8]{figures/figures_1-form_non-simply_laced.pdf}
    }     & $F_4$ &\parbox{8cm}{ $1+ \chi_{1,0,0,0} t + \chi_{2,0,0,0} t^2 + \chi_{3,0,0,0} t^3 + \ldots \\
    = 1 + 52 t + 1053 t^2 + 12376 t^3 + \ldots$} \\ \midrule[1pt]
         \raisebox{-.5\height}{
    \includegraphics[page=19,scale=0.8]{figures/figures_1-form_non-simply_laced.pdf}
    }  &  $A_1 \times C_3$  & \parbox{8cm}{ $1 + t(\chi_{2,0,0} + \phi_{2})   +  t^2(1+ \chi_{4,0,0} + \chi_{0,2,0}  +\chi_{2,0,0} \phi_{2} +\chi_{0,0,2} \phi_{2} + \phi_{4})  + \ldots \\
    = 1+ 24 t + 537 t^2 +\ldots$} \\
       \midrule
         \raisebox{-.5\height}{
    \includegraphics[page=20,scale=0.8]{figures/figures_1-form_non-simply_laced.pdf}
    }  &  $A_2 \times A_2$  & \parbox{8cm}{ $1 + t(\chi_{1,1} + \phi_{1,1})   +  t^2(1+ \chi_{1,1} + \chi_{2,2}  +\chi_{1,1} \phi_{1,1}  +\chi_{2,2} \phi_{1,1} + \phi_{2,2} + \phi_{1,1})  + \ldots \\
    =1+ 16t + 351 t^2 + \ldots$ }\\
          \midrule
             \raisebox{-.5\height}{
    \includegraphics[page=21,scale=0.8]{figures/figures_1-form_non-simply_laced.pdf}
    }  &  $A_3 \times A_1 \subset B_4$  & \parbox{8cm}{ $1 + t(\chi_{2} + \chi_{2} \phi_{0,1,0} + \phi_{1,0,1})   
           +  t^2(1
           + \chi_{4} 
           + \chi_{2} \phi_{2,0,0} +\phi_{0,1,0} 
           + \chi_{2} \phi_{0,1,0}
            + \chi_{4} \phi_{0,1,0}
            +\phi_{0,2,0}
            + \chi_{4} \phi_{0,2,0}
            +\phi_{1,0,1}
            + 2\chi_{2} \phi_{1,0,1}
            +\chi_{2} \phi_{1,1,1}
            +\chi_{2} \phi_{0,0,2}
            + \phi_{2,2})  + \ldots \\ =1+ 36t + 621 t^2 + \ldots$} \\
          \midrule[1pt]
            \raisebox{-.5\height}{
    \includegraphics[page=22,scale=0.8]{figures/figures_1-form_non-simply_laced.pdf}
    } 
    &  $C_3 \times U_1 \subset C_4$  & \parbox{8cm}{ $
    1+t (Q \chi _{1,0,0}+\frac{\chi _{1,0,0}}{Q}+\chi _{0,1,0}+1)
    +t^2 (Q^2 \chi _{2,0,0}+\frac{\chi _{2,0,0}}{Q^2}+Q \chi _{0,0,2}+Q \chi _{1,0,0}+Q \chi _{1,1,0}+\frac{\chi _{0,0,2}}{Q}+\frac{\chi _{1,0,0}}{Q}+\frac{\chi _{1,1,0}}{Q}+\chi _{0,0,2}+2 \chi _{0,1,0}+\chi _{0,2,0}+\chi _{2,0,0}+1)+ \ldots \\
    =1+ 36t + 621 t^2 + \ldots $} \\
          \midrule
     \raisebox{-.5\height}{
    \includegraphics[page=23,scale=0.8]{figures/figures_1-form_non-simply_laced.pdf}
    } 
    &  $C_3 \times U_1 $  & \parbox{8cm}{ $
    1+t (\chi _{0,1,0}+1) +
    t^2 (Q \chi _{1,0,1}+\frac{\chi _{1,0,1}}{Q}+\chi _{0,0,2}+2 \chi _{0,1,0}+\chi _{0,2,0}+\chi _{2,0,0}+1)
    +
    \ldots \\
    =1+ 22t + 369 t^2 + \ldots$} \\
          \midrule
               \raisebox{-.5\height}{
    \includegraphics[page=24,scale=0.8]{figures/figures_1-form_non-simply_laced.pdf}
    } 
    &  $C_3 \times U_1 $  & \parbox{8cm}{ $
    1+t (\chi _{0,1,0}+1)+
    t^2 (Q \chi _{2,0,0}+\frac{\chi _{2,0,0}}{Q}+\chi _{0,0,2}+2 \chi _{0,1,0}+\chi _{0,2,0}+\chi _{2,0,0}+1)
    +
    \ldots  \\
    =1+ 22t + 327 t^2 + \ldots$} \\
       \bottomrule
    \end{tabular}
    \caption{ The $F_4$ Coulomb branch quiver and its $\Z_q$ gaugings. The first row is the standard $F_4$ quiver proposed in \cite{Hanany:2017ooe}. Rows 2 - 4 display different choices of gauging a $\Z_N^t$ of a $\urm(N)$ node in the Coulomb branch quiver for the minimal nilpotent orbit closure of $F_4$. The gaugings on the ``long'' side produce global symmetries given by the balanced set of nodes. For the gauging on the ``short'' side, the global $\sormL(9)$ symmetry is only visible via the subalgebra $\surmL(4)\oplus \surmL(2)$.
    Rows 5 - 7 display the effects of gauging a $\Z_q^t$ inside the topological Cartan factor of the $\urm(1)$ gauge node. For $q=2$, the symmetry algebra is enhance from $\sprmL(3) \times \urmL(1)$ to $\sprmL(4)$; while for $q>2$, the algebra is simply  $\sprmL(3) \times \urmL(1)$. In the Hilbert series expressions,
    $\chi$ and $\phi$ are characters for the non-abelian symmetry factors and $Q$ is a $\urm(1)$ fugacity. }
    \label{tab:min_F4_gauging_topol}
\end{table}

\subsection{Magnetic quivers and gauging discrete topological symmetries}
\label{sec:comment_magQuiv}
Suppose that one is given an unframed unitary magnetic quiver $\Tcal$ with only simply-laced edges (i.e.\ bifundamental hypermultiplets between the unitary gauge nodes). To evaluate the Hilbert series or the index,  it is necessary to remove an overall $\urm(1)$ gauge group factor. In \cite{Bourget:2020xdz}, it was emphasised that choosing this $\urm(1)$ from a $\urm(k)$ gauge node leads to an $\surm(k)$ gauge node, but the magnetic lattice $\Gamma$ is extended to include shifted versions of the form $\bigcup_{i=0}^{k-1} \left( \Gamma + \frac{i}{k}\right)$. This situation can also be understood from a complementary perspective.

Given an unframed unitary magnetic quiver, pick a $\urm(k)$ gauge node and rewrite it as $\urm(k) \cong \frac{\surm(k)\times\urm(1)}{\Z_k}$, with fluxes $(\bm{l},h) \in \bigcup_{i=0}^{k-1} \left( \left(\Z+\frac{i}{k}\right)^{k-1},\;\left(\Z+\frac{i}{k}\right) \right) $. The aim is to remove this $\urm(1)$ factor. As demonstrated in Appendix \ref{app:discrete_gauging}, this rewriting shifts all other magnetic fluxes $\bm{m}$ by the flux $h$ associated to the $\urm(1)$;  as a result, all magnetic fluxes receive the shifts $\Gamma+\frac{i}{k}$ simultaneously. Now, removing this $\urm(1)$ means treating it as a background vector multiplet. Nevertheless, all remaining magnetic fluxes are still subject to the shifts $\Gamma+\frac{i}{k}$. Hence, the Coulomb branch Hilbert series, as well as the index for $\Tcal$, have the form
\begin{align}
    \Fcal_\Tcal =\sum_{(\bm{l},\bm{m})\in \bigcup_{i=0}^{k-1}(\Gamma +\frac{i}{k})} f(\bm{l},\bm{m})
\end{align}
which is message conveyed in \cite{Bourget:2020xdz}.

It turns out that one can refine $\Fcal_\Tcal$ by introducing a $\Z_k$-valued fugacity $z$ as follows:  the $\urm(1)_t$ topological symmetry of the $\urm(k)$ node appears in both the monopole formula and the index through the factor $w_k^{\sum_{a=1}^k m_a}$. Upon rewriting into magnetic fluxes $(\bm{l},h)$ for $\left(\surm(k)\times\urm(1)\right)\slash \Z_k$, this becomes $w_k^{k\cdot h}$. Since $h\in \bigcup_{i=0}^{k-1} (\Z+\frac{i}{k})$, one has $w_k^{k\cdot  n + i}$ for $h= n+\frac{i}{k} \in (\Z+\frac{i}{k})$ and some $n\in \Z$. This means that one can introduce a discrete fugacity $z$ to keep track of the $\Z_k$ centre symmetry, setting $w_k \to z $ such that $w_k^{k\cdot n + i} = z^i$. This fugacity remains even if the $\urm(1)$ is taken to be non-dynamical. One ends up with
\begin{align}
    \Fcal_{\Tcal}(z) =\sum_{i=0}^{k-1} z^{i}  \sum_{(\bm{l},\bm{m})\in (\Gamma +\frac{i}{k})} f(\bm{l},\bm{m}) \,.
\end{align}
It is now clear what happens if this discrete $\Z_k^t$ topological symmetry is gauged: the entire range of the summation collapses to the $i=0$ sector, i.e., the integer lattice
\begin{align}
   \Fcal_{\Tcal\slash \Z_k^t}= \frac{1}{k} \sum_{i=0}^{k-1} \Fcal_\Tcal \left(z=(\zeta_k)^i\right)
    =\sum_{(\bm{l},\bm{m})\in \Gamma } f(\bm{l},\bm{m}) \,.
\end{align}
Consequently, the quiver theory, in which $\urm(k)$ is replaced by an $\surm(k)$ and the magnetic lattice is simply the integer lattice, is obtained from the unframed unitary quiver $\Tcal$ by gauging a discrete $\Z_k^t$ topological symmetry. This $\Z_k$ distinguishes between $\frac{\surm(k)\times\urm(1)}{\Z_k}\cong \urm(k)$ and $\surm(k)\times\urm(1)$. Additionally, the gauging of the $\Z_k^t$ symmetry has introduced a $\Z_k$ 1-form symmetry into $\Tcal\slash \Z_k^t$.

\subsection{Examples from 5d magnetic quivers}
One can demonstrate gauging discrete subgroups of the topological symmetry on known magnetic quivers\footnote{See also \cite{Closset:2020afy,Closset:2020scj,Closset:2021lwy,Nawata:2021nse,Carta:2022spy,Carta:2022fxc} for magnetic quivers of theories with 1-form symmetries.}. It is most suitable to choose quivers whose Coulomb branches have a known Higgs branch realisation.

\paragraph{$E_5$ quiver.}
The infinite coupling magnetic quiver for 5d $\sprm(1)$ SQCD with 4 flavours realises $\orbit{\min}{E_5}\cong \orbit{\min}{D_5}$, which is also the Higgs branch of $\sprm(1)$ with 5 flavours. Thus, one arrives at \cite{Bourget:2020gzi}
\begin{align}
\left[ \raisebox{-.5\height}{
    \includegraphics[page=1]{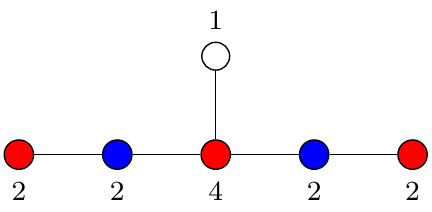}
    }  \right] \slash \Z_2
    \qquad \longleftrightarrow \qquad 
     \raisebox{-.5\height}{
    \includegraphics[page=2]{figures/figures_1-form_magnetic_quiver.pdf}
    } 
    \label{eq:magQuiv_E5}
\end{align}
It is worth recalling that the magnetic lattice for the left-hand side quiver has the form $\Gamma \cup (\Gamma +\frac{1}{2})$ with $\Gamma$ being the standard GNO integer lattice, as can be found in \cite{Bourget:2020xdz} and also see \cite{Bourget:2020gzi,Akhond:2020vhc,Akhond:2021knl,Bourget:2021zyc,Carta:2021whq,Bourget:2021xex,Sperling:2021fcf,Nawata:2021nse,Akhond:2022jts} for examples with orthosymplectic quivers.  The corresponding discrete $\Z_2^t$ topological symmetry of the magnetic quiver can be gauged in the same vein as before. On the level of the magnetic quiver, this just reduces the relevant magnetic lattice to the integer lattice $\Gamma$. Equivalently, one can gauge a $\Z_2^f$ on the $\sprm(2)$ SQCD side, which then gives rise to the following pair of theories
\begin{align}
 \raisebox{-.5\height}{
    \includegraphics[page=1]{figures/figures_1-form_magnetic_quiver.pdf}
    } 
    \qquad \longleftrightarrow \qquad 
     \raisebox{-.5\height}{
    \includegraphics[page=3]{figures/figures_1-form_magnetic_quiver.pdf}
    } 
    \label{eq:magQuiv_E5_gauged}
\end{align}
and it is straightforward to verify that the Coulomb / Higgs branch Hilbert series reproduce the results of \cite[Tab.\ 9]{Bourget:2020xdz}. The global form of the 0-form symmetry is $\psorm(8)\times \urm(1)$.

\paragraph{$E_4$ quiver.}
Similarly, the infinite coupling magnetic quiver for 5d $\sprm(1)$ SQCD with 3 flavours realises $\orbit{\min}{E_4}\cong \orbit{\min}{A_4}$ via its Coulomb branch. Of course, this moduli space admits a known Higgs branch realisation and one arrives at
\begin{align}
 \left[ \raisebox{-.5\height}{
    \includegraphics[page=4]{figures/figures_1-form_magnetic_quiver.pdf}
    } \right] \slash \Z_2
    \qquad \longleftrightarrow \qquad 
     \raisebox{-.5\height}{
    \includegraphics[page=5]{figures/figures_1-form_magnetic_quiver.pdf}
    } 
    \label{eq:magQuiv_E4}
\end{align}
The magnetic lattice for the magnetic quiver is of the form $\Gamma \cup (\Gamma +\frac{1}{2})$, so the associated $\Z_2^t$ symmetry can be gauged. The question then becomes what $\Z_2^f$ symmetry is realised on the SQED side. Through explicit calculations, one verifies that
\begin{align}
 \raisebox{-.5\height}{
    \includegraphics[page=4]{figures/figures_1-form_magnetic_quiver.pdf}
    } 
    \qquad \longleftrightarrow \qquad 
     \raisebox{-.5\height}{
    \includegraphics[page=6]{figures/figures_1-form_magnetic_quiver.pdf}
    } 
    \label{eq:magQuiv_E4_gauged}
\end{align}
reproduces the known Hilbert series  \cite[Tab.\ 10]{Bourget:2020xdz}. The isometry group in this case is $\sorm(6)\times \urm(1)$.

\paragraph{Folded $E_6$ quiver.}
The infinite coupling magnetic quiver for 5d $\sprm(1)$ SQCD with 5 flavours admits a $\Z_2$ outer automorphism. Folding the corresponding magnetic quiver leads to $\orbit{\min}{E_6} \to \orbit{\min}{D_5}$ on the Coulomb branch \cite{Bourget:2021xex}. Since there is a known Higgs branch realisation for $D$-type minimal nilpotent orbit closures, one arrives at
\begin{align}
 \left[ \raisebox{-.5\height}{
    \includegraphics[page=7]{figures/figures_1-form_magnetic_quiver.pdf}
    } \right] \slash \Z_2
    \qquad \longleftrightarrow \qquad 
     \raisebox{-.5\height}{
    \includegraphics[page=2]{figures/figures_1-form_magnetic_quiver.pdf}
    } 
    \label{eq:magQuiv_folded_E6}
\end{align}
where again the left-hand side quiver has  a magnetic lattice of the form $\Gamma \cup (\Gamma +\frac{1}{2})$.
Gauging this $\Z_2^t$ has a by now clear consequence on the magnetic quiver, as the GNO lattice is reduced to the integer lattice. On the $\sprm(1)$ SQCD side, the corresponding $\Z_2^f$ is realised as follows:
\begin{align}
 \raisebox{-.5\height}{
    \includegraphics[page=7]{figures/figures_1-form_magnetic_quiver.pdf}
    } 
    \qquad \longleftrightarrow \qquad 
     \raisebox{-.5\height}{
    \includegraphics[page=8]{figures/figures_1-form_magnetic_quiver.pdf}
    } 
    \label{eq:magQuiv_folded_E6_2}
\end{align}
and one straightforwardly verifies the agreement of the Coulomb branch / Higgs branch  Hilbert series, which is  given by
\begin{align}
\HS=
1&+25 t+400 t^2+3864 t^3+26600 t^4+141672 t^5+621480 t^6+2337280 t^7 \\
&+7763283 t^8+23265515 t^9+63954800 t^{10}+O\left(t^{11}\right) \notag 
\end{align}
and the global symmetry group is $\psurm(5) \times \urm(1)$.

\section{Discussion and conclusions}
\label{sec:conclusion}
In this paper, mirror pairs with non-trivial 1-form symmetry have been studied. Starting from known mirror pairs with trivial 1-form symmetry, gauging of discrete $\Z_q$ subgroups of the 0-form symmetry allowed us to construct new mirror pairs with non-trivial 1-form symmetry.

The main results are as follows:
\begin{enumerate}
    \item It has been shown that theories $\Tcal \slash \Z_q^t$, obtained by gauging a discrete subgroup $\Z_q^t$ of the topological symmetry, may admit quiver descriptions if the discrete subgroup is suitably chosen.
\item The mirror theories $\left( \Tcal \slash \Z_q^t\right)^\vee$ can be constructed using $\Tcal^\vee \slash \Z_q^f$, but the precise choice of $\Z_q^f$ in the flavour symmetry of $\Tcal^\vee$ can be subtle. This paper provides a simple algorithm for specifying $\Z_q^f$.
\item  The global form of the 0-form symmetries of $(\Tcal \slash \Z_q^t,\Tcal^\vee \slash \Z_q^f)$ have been derived using both field theory methods and monopole operators (via the balanced set of nodes), and the resulting symmetry groups have been verified through explicit Hilbert series computations.
\item  The interplay between continuous 0-form and discrete 1-form symmetries has been studied using established field theory techniques and the equivalence classes of lines.
\item  On the technical side, the gauging of discrete subgroups of the topological symmetry on non-simply laced quivers has been proposed and tested on both long and short-side gauge nodes.
\end{enumerate}

\paragraph{A comment on the moduli spaces.}
The maximal branches of the moduli space of vacua in a theory $\Tcal$ are the Coulomb branch $\Coulomb(\Tcal)$ and the Higgs branch $\Higgs(\Tcal)$. These are symplectic singularities that can be resolved when the theory $\Tcal$ is given either an FI parameter (for the Higgs branch) or a mass parameter (for the Coulomb branch). For instance, consider SQED with $N$ hypermultiplets of charge $1$. This theory admits $N-1$ mass parameters that resolve the $\C^2\slash \Z_N$ Coulomb branch, and a single FI parameter that resolves the Higgs branch, specifically the minimal nilpotent orbit closure $\orbit{\min}{\surmL(N)}$. If we gauge a $\Z_q^t$ 0-form symmetry in this theory, the resulting SQED with charge $q$ hypers has the same Higgs branch, but the Coulomb branch is modified to be $\C^2\slash \Z_{N\cdot q}$. However, there are no additional mass parameters in the theory, which means that the singularity cannot be fully resolved even though a symplectic resolution exists.

More generally, one can perform a simple test\footnote{Following \cite{Martelli:2006yb}, the volume of the Sasakian base $S$ of $\Higgs$ or $\Coulomb$ is evaluated via $\mathrm{Vol}(S)=\lim_{t\to1} (1-t)^d \HS(t)$, where $d=\dim_\C (\Higgs \text{ or }\Coulomb)$.} via Hilbert series that shows
\begin{align}
\Tcal \longrightarrow
\begin{cases}
\Tcal\slash \Z_q^t : 
& \lim_{t\to 1} \frac{ \HS_{\Coulomb(\Tcal\slash \Z_q^t)} (t) }{ \HS_{\Coulomb(\Tcal)} (t) } = \frac{1}{q} \\
&\text{or } \\ 
\Tcal\slash \Z_q^f : 
& \lim_{t\to 1} \frac{ \HS_{\Higgs(\Tcal\slash \Z_q^f)} (t) }{ \HS_{\Higgs(\Tcal)} (t) } = \frac{1}{q}
\end{cases}
\end{align}
and the presence of a $\frac{1}{q}$ fraction in the expression suggests (at least locally) that the Coulomb branch $\Coulomb(\Tcal\slash \Z_q^t)$ is a $\Z_q$ orbifold of  $\Coulomb(\Tcal)$, and a similar relationship holds for the Higgs branches $\Higgs(\Tcal\slash \Z_q^f)$ and $\Higgs(\Tcal)$. Again, no additional deformation parameter appears.

In contrast, consider $\Tcal$ to be $\urm(2)$ SQCD with $4$ fundamental flavours. The maximal branches are $\Higgs(\Tcal)= \orbit{(2^2)}{\surmL(4)}$ and $\Coulomb (\Tcal) = \Scal_{(2^2)} \cap \Ncal_{\surmL(4)}$, i.e.\ the Slodowy slice to the $\surmL(4)$ nilpotent orbit defined by partition $(2^2)$. There are 3 masses resolving the Coulomb branch and 1 FI term resolving the Higgs branch. If we gauge the topological $\urm(1)_t$ symmetry in this theory, the resulting theory is $\surm(2)$ SQCD with $4$ fundamental flavours. Then, the Coulomb branch of this theory is $\Coulomb(\Tcal \slash \urm(1)_t)=\C^2\slash D_4$ while the Higgs branch is $\Higgs(\Tcal \slash \urm(1)_t)=\orbit{\min}{\sormL(8)}$. In this case, the Coulomb branch can be resolved by the $3+1$ mass parameters, while the minimal orbit closure of $\sormL(8)$ does not admit a symplectic resolution, which is consistent with the absence of an FI parameter in this theory. These symplectic resolutions can also be studied via Hilbert series techniques, see for instance \cite{Cremonesi:2016nbo,Hanany:2018uzt}.

\paragraph{Generalisations and open questions.}
In this work, a single $\Z_q$ factor of the 0-form symmetry has been gauged. 
One straightforward generalisation is to consider orthosymplectic quivers and gauge several $\Z_2^t$ topological symmetry factors associated to $\sorm(n_i)$ nodes. The resulting theory is simply obtained by replacing the relevant $\sorm(n_i)\to \spin(n_i)$ and the 1-form symmetry is the product group $\prod_i (\Z_2^t)_i$.
Similarly, one could also entertain the thought of gauging several $\Z_{q_i}$ inside distinct topological Cartan factors of, say, $T[\surm(N)]$. It is \textit{a priori} not clear if a simple quiver description exists.  

Another possibility is to gauge a discrete $\Z_q^t$ group embedded into several topological Cartan $\urm(1)$ factors of a Coulomb branch unitary quiver. Inspecting the mirror maps for a fully balanced linear quiver \eqref{eq:mirror_map_min_A-type_orbit} or \eqref{eq:mirror_map_TSUN}, one observes that the effect on the mirror Higgs branch quiver is as follows: the set of fundamental flavours splits into several sets, with each subset being acted upon by $\Z_q^f$ in a distinct fashion. One might hope to find a simple quiver description on the Coulomb branch side, but this is only possible for very specific $q$ values, similar to the choices in this paper. For instance, gauging a $\Z_q^t$ in two adjacent node $\urm(k)$ and $\urm(k+1)$ in a $T[\surm(N)]$ theory, one can expect a quiver-type description with an $\surm(k)$, $\surm(k+1)$ node and two ``new'' $\urm(1)_{1,2}$ gauge factors for $q=k\cdot (k+1)$. While one $\urm(1)_1$ factor behaves similarly to the discussion in this paper, the second $\urm(1)_2$ factor is expected to lead to trifundamental hypermultiplets for $\urm(k-1) \times \surm(k)\times \urm(1)_2$, $\urm(k+2) \times \surm(k+1)\times \urm(1)_2$, and $\surm(k) \times \surm(k+1)\times \urm(1)_2$.  A systematic analysis of these cases is left for future work.

Another aspect of 3d mirror symmetry is the exchange of Wilson and vortex line defects \cite{Assel:2015oxa,Dimofte:2019zzj,Dey:2021jbf,Dey:2021gbi,Nawata:2021nse}. Given the central role of line defects in understanding 1-form and 2-group symmetries, it would be interesting to systematically analyse the exchange of Wilson and vortex lines under mirror symmetry for the theories with 1-form symmetry.

\acknowledgments
We would like to thank Fabio Apruzzi, Lakshya Bhardwaj, Mathew Bullimore, Andrea Ferrari, Heeyeon Kim, Noppadol Mekareeya, Matteo Sacchi, and Sakura Sch\"afer-Nameki for discussions.
The research of S.N.\ is supported by the National Science Foundation of China under Grant No.\ 12050410234 and Shanghai Foreign Expert grant No.\ 22WZ2502100.
M.S.\ is grateful to Ryo Suzuki for use of his computing facilities. M.S.\ is also grateful to Rudolph Kalveks for
invaluable help with \texttt{Mathematica}.
The research of Z.Z. is supported by the ERC Consolidator Grant \# 864828 “Algebraic Foundations of Supersymmetric Quantum Field Theory” (SCFTAlg).

%
\appendix
\section{Notations and conventions}
\label{app:notation}

\begin{table}[ht]
    \centering
\begin{subtable}[b]{0.3\textwidth}
\centering
\ra{1.5}
\begin{tabular}{cc}
\toprule
  \textbf{node}  & \textbf{vector}  \\ \midrule
  \raisebox{-.5\height}{\includegraphics[page=1]{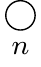} }   &  $\urm(n)$ \\
  \raisebox{-.5\height}{\includegraphics[page=2]{figures/figures_1-form_notation.pdf} }   &  $\surm(n)$ \\
  \raisebox{-.5\height}{\includegraphics[page=3]{figures/figures_1-form_notation.pdf} }   &  $\sorm(n)$ \\
  \raisebox{-.5\height}{\includegraphics[page=4]{figures/figures_1-form_notation.pdf} }   &  $\spin(n)$\\
  \raisebox{-.5\height}{\includegraphics[page=5]{figures/figures_1-form_notation.pdf} }   &  $\sprm(n)$ \\
  \raisebox{-.5\height}{\includegraphics[page=6]{figures/figures_1-form_notation.pdf} }   &  $\Z_q$  \\ \bottomrule
\end{tabular}
\caption{}
\label{subtab:gauge_nodes}
\end{subtable}
\begin{subtable}[b]{0.65\textwidth}
\centering
\ra{1.5}
\begin{tabular}{cc}
\toprule
   \textbf{edge}  & \textbf{hyper} \\ \midrule
     \raisebox{-.5\height}{\includegraphics[page=7]{figures/figures_1-form_notation.pdf} }   & bifundamental $\bm{n} \otimes \overline{\bm{k}}$ \\
      \raisebox{-.5\height}{\includegraphics[page=8]{figures/figures_1-form_notation.pdf} }   & bifundamental $\bm{n} \otimes [0,\ldots,0,1]_{A}$ \\
       \raisebox{-.5\height}{\includegraphics[page=9]{figures/figures_1-form_notation.pdf} }   & half-hyper $[1,0,\ldots,0]_{D/B} \otimes [1,0,\ldots,0]_{C}$ \\
       \raisebox{-.5\height}{\includegraphics[page=10]{figures/figures_1-form_notation.pdf} }   & half-hyper in vector $\times $ vector \\ \raisebox{-.5\height}{\includegraphics[page=11]{figures/figures_1-form_notation.pdf} }   & $N$ copies of bifundamental \\
        \raisebox{-.5\height}{\includegraphics[page=12]{figures/figures_1-form_notation.pdf} }   &  $N$ copies of fundamental  \\ 
         \raisebox{-.5\height}{\includegraphics[page=13]{figures/figures_1-form_notation.pdf} }   &  fundamental of $\urm(n)$ but charge $Q$ of $\urm(1)$  \\\bottomrule
\end{tabular}
\caption{}
\label{subtab:links}
\end{subtable}
    \caption{Notation for nodes and links in the quiver diagrams.}
    \label{tab:notation}
\end{table}

A quiver diagram, composed of nodes and edges, encodes a 3d $\Ncal=4$ theory as follows:
\begin{compactitem}
\item Gauge nodes $\bigcirc$ denote dynamical vector multiplets, while flavour nodes $\Box$ denote background vector multiplets. The notations are summarised in Table \ref{subtab:gauge_nodes}.
\item An edge between two nodes corresponds to a hypermultiplet $H=(X,Y^\dagger)$, with $X,Y$ two $\Ncal=2$ chiral multiplets. The notation is summarised in Table \ref{subtab:links}.

\item An exception are so-called \emph{non-simply laced edges} in a quiver theory. Between unitary gauge node, such an edge has been proposed purely on the level of the conformal dimension of the monopole formula \cite{Cremonesi:2014xha}
\begin{align}
\raisebox{-.5\height}{\includegraphics[page=14]{figures/figures_1-form_notation.pdf} } 
\quad \longleftrightarrow \quad 
\frac{1}{2} \sum_{i=1}^n \sum_{j=1}^k |m_{1,i} - \kappa \cdot m_{2,j}|
\end{align}
and it is to stress that this does not correspond to a representation of the gauge groups. For the special case of $\urm(n=1)$, such a non-simply laced edge is effectively the same as a $\urm(1)$ gauge group with a charge $\kappa$ hypermultiplet.

Between orthosymplectic nodes, the conformal dimension has been proposed in \cite{Bourget:2021xex}
\begin{align}
\raisebox{-.5\height}{\includegraphics[page=15]{figures/figures_1-form_notation.pdf} } \quad \longleftrightarrow \quad 
\frac{1}{2\cdot 2} \sum_{\rho \in [1,0,\ldots,0]_{B/D}}
\sum_{\lambda \in [1,0,\ldots,0]_{C}}
|\rho(\bm{m}) - \kappa \cdot \lambda(\bm{n})|
\end{align}
with $\bm{m}$, $\bm{n}$ the magnetic fluxes which are evaluated on the weights $\rho$, $\lambda$, respectively.

\end{compactitem}

\subsection{Hilbert series}
\label{app:Hilbert_series}
\subsubsection{Monopole formula}
The Hilbert series for the 3d $\Ncal=4$ Coulomb branch is known as the monopole formula \cite{Cremonesi:2013lqa}.
Schematically, the Hilbert series is computed as a sum over magnetic fluxes $\bm{m}$ valued in the GNO lattice $\Gamma$ of the gauge group $G$.
\begin{align}
    \HS_{\Coulomb}= \sum_{\mathbf{m} \in \Gamma \slash \Wcal } P(t, \mathbf{m}) w^{\mathbf{m}} t^{\Delta(\mathbf{m})}
\end{align}
and $\Wcal$ denotes the Weyl group of $G$. A bare monopole operator is characterised by the flux $\bm{m}$ as well as its conformal dimension $\Delta(\bm{m})$, which coincides with the third component of the $\surm(2)_R$ spin. The factors $P(t,\bm{m})$ dress a bare monopole operator by gauge invariants formed by the adjoint chiral multiplet of the residual gauge group $H_{\bm(m)}$. Lastly, $w$ denotes the fugacity of the topological symmetry, assuming that $G$ contains $\urm(1)$ factors.

\subsubsection{Higgs branch Hilbert series}
The Higgs branch Hilbert series \cite{Benvenuti:2006qr,Feng:2007ur,Gray:2008yu} for the 3d $\Ncal=4$ quiver gauge theory relevant here is schematically obtained by a Molien-Weyl integral of the form
\begin{align}
    \HS_{\Higgs}= \int_{G} \diff \mu_G \frac{\PE[\chi_{\mathrm{Adj}}^G \ t ]}{ \PE[ \chi_{\mathcal{R}}^G \cdot \chi_{\mathcal{F}}^{F}\ t^{\frac{1}{2}}]  }
\end{align}
where the numerator contains the character $\chi_{\mathrm{Adj}}^G$ of the adjoint representation of the gauge group, while the denominator contains all matter fields characterised by their representations $\mathcal{R}$ under the gauge group $G$ and the representations $\mathcal{F}$ under the flavour symmetry $F$.

\subsubsection{Gauging a discrete 0-form symmetry.}
Suppose one is given a generating function $\HF(z|t)$ which is a power series in $t$ with coefficients that are Laurent polynomials in a $\urm(1)$ fugacity $z$. Next, embed a $\Z_q\hookrightarrow \urm(1)$ via $(\zeta_q)^p  = e^{\frac{2\pi \im p}{q}}$ with $p=0,1,\ldots,q-1$. Gauging this discrete $\Z_q$ 0-form symmetry is realised in terms of the generating function via a discrete Molien-Weyl sum  
\begin{align}
    \frac{1}{q}\sum_{p=0}^{q-1} 
    \HF\left( (\zeta_q)^p  \cdot y^{\frac{1}{q}} | t \right)
\end{align}
where $y$ is the fugacity for the residual $\urm(1)\slash \Z_q \cong \urm(1)$ symmetry.

\subsection{Superconformal index}
The 3d superconformal index can be computed as partition function on $S^2\times S^1$ via localisation techniques, see \cite{Bhattacharya:2008zy,Bhattacharya:2008bja,Kim:2009wb,Imamura:2011su,Kapustin:2011jm,Dimofte:2011py,Razamat:2014pta} for details. Schematically, one arrives at 
\begin{align}
Z= \sum_{\bm{m}} \frac{1}{|\Wcal_{\bm{m}}|} \oint_{\mathbb{T}^{\rk(G)}}
\prod_{i=1}^{\rk(G)} \frac{\diff s_i}{2\pi \im s_i}
I_{\mathrm{cl}} \cdot 
I_{\mathrm{vec}}\cdot
I_{\mathrm{matter}}
\end{align}
where $\bm{s}$ denotes the gauge fugacities, which are valued in a maximal torus of the gauge group $G$. The magnetic fluxes $\bm{m}$ take values in the GNO-lattice of $G$. A flux $\bm{m}$ breaks $G$ to the residual gauge group $H_{\bm{m}}$ (the stabiliser subgroup of $\bm{m}$ inside $G$) with Weyl group $\Wcal_{H_{\bm{m}}} \equiv \Wcal_{\bm{m}}$.
The integration contour is chosen to be the unit circle $\mathbb{T}$ for each $s_i$. 
The integrand is composed of classical contributions and the 1-loop determinants of the supermulitpelts. For concreteness, the $G=\urm(N)$ case is reviewed:

The classical contribution is given by
\begin{align}
    I_{\mathrm{cl}}^{\urm(N)}(w,\bm{m};n) = 
    \prod_{a=1}^{N} \left(s_{a}\right)^{n}
    w^{\sum_{a=1}^N m_a}  
\end{align}
with $w$ the fugacity of the topological $\urm(1)_t$ symmetry.

The $\Ncal=2$ multiplets have the following 1-loop determinants:
\begin{compactitem}
\item 3d $\Ncal=2$ Chiral multiplet of R-charge $r$ coupled with unit charge to a gauge field:
\begin{align}
    I_{\mathrm{chi}}^{r}(z,m|\ x) &=
    \left(x^{1-r} z^{-1}\right)^{\frac{|m|}{2}}
    \prod_{j=0}^\infty \frac{1-(-1)^m z^{-1} x^{|m|+2-r +2j}}{1-(-1)^m z x^{|m|+r +2j}} \\
    &=    \left(x^{1-r}  z^{-1}\right)^{\frac{|m|}{2}}
    \frac{
    \left( (-1)^m z^{-1} x^{|m| +2-r};\ x^2 \right)_\infty 
    }{
    \left( (-1)^m z x^{|m| +r};\ x^2 \right)_\infty
    } \notag
\end{align}
with a $\urm(1)$ holonomy $z$ around $S^1$ and the $\Z$-valued magnetic flux $m$ on $S^2$. Here, the definition $(z; \ q)_\infty = \prod_{j=0}^\infty (1-z q^j)$ has been used.
\item 3d $\Ncal=4$ Hypermultiplet transforming as bifundamental of $\urm(N)\times\urm(M)$
\begin{align}
I_{\mathrm{hyp}}^{\urm(N)\times\urm(M)}(\bm{s_1},\bm{m_1}; \bm{s_2},\bm{m_2}|\ x)
=
\prod_{a=1}^N \prod_{b=1}^M
&I_{\mathrm{chi}}^{\frac{1}{2}}
\left(s_{1,a} s_{2,b}^{-1} , m_{1,a} - m_{2,b}|\ x \right) \\
&\cdot 
I_{\mathrm{chi}}^{\frac{1}{2}}
\left(s_{1,a}^{-1} s_{2,b} , m_{2,b} - m_{1,a}|\ x \right) \notag
\end{align}
\item 3d $\Ncal=2$ vector multiplet for a $\urm(N)$ gauge group:
\begin{align}
    I_{\mathrm{vec}}^{\urm(N)}(\bm{s},\bm{m}|\ x) =
    \prod_{a<b} x^{-|m_a-m_b|}
    &\left( 1-(-1)^{m_a-m_b} s_a s_b^{-1} x^{|m_a -m_b|} \right)\\
   &\cdot  \left( 1-(-1)^{m_a-m_b} s_a^{-1} s_b x^{|m_a -m_b|} \right) \notag \,.
\end{align}
\item 3d $\Ncal=4$ vector multiplet has the same 1-loop determinant as the $\Ncal=2$ vector multiplet, because the contribution of the adjoint-valued chiral multiplet is trivial.
\end{compactitem}

\subsection{Centre symmetries of classical Lie algebras}
\label{app:congruency}
Following \cite{lemire1980congruence,slansky1981group}, a representation labelled by Dynkin labels $[n_1,\ldots,n_r]$ lies in a specific congruence class of the centre, as detailed in Table \ref{tab:congruency_class_algebras}.

\begin{table}[ht]
    \centering
    \begin{tabular}{llc}
    \toprule
         algebra & centre $\Zcal$ & congruence\\ \midrule 
        $A_r$ & $\Z_{r+1}$ & $\sum_{k=1}^r k \cdot n_k \bmod r+1$  \\
        $B_r$ & $\Z_2$ & $n_r \bmod 2$ \\
        $C_r$ & $\Z_2$ & $ \sum_{j=0}^{\frac{r-1}{2}} n_{2j+1} \bmod 2$ \\
        $D_r$, $r$ even & $\Z_2 \times \Z_2$ & 
        $\begin{pmatrix} n_{r-1} + n_r \bmod 2 \\
        \sum_{j=0}^{\frac{r-4}{2}} n_{2j+1} + 
        \frac{r-2}{2} n_{r-1} + \frac{r}{2} n_r \bmod 2
        \end{pmatrix}$ \\
        $D_r$, $r$ odd & $\Z_4 $ &
        $ \sum_{j=0}^{\frac{r-3}{2}} 2n_{2j+1} + 
        (r-2) n_{r-1} + r n_r \bmod 4$ \\ \bottomrule
    \end{tabular}
    \caption{Centre symmetries of classical Lie algebras}
    \label{tab:congruency_class_algebras}
\end{table}

\section{Discrete gauging of \texorpdfstring{$T[\surm(N)]$}{TSUN}}
\label{app:discrete_gauging}
In this appendix, the monopole formula and the superconformal index are used to determine the theories obtained by gauging a discrete subgroup of the topological symmetry. Before proceeding with the details, the logic of the argument is summarised.

Start from a given theory $\Tcal$ and consider a function $\Fcal(\bm{w},\ldots)$, like a Hilbert series or index, depending on topological symmetry fugacities $\bm{w}$ and possibly other variables.
\begin{compactenum}[(i)]
\item Pick a gauge node $\urm(k)$ with topological fugacity $w$ and introduce a $\Z_q$-valued variable $\zeta_q$ via $w = \zeta_q \cdot f(w,\ldots)$, where $f$ is some function of the fugacities. For a $\urm(k)$, the topological fugacity enters via $w^{\sum_{i=1}^k m_i}$, with $m_i$ the magnetic fluxes.
\item Gauging $\Z_q$ on the level of $\Fcal(\bm{w},\ldots)$ is realised by
\begin{align}
    \widetilde{\Fcal} = \frac{1}{q} \sum_{p=0}^{q-1} \Fcal(\bm{w}, w= (\zeta_q)^p \cdot f(w),\ldots)
\end{align}
and to perform the $\Z_q$ average, one rewrites $w^{\sum_{i=1}^k m_i} =w^{k\cdot h}$ where $h$ is $\urm(1)$ magnetic flux (for normalisation and sign, see below).
\item This suggests to rewrite $\urm(k) = \frac{\surm(k)\times \urm(1)}{\Z_k} $ where $\Z_k$ acts via $(\diag(\zeta_k,\ldots,\zeta_k)\, , \; \zeta_k^{-1})$ on $\surm(k)\times \urm(1)$. Denote the $\surm(k)$ fluxes as $\bm{l} \in \Z^{k{-}1}$. For $\surm(k)\times\urm(1)$ the magnetic fluxes take value in $(\bm{\ell},h) \in \Z^{k}$; however, the additional $\Z_k$ quotient enlarges the magnetic lattice \cite{Bourget:2020xdz} 
\begin{align} \label{eq:UN_as_SUNxU1}
    \urm(k) = \frac{\surm(k)\times \urm(1)}{\Z_k}
    \qquad \text{with}\qquad
    (\bm{\ell},h) \in \bigcup_{i=0}^{k{-}1} \left( \Z + \frac{i}{k} \right)^{k} \,.
\end{align}
\item After a change of variables (see below), the $\Z_q$ gauging reduces to
\begin{align}
       \widetilde{\Fcal} = \frac{1}{q} \sum_{p=0}^{q-1}   
       \sum_{h \in \bigcup_{i=0}^{k{-}1} \left( \Z+\frac{i}{k} \right) }
      (\zeta_q)^{p\cdot k \cdot h} \ \widehat{\Fcal}(\bm{w} ,\ldots)
\end{align}
using $\Fcal = \sum_h \widehat{\Fcal}$, schematically. Utilising 
\begin{align}\label{eq:discrete_sum}
    \frac{1}{q} \sum_{p=0}^{q-1}  (\zeta_q)^{p \cdot (k\cdot h)}  = 
    \begin{cases}
    1 \,, & q | (k\cdot h)  \\
    0 \,, &  \text{else}
    \end{cases}
\end{align}
and restricting to $q=a\cdot k$ with $a\in \N$, the summation range of $h$ collapses from $\bigcup_{i=0}^{k{-}1} \left( \Z+\frac{i}{k} \right)$ to $a \cdot \Z$. But this leads to the collapse of the summation range of all other magnetic fluxes onto the integers too.
\item Lastly, one obtains an expression of $\widetilde{\Fcal}$ written as a sum over integer-valued magnetic fluxes. One can then read off a Lagrangian description for the theory $\Tcal \slash \Z_q$.
\end{compactenum}

\subsection{Gauging discrete subgroup of topological symmetry}
\label{app:discrete_gauge_TSUN}
Consider the $T[\surm(N)]$ quiver
\begin{align}
    \raisebox{-0.5\height}{
    \includegraphics[page=1]{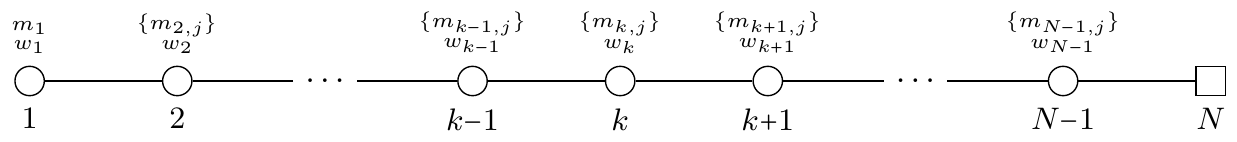}
} \label{eq:quiver_TSUN}
\end{align}
and a variation thereof
\begin{align}
    \raisebox{-0.5\height}{
    \includegraphics[page=2]{figures/figures_1-form_gauging_in_index.pdf}
}\label{eq:quiver_TSUN_gauge}
\end{align}
The aim is to show that 
\begin{align}
    \HS_{\eqref{eq:quiver_TSUN_gauge}} =& \frac{1}{q} \sum_{p=0}^{q-1} \HS_{\eqref{eq:quiver_TSUN}}(w_1,\ldots,w_{N{-}1})\big|_{w_k\to \# \cdot (\zeta_q)^p}
    \quad \text{for }q=k \,,\\
    &\text{with } \qquad (\zeta_q)^p= e^{2\pi i \frac{p}{q}} \in \Z_q \hookrightarrow \mathrm{PSU}(N)_{\mathrm{topol}}\notag \,.
\end{align}
To begin with, rewrite the $\urm(k)$ fluxes in $\left( \urm(1)\times\surm(k) \right) \slash \Z_k$. Use the following map
\begin{align}
    \mathrm{diag}(m_{k,1},m_{k,2},\ldots,m_{k,k}) =& 
    -h \cdot \mathbf{1}
    + \sum_{i=1}^{k{-}1} l_i \cdot  E_i\\
  \text{with }   \mathbf{1}=  \mathrm{diag}(1,\ldots,1) \quad \text{and} \quad 
      E_i =& \underbrace{\mathrm{diag}(0,\ldots,0,1,-1,0,\ldots,0)}_{\text{$+1$ at $i$-th slot, $-1$ at ${i+1}$-th slot}} \,.
      \notag
\end{align}
where $\mathbf{1}$ is the Cartan generator for the diagonal $\urm(1)\subset \urm(k)$ and $\{E_i\}$ are the Cartan generators for $\surm(k)$. One finds
\begin{subequations}
\label{eq:fluxes_U_to_SU}
\begin{align}
m_{k,1} &= -h + l_1
\;, \quad 
m_{k,j} = -h + l_j - l_{j-1}
\; \text{ for } 1<j<k 
\;, \qquad 
m_{k,k} = -h - l_{k{-}1} \\
h &= -\frac{1}{k} \sum_{i=1}^k m_{k,i} 
\; , \qquad
l_j = \frac{1}{k} \left( (k{-}j) \sum_{i=1}^j m_i 
- j \sum_{i=j+1}^k m_i 
\right)
\end{align}    
\end{subequations}
And one finds for contributions of the fundamental $\urm(k)$ weights
\begin{subequations}
\label{eq:weights_UN_to_SUN}
\begin{align}
    m_{k,i} &= -h + \mu_i(\{l_j\}) 
    \equiv  -h + \mu_i (\bm{l}) \\
    \mu_1 &=e_1\;, \qquad 
    \mu_j = e_j -e_{j-1} \text{ for } 1<j<k \;, \qquad 
    \mu_k = -e_k \
\end{align}    
\end{subequations}
with $\mu_i(\bm{l})$ are the weights\footnote{Here, the Dynkin/weight basis \cite{Feger:2012bs,Feger:2019tvk}, is chosen; i.e.\ the fundamental weights $\{e_i\}$ define the basis.} of the fundamental $\surm(k)$ representation evaluated on the fluxes $\bm{l}$.
The summation range for $\{m_{k,i}\}$ is $\Z^k \cap \{m_{k,1}\geq m_{k,2}\geq\ldots \geq m_{k,k}\}$, due to the Weyl group action, and translates into 
\begin{align}
    (h,l_1,\ldots,l_{k{-}1}) \in \bigcup_{p=0}^{k{-}1} \left( \left(\Z+ \frac{p}{k}\right),\left(\Z+ \frac{p}{k}\right)^{k{-}1}\slash \Wcal_{A_{k{-}1}}\right) 
    \label{eq:sum_range_SUxU1}
\end{align}
with $ \Wcal_{A_{k{-}1}}$ the Weyl group action of $\surm(k)$.
Moreover, one straightforwardly verifies
\begin{align}
    \sum_{i<j} |m_{k,i}-m_{k,j}| =& \sum_{i<j} |l_i - l_{i-1} -l_j + l_{j-1}| = \sum_{\alpha \in \Phi^+(\surmL(k))} |\alpha(\bm{l}) | \\
    P_{\urm(k)}(\{m_{k,i}\}) =&P_{\urm(1)}(h) \cdot P_{\surm(k)}(\{l_{i}\})
\end{align}
where $\Phi^+(\surmL(k))$ denotes the set of positive roots, here expressed in the weight basis\footnote{Recall, in weight basis the simple roots are given by $\alpha_i = \sum_k C_{ik} e_k$ with $e_k$ the fundamental weights and $C_{ik}$ the Cartan matrix. Thus, $\alpha_i = 2 e_i -e_{i-1} -e_{i+1}$ for $1<i<k$, which agrees with $\alpha(\bm{l})= l_i - l_{i-1} -l_j + l_{j-1}$  for $j=i+1$, i.e.\ the simple roots. The remaining expressions are positive but non-simple roots.}.

Next, consider the hypermultiplet with the adjacent gauge nodes
\begin{align}
    2 \Delta_{k{-}1,k} =& 
    \sum_{i=1}^{k{-}1} \sum_{j=1}^{k} 
    |m_{k{-}1,i}-m_{k,j}|
    =\sum_{i=1}^{k{-}1} \sum_{j=1}^{k} 
    |\underbrace{m_{k{-}1,i}+h}_{\equiv n_{k{-}1,i} }-\mu_i (\bm{l})| \notag \\
    =& \sum_{i=1}^{k{-}1} \sum_{j=1}^{k} 
    |n_{k{-}1,i}-\mu_i (\bm{l})| \\
    2 \Delta_{k,k+1} =& 
    \sum_{i=1}^{k} \sum_{j=1}^{k+1} 
    |m_{k,i}-m_{k+1,j}|
    =\sum_{i=1}^{k} \sum_{j=1}^{k+1} 
    |\mu_i (\bm{l}) \underbrace{- h - m_{k{-}1,i}}_{\equiv -n_{k+1,i}}| \notag \\
    =& \sum_{i=1}^{k} \sum_{j=1}^{k+1} 
    |\mu_i (\bm{l}) - n_{k+1,i}|
\end{align}
A similar argument applies for all other gauge nodes $a\neq k$, such that one redefines
\begin{align}
    n_{a,i} = m_{a,i} +h \qquad \forall a\in\{1,2,\ldots,k{-}1,k+2,\ldots,N{-}1\} \,.
\end{align}
which also implies a modification of the summation ranges of the magnetic fluxes
\begin{align}
    (n_{a,1},\ldots,n_{a,a}) \in  \left( \bigcup_{p=0}^{k{-}1} \left(\Z +\frac{p}{k} \right)^a \right) 
    \cap
    \left\{ n_{a,1} \geq n_{a,2} \geq \ldots \geq n_{a,a}\right\} 
    \label{eq:other_sum_range}
\end{align}
wherein intersection ensures the restriction to the dominant Weyl chamber. It is important to stress that all summation ranges are linked in their shift $\frac{i}{k}$, i.e.\ schematically
\begin{align}
    (h,\bm{l},\{\bm{n}_a\}_{a\neq k}) \in \widetilde{\Gamma} \coloneqq \bigcup_{p=0}^{k{-}1} \left( \Gamma + \frac{p}{k} \right)
\end{align}
where $\Gamma$ denotes the underlying integer lattice.

Next, focus on the fundamental flavour contributions
\begin{align}
    2\Delta = \sum_{i=1}^{N{-}1} N\cdot |m_{N{-}1,i}| = \sum_{i=1}^{N{-}1} N\cdot |n_{N{-}1,i} -h|
\end{align}
which is exactly the contribution of $N$ copies of $\urm(N{-}1)\times\urm(1)$ bifundamentals.
Collecting all the results, one ends up with
\begin{align}
    \HS=& \frac{1}{q} \sum_{p=0}^{q-1}
    \prod_{a=1}^{N{-}1} \sum_{\{m_{a,i}\}} P_{\urm(a)}(\{m_{a,i}\}) w_{a}^{\sum_{i=1}^a m_{a,i}} t^{2\Delta_{a,a+1}} \big|_{w_k \to \# \cdot (\zeta_q)^p} \notag\\
   =& \frac{1}{q} \sum_{p=0}^{q-1}
 \sum_{(h,\{\bm{l}\},\{\bm{n}_{a}\}) \in \widetilde{\Gamma}} 
 \prod_{a\neq k}^{N{-}1}  P_{\urm(a)}(\bm{n}_{a}) w_{a}^{\sum_{i=1}^a n_{a,i}} 
    \cdot P_{\surm(k)}(\bm{l})
    \cdot P_{\urm(1)}(h)
     \cdot t^{2\Delta_{\eqref{eq:quiver_TSUN_gauge}}} \notag \\
    &\qquad \cdot \left(\prod_{a\neq k} w_{a}^{a } \cdot w_k^k \right)^{-h}\big|_{w_k \to \# \cdot (\zeta_q)^p}
\end{align}
and defines the map $\#$ suitably
\begin{align}\label{eq:relable_topo_discrete_gauge}
    \# = \left(\frac{y^{-1}}{\prod_{a\neq k} w_{a}^{a }}\right)^\frac{1}{k}
\end{align}
such that 
\begin{align}
   \HS= &\frac{1}{q} \sum_{p=0}^{q-1}
 \sum_{(h,\{\bm{l}\},\{\bm{n}_{a}\})\in \widetilde{\Gamma} \slash \Wcal} 
 \prod_{a\neq k}^{N{-}1}  P_{\urm(a)}(\bm{n}_{a}) w_{a}^{\sum_{i=1}^a n_{a,i}} 
    \cdot P_{\surm(k)}(\bm{l})
    \cdot P_{\urm(1)}(h)
     \cdot t^{2\Delta_{\eqref{eq:quiver_TSUN_gauge}}} \notag \\
    &\qquad \cdot \left(y \ (\zeta_q)^{-k\cdot p} \right)^{h}
\end{align}
Recalling \eqref{eq:discrete_sum} and $h\in \bigcup_{i=0}^{k{-}1} (\Z+\frac{i}{k})$, choosing $q=k$
implies that only $h\in\Z$ has non-vanishing contributions. This then also reduces all other summation ranges $\widetilde{\Gamma}$, cf.\  \eqref{eq:sum_range_SUxU1} and \eqref{eq:other_sum_range}, to the integer part $\Gamma$. Therefore
\begin{align}
   \HS= &
 \sum_{(h,\{\bm{l}\},\{\bm{n}_{a}\}) \in \Gamma \slash \Wcal} 
 \prod_{a\neq k}^{N{-}1}  P_{\urm(a)}(\bm{n}_{a}) w_{a}^{\sum_{i=1}^a n_{a,i}} \cdot y^{h}
    \cdot P_{\surm(k)}(\bm{l})
    \cdot P_{\urm(1)}(h)
     \cdot t^{2\Delta_{\eqref{eq:quiver_TSUN_gauge}}} 
\end{align}
which is indeed the monopole formula for $\eqref{eq:quiver_TSUN_gauge}$.

\paragraph{Corollary.}
Consider the $T[\surm(N)]$ quiver \eqref{eq:quiver_TSUN}, pick a $\urm(k)$ node with $k>1$ and gauge a discrete subgroup $\Z_d \subset \urm(1)_t$ of the $k$-th Cartan subgroup of the topological symmetry, provided $d|k$. One can repeat all steps as above, i.e.\ rewriting all contributions as $\urm(k) \cong \left(\urm(1) \times \surm(k) \right) \slash \Z_k$. The only step that requires modifications is \eqref{eq:discrete_sum}. Recall $h \in \bigcup_{i=0}^{k{-}1} \left(\Z + \frac{i}{k} \right)$  and non-trivial contributions arise for $d | (k \cdot h)$. Since $d|k$, relevant fluxes $h$ need to satisfy $\frac{k}{d} \cdot h \in \Z$, the summation range after $\Z_d$ gauging becomes $h \in \bigcup_{i=0}^{\frac{k}{d}-1} \left(\Z + \frac{i}{\frac{k}{d}} \right)$. Therefore, the resulting theory is given by \eqref{eq:quiver_TSUN_gauge}, but the magnetic fluxes take values in $   (h,\bm{l},\{\bm{n}_a\}_{a\neq k}) \in  \bigcup_{p=0}^{\frac{k}{d}-1} \left( \Gamma + \frac{d}{k} \cdot p \right)$. See also \cite{Bourget:2020xdz} for a related discussion.


\subsection{Gauging discrete subgroup of topological symmetry revisited}
\label{app:discrete_gauging_index}
In this appendix, the aim is to consider a more general choice of discrete subgroup and to gain further evidence on the resulting theory.
Starting from
\begin{align}
    \raisebox{-0.5\height}{
    \includegraphics[page=3]{figures/figures_1-form_gauging_in_index.pdf}
}
\end{align}
define the superconformal index of $T[\surm(N)]$ as 
\begin{align}
    Z_{T[\surm(N)]}
    =&
    \prod_{\ell=1}^{N{-}1}  \left[ 
    \sum_{\bm{m_{\ell}} \in \Z^{\ell}} 
    \;
    \frac{1}{|\Wcal_{\bm{m}_\ell}|}
    \;
    \oint_{\mathbb{T}^\ell} \prod_{a=1}^{\ell} 
\frac{\diff s_{\ell,a}}{ 2\pi i s_{\ell,a}}
    \right]
    I_{T[\surm(N)]} \left( \{\bm{s_\ell}\}_{\ell=1}^{N{-}1};
    \{\bm{m_\ell}\}_{\ell=1}^{N{-}1}\right) \\
    I_{T[\surm(N)]} =&
    \prod_{\ell=1}^{N-2}
    I_{\mathrm{hyp}}^{\urm(\ell)\times\urm(\ell+1)}(\bm{s_{\ell}},\bm{m_{\ell}}; \bm{s_{\ell+1}},\bm{m_{\ell+1}}|\ x)\\
    &\quad \cdot
    I_{\mathrm{hyp}}^{\urm(N{-}1)\times\urm(N)}(\bm{s_{N{-}1}},\bm{m_{N{-}1}}; \bm{y},\bm{k}|\ x) \notag \\
    &\qquad \cdot \prod_{\ell=1}^{N{-}1} 
    I_{\mathrm{cl}}^{\urm(\ell)}(w_\ell,\bm{m_\ell};n_\ell) \; 
    I_{\mathrm{vec}}^{\urm(\ell)}(\bm{s_\ell},\bm{m_\ell}|\ x) \notag
\end{align}
and repeat the analogous step as in the monopole formula. Pick a gauge node $\urm(k)$ and relabel the magnetic fluxes $\bm{m_k}$ into a $\left(\urm(1)\times \surm(k)\right) \slash \Z_k$ fluxes $(h, \bm{l})$, see \eqref{eq:fluxes_U_to_SU} and \eqref{eq:UN_as_SUNxU1}. Likewise, the $\urm(k)$ gauge fugacities $\bm{s_k}$ are transformed into $\urm(1)\times \surm(k)$ fugacities $(S,\bm{\sigma})$ via
\begin{align}
    \begin{cases}
     s_{k,1} =& S^{-1}\cdot \sigma_1 \;, \\
     s_{k,i} =& S^{-1} \cdot \sigma_{i} \cdot \sigma_{i-1}^{-1}\,,  \quad 1<i<k \;,\\
     s_{k,k} =& S^{-1} \cdot \sigma_{k{-}1}^{-1} \;.
    \end{cases}
\end{align}
Consider the vector multiplet
\begin{align}
  I_{\mathrm{cl}}^{\urm(k)}(w_k,\bm{s_k},\bm{m_k};n_k) 
    =& \left(S^{-k}\right)^{n_k}
    w_k^{-k \cdot h}  \notag \\
    =& I_{\mathrm{cl}}^{\urm(1)}(w_k^k,S^{-k},-h;n_k)
    \\
    I_{\mathrm{vec}}^{\urm(k)}(\bm{s_k},\bm{m_k}|\ x) 
    =&
     \prod_{a<b} x^{-|l_a-l_b|}
    \left( 1-(-1)^{l_a-l_b} \mu(\sigma)_a \mu(\sigma)_b^{-1} x^{|l_a -l_b|} \right) \notag \\
   &\quad \cdot  \left( 1-(-1)^{l_a-l_b} \mu(\sigma)_a^{-1} \mu(\sigma)_b x^{|l_a -l_b|} \right)  \notag \\
   =& I_{\mathrm{vec}}^{\surm(k)}(\bm{\sigma},\bm{l}|\ x) 
\end{align}
which reduces to the $\surm(k)$ vector multiplet contribution. Using the $\mu(\sigma)_a \mu(\sigma)_b^{-1} \equiv \alpha(\sigma)_{a,b}$ yields the root contributions, see also \eqref{eq:weights_UN_to_SUN} and recall the use of weight/Dynkin basis. Note also that the $\urm(1)$ vector multiplet part receives a fitting classical contribution $ I_{\mathrm{cl}}^{\urm(1)}(w_k^k,S^{-k},-h;n_k)$; while such a term is absent for the $\surm(k)$ part of the vector.

Next, inspect the hypermultiplet contributions 
\begin{align}
I_{\mathrm{hyp}}^{\urm(k)\times\urm(k+1)}(\bm{s_k},\bm{m_k}; \bm{s_{k+1}},\bm{m_{k+1}}|\ x)
=&
\prod_{a=1}^k \prod_{b=1}^{k+1}
I_{\mathrm{chi}}^{\frac{1}{2}}
\left( S^{-1}\ \mu(\sigma)_{a} s_{k+1,b}^{-1} , -h + \mu(\bm{l})_{a} - m_{k+1,b}|\ x \right)  \notag \\
&\qquad \cdot 
I_{\mathrm{chi}}^{\frac{1}{2}}
\left(S\ \mu(\sigma)_{a}^{-1} s_{k+1,b} , m_{k+1,b}  h -\mu(\bm{l})_{a}|\ x \right) \notag\\
=&
\prod_{a=1}^k \prod_{b=1}^{k+1}
I_{\mathrm{chi}}^{\frac{1}{2}}
\left(  \mu(\sigma)_{a} \tilde{s}_{k+1,b}^{-1} ,  \mu(\bm{l})_{a} - \tilde{m}_{k+1,b}|\ x \right) \notag \\
&\qquad \cdot 
I_{\mathrm{chi}}^{\frac{1}{2}}
\left( \mu(\sigma)_{a}^{-1} \tilde{s}_{k+1,b} , \tilde{m}_{k+1,b}  -\mu(\bm{l})_{a}|\ x \right) \notag\\
  =&
  I_{\mathrm{hyp}}^{\surm(k)\times\urm(k+1)}(\bm{\sigma},\bm{l}; \tilde{\bm{s}}_{k+1},\tilde{\bm{m}}_{k+1}|\ x)
\end{align}
wherein the fluxes and fugacities at the $k+1$-st node have been redefined as follows:
\begin{align}
 \tilde{m}_{k+1,b}  \coloneqq m_{k+1,b}  +h 
 \;,\qquad 
  \tilde{s}_{k+1,b} \coloneqq s_{k+1,b}\ S \,.
\end{align}
Similarly for the other connected hypermultiplet:
\begin{align}
I_{\mathrm{hyp}}^{\urm(k{-}1)\times\urm(k)}(\bm{s}_{k{-}1},\bm{m}_{k{-}1}; \bm{s}_k,\bm{m}_k|\ x)
=&
\prod_{a=1}^N \prod_{b=1}^M
I_{\mathrm{chi}}^{\frac{1}{2}}
\left(s_{k{-}1,a} \mu(\sigma)_{b}^{-1} S, m_{k{-}1,a} +h -\mu(\bm{l})_{b}|\ x \right) \notag  \\
&\qquad \cdot 
I_{\mathrm{chi}}^{\frac{1}{2}}
\left(s_{k{-}1,a}^{-1} \mu(\sigma)_{2,b} S^{-1} , -h +\mu(\bm{l})_{b} - m_{k{-}1,a}|\ x \right) \notag \\
=&
\prod_{a=1}^N \prod_{b=1}^M
I_{\mathrm{chi}}^{\frac{1}{2}}
\left(\tilde{s}_{k{-}1,a} \mu(\sigma)_{b}^{-1}  , \tilde{m}_{k{-}1,a}  -\mu(\bm{l})_{b}|\ x \right) \notag  \\
&\qquad \cdot 
I_{\mathrm{chi}}^{\frac{1}{2}}
\left(\tilde{s}_{k{-}1,a}^{-1} \mu(\sigma)_{2,b} , \mu(\bm{l})_{b} - \tilde{m}_{k{-}1,a}|\ x \right) \notag\\
  =&
  I_{\mathrm{hyp}}^{\urm(k{-}1)\times\surm(k)}( \tilde{\bm{s}}_{k{-}1},\tilde{\bm{m}}_{k{-}1}; \bm{\sigma},\bm{l}|\ x)
\end{align}
using a similar redefinition
\begin{align}
 \tilde{m}_{k{-}1,a}  \coloneqq m_{k{-}1,a}  +h 
 \;,\qquad 
  \tilde{s}_{k{-}1,a} \coloneqq s_{k{-}1,a}\ S \,.
\end{align}
Repeating the same arguments for the other bifundamental hypermultiplets yields ($\ell\neq k,k\pm1$)
\begin{align}
    I_{\mathrm{hyp}}^{\urm(\ell)\times\urm(\ell+1)}(\bm{s}_{\ell},\bm{m}_{\ell}; \bm{s}_{\ell+1},\bm{m}_{\ell+1}|\ x) =
    I_{\mathrm{hyp}}^{\urm(\ell)\times\urm(\ell+1)}(\tilde{\bm{s}}_{\ell},\tilde{\bm{m}}_{\ell}; \tilde{\bm{s}}_{\ell+1},\tilde{\bm{m}}_{\ell+1}|\ x)
\end{align}
with
\begin{align}
 \tilde{m}_{\ell,a}  \coloneqq m_{\ell,a}  +h 
 \;,\qquad 
  \tilde{s}_{\ell,a} \coloneqq s_{\ell,a}\ S \,.
\end{align}
It follows straightforwardly that the vector multiplet contributions behave as ($\ell\neq k$)
\begin{align}
      I_{\mathrm{cl}}^{\urm(\ell)}(w_\ell,\bm{s}_{\ell},\bm{m}_\ell;n_\ell)
       =&   I_{\mathrm{cl}}^{\urm(\ell)}(w_\ell,\tilde{\bm{s}}_{\ell},\tilde{\bm{m}}_\ell;n_\ell) 
       \cdot w_\ell^{-\ell\cdot h}  \cdot S^{-\ell \cdot n_\ell}\\
    I_{\mathrm{vec}}^{\urm(\ell)}(\bm{s}_{\ell},\bm{m}_{\ell}|\ x) 
    =&
    I_{\mathrm{vec}}^{\urm(\ell)}(\tilde{\bm{s}}_{\ell},\tilde{\bm{m}}_{\ell}|\ x)  
\end{align}
Lastly, consider the fundamental hypermultiplet
\begin{align}
     I_{\mathrm{hyp}}^{\urm(N{-}1)\times\urm(N)}(\bm{s_{N{-}1}},\bm{m_{N{-}1}}; \bm{y},\bm{k}|\ x)
    =& \prod_{a=1}^{N{-}1} \prod_{b=1}^N
I_{\mathrm{chi}}^{\frac{1}{2}}
\left(\tilde{s}_{N{-}1,a}\ S^{-1}\ y_{b}^{-1} , \tilde{m}_{N{-}1,a}-h - k_{b}|\ x \right) \notag \\
&\qquad \cdot 
I_{\mathrm{chi}}^{\frac{1}{2}}
\left(\tilde{s}_{N{-}1,a}^{-1}\ S  \  y_{b} , k_{b} - \tilde{m}_{N{-}1,a} + h|\ x \right) \\
=&I_{\mathrm{hyp}}^{\urm(N{-}1)\times\urm(N)}(\tilde{\bm{s}}_{N{-}1} S^{-1},\tilde{\bm{m}}_{N{-}1} - h; \bm{y},\bm{k}|\ x) \notag 
\end{align}
which becomes the contribution of a $N$ copies of a bifundamental hypermultiplet between $\urm(N{-}1)_{\tilde{\bm{s}}_{N{-}1}, \tilde{\bm{m}}_{N{-}1}}$ and $\urm(1)_{S,h}$.

\paragraph{Summation range.}
Originally, the $\urm(k)$ fluxes $\bm{m}_k$ are valued in $\Z^k$. The rewriting forces the summation range to be
$\left(h,l_1,\ldots,l_{k{-}1}\right) \in \bigcup_{i=0}^{k{-}1} \left(\Z +\frac{i}{k}\right)^k $, see \eqref{eq:UN_as_SUNxU1} and \eqref{eq:sum_range_SUxU1}.
Further, by redefining all other fluxes, one finds
\begin{align}
    \tilde{\bm{m}}_\ell \in  \bigcup_{i=0}^{k{-}1} \left(\Z +\frac{i}{k}\right)^\ell \qquad \forall \ell \neq k \,,
\end{align}
analogous to \eqref{eq:other_sum_range}.
\paragraph{Contour integral.}
Consider the Jacobian of the gauge fugacity transformation
\begin{alignat}{3}
   J\left(\bm{s}_{\ell} \to \tilde{\bm{s}}_\ell  \right)=& S^{-\ell}  &
    \qquad  &\Longrightarrow 
    \qquad &
    \oint \prod_{a=1}^\ell \frac{\diff s_{\ell,a}}{2\pi \im s_{\ell,a}} =&\oint \prod_{a=1}^\ell \frac{\diff \tilde{s}_{\ell,a}}{2\pi \im \tilde{s}_{\ell,a}} \,, \\
     J\left(\bm{s}_{k} \to (S,\bm{\sigma})  \right)=& \frac{k \cdot S^{1-k}}{\prod_{l=1}^{k{-}1} \sigma_l}  &
    \qquad  &\Longrightarrow 
    \qquad &
    \oint\prod_{a=1}^k \frac{\diff s_{\ell,a}}{2\pi \im s_{\ell,a}} =& 
    \oint
    \frac{\diff S}{2 \pi \im S} \cdot \prod_{j=1}^{k{-}1} \frac{\diff \sigma_{j}}{2\pi \im \sigma_{j}} \,.
\end{alignat}
such that the contour integration transitions nicely into the new variables, again integrated along the unit circle.
\paragraph{Discrete gauging.}
Next, gauge a $\Z_q$ subgroup of the $\urm(1)_t$ Cartan subgroup of the topological symmetry $\psurm(N)$. To do so, one performs a discrete Molien-Weyl sum
\begin{align}
    Z_{T[\surm(N)]}(w_k,\ldots) \to 
\frac{1}{q}   \sum_{p=0}^{q-1}Z_{T[\surm(N)]}(w_k,\ldots)\big|_{w_k\to f \cdot (\zeta_q)^p} 
\qquad \zeta_q =e^{ \frac{2\pi \im }{q}}
\end{align}
with $f$ to be determined. Following the rewriting induced on the $k$-th node $\urm(k) \to \urm(1) \times \surm(k)$, the affected terms are
\begin{align}
    \frac{1}{q} \sum_{p=0}^{q-1} \prod_{\ell \neq k} (w_\ell)^{-\ell \cdot h} 
    \cdot (w_k)^{-k\cdot h} F(\ldots)
    \to 
    \frac{1}{q} \sum_{p=0}^{q-1} \prod_{\ell \neq k} (w_\ell)^{- \ell \cdot h} 
    \cdot (f (\zeta_q)^p )^{-k\cdot h} F(\ldots) \,.
\end{align}
Using the same argument as in \eqref{eq:discrete_sum} and 
recalling that $h\in \bigcup_{j=0}^{k{-}1} (\Z +\frac{j}{k})$, one might want to consider the case $q =a \cdot k$ with $a\in \N$. Then $(a\cdot k)| (k\cdot h)$ implies $a | h$, which in particular requires that $h$ is integer-valued. This implies the collapse of the entire summation range onto the integers. Moreover, the $h$ summation is further restricted to $h\in a \Z$. To be explicit
\begin{align}
    \begin{cases}
     (h,l_1,\ldots,l_{k{-}1}) \in \bigcup_{i=0}^{k{-}1} \left(\Z +\frac{i}{k}\right)^k  \\
  \tilde{\bm{m}}_\ell \in  \bigcup_{i=0}^{k{-}1} \left(\Z +\frac{i}{k}\right)^\ell \qquad \forall \ell \neq k \,.
    \end{cases}
    \; \xrightarrow{\; \Z_{k\cdot a} \;} \;
    \begin{cases}
     (h,l_1,\ldots,l_{k{-}1}) \in  a\Z \oplus \left(\Z \right)^{k{-}1}  \\
  \tilde{\bm{m}}_\ell \in   \left(\Z\right)^\ell \qquad \forall \ell \neq k \,.
    \end{cases}
\end{align}
Defining $h = a \cdot \tilde{h}$ with $\tilde{h}\in \Z$
and focusing on the terms involving $\tilde{h}$
\begin{align}
    &\prod_{\substack{\ell=1\\ \ell \neq k}}^{N{-}1} w_\ell^{-\ell\cdot {a\cdot \tilde{h}}}
     \cdot \widehat{S}^{-\ell \cdot n_\ell} \;
    I_{\mathrm{cl}}^{\urm(1)}({f}^k,\widehat{S}^{-k},{-a\cdot \tilde{h}};n_k)  
    =
      \prod_{\ell=1}^{N{-}1} \left(\widehat{S}^{-\ell} \right)^{n_\ell}
    \left({f}^k 
    \cdot 
    \prod_{\substack{\ell=1\\ \ell \neq k}}^{N{-}1} w_{\ell}^{\ell}
    \right)^{ {-a\cdot \tilde{h}}} 
\end{align}
suggests to redefine the $\urm(1)$ gauge fugacity and topological fugacity as follows:
\begin{align}
    S=&\widetilde{S}^a 
    \quad \text{and} \quad
    f^k = \left( \frac{{Q^{-1}}}{\prod_{\substack{\ell=1\\ \ell \neq k}}^{N{-}1} w_{\ell}^{\ell}}   \right) \,,
\end{align}
which coincides with \eqref{eq:relable_topo_discrete_gauge}. As a consequence,
\begin{align}
      \prod_{\ell=1}^{N{-}1} \left( S^{-\ell} \right)^{n_\ell}
      =S^{-\sum_{\ell=1}^{N{-}1} \ell n_\ell}
      =&\widetilde{S}^{-a \cdot \sum_{\ell=1}^{N{-}1} \ell n_\ell} \,,\\
      \prod_{\substack{\ell=1\\ \ell \neq k}}^{N{-}1} w_\ell^{-\ell\cdot {a\cdot \tilde{h}}}
     \cdot S^{\ell \cdot n_\ell} \;
    I_{\mathrm{cl}}^{\urm(1)}({f}^k,\widehat{S}^{-1},{-a\cdot \tilde{h}};k\cdot n_k)  
    =&\widetilde{S}^{-a\cdot \sum_{\ell=1}^{N{-}1} \ell n_\ell}
    Q^{a \cdot \tilde{h}}  \notag \\
    =&I_{\mathrm{cl}}^{\urm(1)}({Q},\tilde{S},{a\cdot \tilde{h}};a \cdot \sum_{\ell=1}^{N{-}1} \ell n_\ell) \,.
\end{align}
Putting all pieces together, one obtains 
\begin{align}
\tilde{Z} =& \frac{1}{a \cdot k} \sum_{p=0}^{a\cdot k{-}1} Z_{T[\surm(N)]} (w_k,\ldots) \big|_{w_k \to f \cdot (\zeta_q)^p } \notag\\
=&
\sum_{\bm{l}\in\Z^{k{-}1}}
\frac{1}{|\Wcal_{\bm{l}}|}
\oint \prod_{j=1}^{k{-}1} \frac{\diff \sigma_j}{2 \pi i \sigma_j}
\; 
\sum_{{\tilde{h}\in \Z}}
\oint \frac{ \diff \tilde{S}}{2\pi i \tilde{S}} 
\;
\prod_{\substack{\ell= 1 \\ \ell\neq k}}^{N{-}1}
\left[
\sum_{\tilde{\bm{m}}_\ell \in \Z^\ell} 
\frac{1}{|\Wcal_{\bm{m}}|}
\oint \prod_{j=1}^{\ell} \frac{\diff \tilde{s}_{\ell,j}}{ 2\pi i \tilde{s}_{\ell,j}} \right] \\
&\cdot \prod_{\substack{\ell=1\\ \ell \neq k{-}1,k}}^{N-2} I_{\mathrm{hyp}}^{\urm(\ell)\times\urm(\ell+1)}(\tilde{\bm{s}}_{\ell},\tilde{\bm{m}}_{\ell}; \tilde{\bm{s}}_{\ell+1},\tilde{\bm{m}}_{\ell+1}|\ x)
\notag \\
& 
\cdot I_{\mathrm{hyp}}^{\urm(k{-}1)\times\surm(k)}( \tilde{\bm{s}}_{k{-}1},\tilde{\bm{m}}_{k{-}1}; \bm{\sigma},\bm{l}|\ x)
\cdot
I_{\mathrm{hyp}}^{\surm(k)\times\urm(k+1)}(\bm{\sigma},\bm{l}; \tilde{\bm{s}}_{k+1},\tilde{\bm{m}}_{k+1}|\ x) 
\notag\\
&\cdot 
I_{\mathrm{hyp}}^{\urm(N{-}1)\times\urm(N)}(\tilde{\bm{s}}_{N{-}1} {\tilde{S}^{a}},\tilde{\bm{m}}_{N{-}1} + {a\cdot \tilde{h}}; \bm{y},\bm{k}|\ x)
\notag \\
&\cdot \prod_{\substack{\ell=1\\ \ell \neq k}}^{N{-}1}
I_{\mathrm{vec}}^{\urm(\ell)}(\tilde{\bm{s}}_{\ell},\tilde{\bm{m}}_{\ell}|\ x) 
\; \cdot 
 I_{\mathrm{cl}}^{\urm(\ell)}(w_\ell,\tilde{\bm{s}}_{\ell},\tilde{\bm{m}}_\ell;n_\ell)
 \notag \\
&  \cdot I_{\mathrm{vec}}^{\surm(k)}(\bm{\sigma},\bm{l}|\ x) 
\cdot
I_{\mathrm{cl}}^{\urm(1)}({Q},\tilde{S},{a\cdot \tilde{h}};\frac{1}{k}\sum_{\ell=1}^{N{-}1} \ell n_\ell) \notag 
\end{align}
which is the superconformal index for 
\begin{align}
    \raisebox{-0.5\height}{
    \includegraphics[page=4]{figures/figures_1-form_gauging_in_index.pdf}
}
\end{align}
wherein the $N$ copies of hypermultiplets between $\urm(N{-}1)$ and $\urm(1)$ transform as $(\boldsymbol{N{-}1})\times (-a)$, i.e.\ the $\urm(1)$ charge is $a$.


\section{Mirror maps}
\label{app:mirror_map}
Suppose that there is a $G$-type global symmetry with $G$ being a classical or exceptional Lie algebra. The root space fugacities $w_i$ are related to the weight space fugacities $\eta_i$ by the Cartan matrix
\begin{align}
w_i = \prod_j \eta_j^{C_{ij}} 
\qquad 
\eta_i = \prod_j w_j^{C^{-1}_{ij}}  \,.
\end{align}
Most relevant here is the closed formula for the inverse of the $A_{N{-}1}$ Cartan matrix 
\begin{align}
C^{-1}_{ij} = \min(i,j) - \frac{i\cdot j}{N}
\qquad \text{for }
i,j=1,\ldots,N{-}1
\label{eq:inv_Cartan_A}
\end{align}
given in \cite{Wei_2017}.
\subsection{SQED and its abelian mirror quiver}
\label{app:mirror_map_SQED}
\subsubsection{Standard mirror map}
Suppose that the abelian mirror has topological fugacities $w_i$ with $i=1,\ldots,N{-}1$, i.e.\ root space fugacities for $\surmL(N)$. The mirror theory, SQED with $N$ fundamental flavours, has $\urm(N)$ fugacities $y_f$ with $f=1,\ldots,N $. Both sets of fugacities are related to the $\surmL(N)$ weight space fugacities $\eta_i$ via the following two maps
\begin{align}
\eta_i &= \prod_{j=1}^{N{-}1} z_j^{C_{ij}^{-1}} 
\qquad \text{with $C_{ij}$ the $A_{N{-}1}$ Cartan matrix} \\
y_f &= \prod_{i=1}^{N{-}1} \eta_i^{ M^{[N]}_{fi} }
\qquad \text{with }
M_{fi}^{[N]} = \delta_{f,i} - \delta_{f,i+1} \quad 
\text{for } 
\begin{cases}
i&=1,\ldots,N{-}1 \;,\\
f&=1,\ldots,N \,.
\end{cases}
\end{align}
and the inverse Cartan matrix is given in \eqref{eq:inv_Cartan_A}. 
The combination 
\begin{align}
y_f = \prod_{j=1}^{N{-}1} w_j^{\sum_{i=1}^{N{-}1} M_{fi}^{[N]} C^{-1}_{ij}} 
\quad \text{with }
\sum_{i=1}^{N{-}1} M_{fi}^{[N]} C^{-1}_{ij}
= \min(f,j) - \min(f-1,j) - \frac{j}{N}
\end{align}
can be simplified by observing 
\begin{align}
    \min(f,j) - \min(f-1,j) 
    =\begin{cases}
    1\, ,& j>f-1 \\
    0\, , & j \leq f-1
    \end{cases}
\end{align}
for  $f=1,\ldots, N$ and  
    $j = 1,\ldots,N{-}1$.
Hence, the mirror map becomes
\begin{align}
y_f = \prod_{j=1}^{N{-}1} w_j^{\sum_{i=1}^{N{-}1}  M^{[N]}_{fi} C_{ij}^{-1}}
= \prod_{i=1}^{f-1} w_i^{-\frac{i}{N}}
\prod_{j=f}^{N{-}1} w_j^{1-\frac{j}{N}}
\label{eq:mirror_map_min_A-type_orbit}
\end{align}
for $f=1,\ldots,N$.

\subsubsection{Mirror map after gauging}
\label{app:mirror-map_SQED_after}
Suppose that one gauges a $\Z_q$ on the $\urm(1)_t$ generated by $w_k$ and employs the following parameter map 
\begin{align}
\label{eq:abelian_quiver_fugacities}
\begin{cases}
    w_i = \prod_{j=1}^{k-1} x_j^{C_{ij}^{A_{k-1}}} \;, \quad i=1,2,\ldots, k-1 \\
    w_k = \frac{Q}{(x_{k-1} \ u_1)^q}
    \\
    w_{i+k} = \prod_{j=1}^{N{-}k-1} u_j^{C_{ij}^{A_{N{-}k-1}}} \;, \quad i=1,2,\ldots, N{-}k-1
\end{cases}
\end{align}
using the Cartan matrices of $A_{k-1}$ and $A_{N{-}k-1}$, respectively.
Using \eqref{eq:mirror_map_min_A-type_orbit} for $A_{k-1}$ and $A_{N{-}k-1}$, the map for $w_k$ can be expressed as
\begin{align}
w_k = \zeta_q \frac{Q^{\frac{1}{q}}}{
\prod_{i=1}^{k-1} w_i^{\frac{i}{k}}
\prod_{j=k+1}^{N{-}1} w_{j}^{1-\frac{j-k}{N{-}k}}
}  \,.
\end{align}
A straightforward computation yields the new mirror map
\begin{subequations}
\label{eq:mirror_map_SQED_after_gauging}
\begin{align}
y_f &=
\begin{cases}
\zeta_q^{1-\frac{k}{N}} Q^{\frac{N{-}k}{q\cdot N}}
\cdot 
\prod_{i=1}^{f-1} w_i^{-\frac{i}{k}}
\cdot
\prod_{j=f}^{k-1} w_j^{1-\frac{j}{k}} \;,
& f \leq k \\
    \zeta_q^{-\frac{k}{N}} Q^{\frac{-k}{N\cdot q}}
   \cdot
   \prod_{i=1}^{f-k-1} w_{k+i}^{-\frac{i}{N{-}k}}
   \cdot
    \prod_{j=f-k}^{N{-}k-1} w_{k+j}^{1-\frac{j}{N{-}k}}
\;, & f \geq k+1 
\end{cases} \\
&=
\begin{cases}
\zeta_q^{1-\frac{k}{N}} Q^{\frac{N{-}k}{q\cdot N}}
\cdot 
\prod_{i=1}^{k-1} x_i^{M^{[k]}_{fi}} \;,
& f \leq k \\
    \zeta_q^{-\frac{k}{N}} Q^{\frac{-k}{N\cdot q}}
   \cdot
   \prod_{i=1}^{N{-}k-1} u_i^{M^{[N{-}k]}_{fi}}
\;, & f \geq k+1 
\end{cases} 
\end{align}    
\end{subequations}
using the parametrisation \eqref{eq:abelian_quiver_fugacities} for the $A_{k-1}$ fugacities $x_i$ and the $A_{N{-}k-1}$ fugacities $u_i$.

\paragraph{Remark.}
The map \eqref{eq:mirror_map_SQED_after_gauging} assigns $\zeta_q$ charges to each $y_f$; however, one can remove any overall $\urm(1)$ phase by a gauge transformation. This leads to two convenient choices: either the first $k$ fundamental flavours are charged under $\Z_q$
\begin{align}
y_f&=
\begin{cases}
\zeta_q \cdot Q^{\frac{N{-}k}{q\cdot N}}
\cdot 
\prod_{i=1}^{k-1} x_i^{M^{[k]}_{fi}} \;,
& f \leq k \\
     Q^{\frac{-k}{N\cdot q}}
   \cdot
   \prod_{i=1}^{N{-}k-1} u_i^{M^{[N{-}k]}_{fi}}
\;, & f \geq k+1 
\end{cases} 
\label{eq:mirror_map_SQED_after_gauging_rotated}
\end{align}
by rotating via $\zeta_q^{\frac{k}{N}}$.
Alternatively, one rotates via $\zeta_q^{-1+\frac{k}{N}}$ such that only the last $N{-}k$ fundamental flavours are non-trivially charged under $\Z_q$.
In principle, one could also extend the overall rotation to include $Q$, but there is no need to do so.

\subsection{\texorpdfstring{$T[\surm(N)]$ theories}{TSUN theories}}
\label{app:mirror_map_TSUN}
One can construct the mirror map explicitly.
\subsubsection{Standard mirror map}
Denote the Coulomb branch root space fugacities of $T[\surm(N)]$ by $\{w_i\}_{i=1}^{N{-}1}$. These can be mapped to the Coulomb branch weight space fugacities via the $A_{N{-}1}$ Cartan matrix $C_{ij}$:
\begin{align}
    w_i = \prod_{j} \omega_j^{C_{ij}} \,.
\end{align}
The Higgs branch $\urm(N)$ flavour fugacities are $\{y_a\}_{a=1}^N$, which are reduced to $\surm(N)$ fugacities $\{\eta_i\}_{i=1}^{N{-}1}$ via
\begin{align}
\label{eq:transformation_U2SU}
y_f = \prod_{i=1}^{N{-}1} \eta_i^{M_{fi}^{[N]}} 
\quad \text{ with } \quad 
M_{fi}^{[N]} = \delta_{f,i} - \delta_{f,i+1} \quad 
\text{for } 
\begin{cases}
i&=1,\ldots,N{-}1 \;,\\
f&=1,\ldots,N \,.
\end{cases}
\end{align}
The self-mirror property of $T[\surm(N)]$ is the reflection in the exchange $\omega_i \leftrightarrow \eta_i$.

The aim is to express the natural flavour fugacities $\{y_a\}$ of the theory in terms of the Coulomb branch fugacities of the mirror. The first step is 
\begin{align}
    (y_1,y_2,\ldots,y_N) &\to \left(\eta_1, \frac{\eta_2}{\eta_1},\ldots,\frac{1}{\eta_{N{-}1}} \right) \notag \\
    &\xrightarrow{\omega_i \leftrightarrow \eta_i}
     \left(\omega_1,\frac{\omega_2}{\omega_1},\ldots,\frac{1}{\omega_{N{-}1}} \right)  \notag \\
     &\xrightarrow{\omega_i \to \prod_j w_j^{C^{-1}_{ij}}}
     \left(f_1(w_i),f_2(w_i),\ldots,f_{N}(w_i) \right) \,.
     \label{eq:transformation_topo2flavour}
\end{align}
This map can be made explicit by using 
the inverse of the $A_{N{-}1}$ Cartan matrix \eqref{eq:inv_Cartan_A}. 
Analogous to the abelian case, the combined map reads
\begin{align}
y_f &= \prod_{j=1}^{N{-}1} w_j^{\sum_{i=1}^{N{-}1} M_{fi}^{[N]} C^{-1}_{ij}} 
\quad \text{with }
\sum_{i=1}^{N{-}1} M_{fi}^{[N]} C^{-1}_{ij}
= \min(f,j) - \min(f-1,j) - \frac{j}{N} \notag \\
&= \prod_{i=1}^{f-1} w_i^{-\frac{i}{N}}
\ \prod_{j=f}^{N{-}1} w_j^{1-\frac{j}{N}} \,,
\label{eq:mirror_map_TSUN}
\end{align}
which is the explicit form of \eqref{eq:transformation_topo2flavour}.

\subsubsection{Mirror map after gauging}
The next step is utilising the parameter map \eqref{eq:relable_topo_discrete_gauge} established in Appendix \ref{app:discrete_gauge_TSUN} 
\begin{align}
w_k = \zeta_q
     \left( \frac{v^{-1}}{\prod_{\substack{i=1 \\ i\neq k}}^{N{-}1} w_i^i  } \right)^{\frac{1}{k}} \,.
\end{align}
Lastly, to make contact with the global symmetries, one uses the Cartan matrix for $A_{k-1}$ and $A_{N{-}1-k}$ in a standard fashion
\begin{align}
\begin{aligned}
w_i &= \prod_{j=1}^{k-1} x_j^{C^{A_{k-1}}_{ij} } \;, \quad i=1,\ldots,k-1
\;,  \\
w_{i+k} &= \prod_{j=1}^{N{-}1-k} u_j^{C^{A_{N{-}1-k}}_{ij} } \;, \quad i=1,\ldots,N{-}1-k \,,
\end{aligned}
\end{align}
and one needs to redefine 
\begin{align}
    v = \frac{Q^{-1}}{\left(u_{N{-}1-k}\right)^N }
    = \frac{Q^{-1}}{
    \prod_{r=k}^{N{-}1} w_r^{\frac{r-k}{N{-}k}\cdot  N}
    }
    \,.
\end{align}
Applying this to \eqref{eq:mirror_map_TSUN}, one finds 
\begin{subequations}
\label{eq:mirror_map_TSUN_after_gauging}
   \begin{align}
y_f
&=
\begin{cases}
\zeta_q^{1-\frac{k}{N}}
\cdot
Q^{\frac{N{-}k}{k\cdot N}} 
\cdot
\prod_{i=1}^{f-1} w_i^{-\frac{i}{k}} 
\cdot
\prod_{j=f}^{k-1} w_j^{1-\frac{j}{k}}
 \;, & \text{ for } f\leq k \\
\zeta_q^{-\frac{k}{N}} \cdot 
Q^{-\frac{1}{N}}
\cdot
\prod_{i=1}^{f-k-1}
w_{i+k}^{-\frac{i}{N{-}k}}
\cdot
\prod_{j=f-k}^{N{-}k-1}
w_{j+k}^{1-\frac{j}{N{-}k}}
\;,
 & \text{ for } f\geq k+1
\end{cases}
\\
&= 
\begin{cases}
\zeta_q^{1-\frac{k}{N}}
\cdot
Q^{\frac{N{-}k}{k\cdot N}} 
\cdot
\prod_{j=1}^{k-1} x_j^{M_{fj}^{[k]}}
 \;, & \text{ for } f\leq k \\
 \zeta_q^{-\frac{k}{N}} \cdot 
Q^{-\frac{1}{N}}
\cdot
\prod_{j=1}^{N{-}k-1} u_j^{M_{fj}^{[N{-}k]}}
 & \text{ for } f\geq k+1
\end{cases}
\end{align} 
\end{subequations}
which displays the split into $A_{k-1}$ fugacities $x_j$ and $A_{N{-}k-1}$ fugacities $u_j$.

\paragraph{Remark.}
Analogously to the SQED case, one can simplify the $\zeta_q$ dependence in \eqref{eq:mirror_map_TSUN_after_gauging} by a suitable overall $\urm(1)$ rotation. A convenient choice is then given by
\begin{align}
y_f&= 
\begin{cases}
\zeta_q
\cdot
Q^{\frac{N{-}k}{k\cdot N}} 
\cdot
\prod_{j=1}^{k-1} x_j^{M_{fj}^{[k]}}
 \;, & \text{ for } f\leq k \\
Q^{-\frac{1}{N}}
\cdot
\prod_{j=1}^{N{-}k-1} u_j^{M_{fj}^{[N{-}k]}}
 & \text{ for } f\geq k+1
\end{cases}
\label{eq:mirror_map_TSUN_after_gauging_rotated}
\end{align}
such that only the first $k$ fundamental flavours are charged under $\Z_q$.
\subsection{\texorpdfstring{Examples for $T_\rho^\sigma[\surm(N)]$}{Example for T rho sigma}}
\label{app:mirror_map_T-rho_sigma}

\subsubsection{Example 1}
Consider the example $T_\rho^\sigma[\surm(15)]$ with $\sigma=[3^2,2^2,1^2]$ and $\rho=[6,4,3,1^2]$ of Section \ref{sec:T_sigma_rho}.
Using the labelling 
\begin{align}
    \raisebox{-0.5\height}{
    \includegraphics[page=16]{figures/figures_1-form_T_rho_sigma.pdf}
}
\quad \longleftrightarrow \quad 
 \raisebox{-0.5\height}{
    \includegraphics[page=17]{figures/figures_1-form_T_rho_sigma.pdf}
}
\end{align}
the mirror map is given by
    \begin{align}
    \begin{cases}
y_1 &= w_1^{4/5} w_2^{3/5} w_3^{2/5} \sqrt[5]{w_4}
\;, \\
y_2&= \frac{w_2^{3/5} w_3^{2/5} \sqrt[5]{w_4}}{\sqrt[5]{w_1}}  
\;,   
    \end{cases}
    \qquad
\begin{cases}
Q_1 &= \frac{w_3^{2/5} \sqrt[5]{w_4}}{\sqrt[5]{w_1} w_2^{2/5}}
\;, \\
Q_2 &= \frac{\sqrt[5]{w_4}}{\sqrt[5]{w_1} w_2^{2/5} w_3^{3/5}}
\;, \\
Q_3 &= \frac{1}{\sqrt[5]{w_1} w_2^{2/5} w_3^{3/5} w_4^{4/5}}
\end{cases}
\label{eq:mirror_map_TSU15}
\end{align}
such that $\prod_{i=1}^2 y_i \cdot \prod_{j=1}^3 Q_j =1 $ holds.

\paragraph{Gauging on $w_2$.}
The example of Figure \ref{fig:T-rho-sigma-ex2} is realised by a $\Z_2$ gauging on $w_2$. Inspecting the mirror map \eqref{eq:mirror_map_TSU15} shows that one has two options for the $\Z_2^f$ gauging in the mirror
\begin{compactitem}
\item The $y_1,y_2$ are charged as $\zeta_2$, while the $Q_{1,2,3}$ are trivial under $\Z_2$.
\item The $y_{1,2}$ are trivial under $\Z_2$, and the $Q_{1,2,3}$ transform with $\zeta_2$.
\end{compactitem}
This reflects the two choices in Figure \ref{fig:T-rho-sigma-ex2}.

\paragraph{Gauging on $w_3$.}
Turning to Figure \ref{fig:T-rho-sigma-ex1}, one performs a $\Z_2$ gauging on $w_3$. The mirror map \eqref{eq:mirror_map_TSU15} indicates two options for the $\Z_2^f$ gauging in the mirror
\begin{compactitem}
\item The $y_1,y_2,Q_1$ are charged as $\zeta_2$, while the $Q_{2,3}$ are trivial under $\Z_2$.
\item The $y_{1,2},Q_1$ are trivial under $\Z_2$, and the $Q_{2,3}$ transform with $\zeta_2$.
\end{compactitem}
Again, this confirms the two choices in Figure \ref{fig:T-rho-sigma-ex1}.

\subsubsection{Example 2}
The labelling for the $T_\rho^\sigma[\surm(9)]$ example with $\rho=(3,2^3)$ and $\sigma=(3^2,1^3)$ of Section \ref{sec:T_sigma_rho} is defined by
\begin{align}
  \raisebox{-.5\height}{
    \includegraphics[page=22]{figures/figures_1-form_T_rho_sigma.pdf}
    }
\qquad \longleftrightarrow\qquad
    \raisebox{-.5\height}{
    \includegraphics[page=23]{figures/figures_1-form_T_rho_sigma.pdf}
    }
    \label{eq:T-sigma-rho-ex3_labels}
\end{align}
and the mirror map is given by
\begin{align}
y_1 = \frac{w_2^{\frac{1}{2}} w_3^{\frac{1}{4}} }{ w_1^{\frac{1}{4}} } 
\;, \quad 
y_2 = \frac{w_3^{\frac{1}{4}} }{ w_1^{\frac{1}{4}} w_2^{\frac{1}{2}}  }
\;, \quad 
y_3 = \frac{1 }{ w_1^{\frac{1}{4}}  w_2^{\frac{1}{2}} w_3^{\frac{3}{4}}  }
\;, \quad 
Q =  w_1^{\frac{3}{4}}  w_2^{\frac{1}{2}} w_3^{\frac{1}{4}}  
\;.
\label{eq:mirror_map_T-sigma-rho_ex3}
\end{align}
Analogous to the example above, gauging the $\Z_2^f$ associated to $w_3$ has two convenient realisations in the mirror theory: either $(y_1,y_2,Q)$ transform non-trivial under $\Z_2$ and $y_3$ is trivial or vice versa. 
\subsection{\texorpdfstring{$\sprm(k)$ SQCD and its $D$-type unitary mirror quiver}{Spk SQCD and its D-type unitary mirror quiver}}
The closed formula for the inverse Cartan matrix of $D_N$ is provided in \cite{Wei_2017}.
\subsubsection{Standard mirror map}
For the balanced $D_N$-type unitary quiver, the mirror map to the flavour symmetry of the $\sprm(k)$ SQCD mirror with $N$ fundamental flavours is given by
\begin{compactitem}
\item Let $y_i$ denote the $\urm(N)$ flavour fugacities.
\item Denote by $x_i$ the $\sorm(2N)$ weight space fugacities. The relation between both is established via
\begin{align}
y_i= \begin{cases}
x_1 \;,&\text{for } i=1 \\
\frac{x_i}{x_{i-1}} \;,  &\text{for } 1<i<N-2 \\
\frac{x_{N{-}1} x_N }{x_{N-2} } \; , &\text{for } i=N{-}1 \\
\frac{x_{N{-}1} }{x_{N} }\; , &\text{for } i=N
\end{cases}
\end{align}
\item Denote by $w_i$ the root space fugacities of $D_N$, which are related to the weight space fugacities $x_i$ via the $D_N$ Cartan matrix
\begin{align}
w_i = \prod_{j} x_j^{C_{ij}^{D_N}}
\end{align}
\item Thus, one finds the map between fundamental flavour fugacities $y_i$ and root space fugacities $w_i$ to be
\begin{align}
y_f =
\begin{cases}
\left( \prod_{i=f}^{N-2} w_i  \right)\sqrt{w_{N{-}1} } \cdot \sqrt{w_{N} } \;, &\text{ for } 1\leq f < N{-}1  \\
\sqrt{w_{N{-}1} } \cdot \sqrt{w_{N} } 
\;, &\text{ for } f = N{-}1\\
\frac{\sqrt{w_{N{-}1} }}{\sqrt{w_{N} }}
\;, &\text{ for } f = N
\end{cases}
\label{eq:mirror_map_D-type_Dynkin}
\end{align}
\end{compactitem}

\subsubsection{Mirror map after gauging}
Suppose that one gauges a discrete $\Z_q$ symmetry at the gauge node with topological fugacity $w_l$. Then, analogous to the $T[\surm(N)]$ derivation, the fugacity map to the quiver after gauging is simply given by
\begin{align}
    w_l = \left( \frac{v^{-1}}{ \prod_{\substack{i=1 \\ i \neq l}}^N (w_i)^{N_i} }
    \right)^{\frac{1}{l}}
\end{align}
where $N_i$ denotes the rank of the $i$-th node. The remaining topological fugacities $w_{i \neq l}$ are identified before and after gauging.

Lastly, one needs to redefine $v$ such that $A_{l-1}$ and $D_{N-l}$ representations become manifest. For this, one uses
\begin{align}
v= \frac{Q}{ \text{weight space fugacity at extra $\urm(1)$} }
\label{eq:extra_U1_D-type}
\end{align}
here, the weight space fugacity is either the $A_{l-1}$ fugacity $x_i$, if the extra $\urm(1)$ intersects the balanced $A_{l-1}$ Dynkin diagram at node $i$, or it is the $D_{N-l}$ weight space fugacity $u_j$, if the extra $\urm(1)$ is attached at the $j$-th node of the balanced $D$-type Dynkin diagram. See Appendix \ref{app:Sp_SQCD_D-type_mirror} for examples.

For $l\leq N-2$, the mirror map is given by\footnote{The choice \eqref{eq:extra_U1_D-type} implies that the $\urm(1)_Q$ charges are negative here. This choice is convenient because the charges of $Q$ under the centre symmetries are directly read off from the Coulomb branch quiver, see \eqref{eq:D-type_SU_before_flavour} and \eqref{eq:D-type_SU_behind_flavour}.} 
\begin{align}
\label{eq:mirror_map_D-type_after_gauging}
y_f =
\begin{cases}
Q^{-\frac{1}{l}}
\prod_{i=1}^{f-1} w_i^{-\frac{i}{l}}
\prod_{j=f}^{l-1} w_j^{1-\frac{j}{l}}
&\text{ for } 1 \leq f \leq l
\\
\left( \prod_{i=f}^{N-2} w_i  \right)\sqrt{w_{N{-}1} } \cdot \sqrt{w_{N} } \;, &\text{ for } l+1\leq f < N{-}1  \\
\sqrt{w_{N{-}1} } \cdot \sqrt{w_{N} } 
\;, &\text{ for } f = N{-}1\\
\frac{\sqrt{w_{N{-}1} }}{\sqrt{w_{N} }}
\;, &\text{ for } f = N
\end{cases}
\end{align}
which clearly displays that the first $l$ fundamental flavours transform under $\surmL(l)\times \urm(1)$, and the remaining $N-l$ fundamental flavours transform under $\sormL(2N-2l)$.

\subsection{\texorpdfstring{$\orm(2k)$ SQCD and its $C$-type unitary mirror quiver}{O(2k) SQCD and its C-type unitary mirror quiver}}

Consider the $\orm(2k)$ SQCD with $N$ fundamental hypermultiplets. The mirror theory is a balanced $C$-type Dynkin quiver with $N$ gauge nodes.
\begin{compactitem}
\item Denote the $\urm(N)$ flavour fugacities by $y_i$.
\item Denote the $\sprm(N)$ flavour fugacities by $x_i$. These are related via the transformation
\begin{align}
y_1 = x_1  \quad \text{and} \quad  
y_j = \frac{x_j}{x_{j-1}}
\quad \text{for } j=2,\ldots,n \,.
\end{align}
\item Denote the topological fugacities of the $C$-type quiver by $w_i$. Then the $C_N$ Cartan matrix mediates the transformation between root and weight space fugacities
\begin{align}
w_i = \prod_{j=1}^N x_j^{C_{ij}^{C_N}} \quad \text{for } i=1, \ldots, N\,.
\end{align}
\end{compactitem}
Combining the above leads to the mirror map between the unitary flavour fugacities and the root space topological fugacities
\begin{align}
\label{eq:mirror_map_C-type_Dynkin}
\begin{cases}
y_i =\sqrt{w_N} \ \prod_{j=i}^{N{-}1} w_j  & i=1,\ldots,N{-}1 \\
y_N = \sqrt{w_N} &
\end{cases}
\end{align}

\section{Explicit Hilbert series results}
In this appendix, some exemplary Hilbert series calculations are presented. Matching the Hilbert series can be used as the stringent test to check the dualities and find the global topological or flavour symmetry groups.

\subsection{Linear Abelian quiver}
\label{app:abelian_quiver}
Consider the abelian quiver gauge theory
\begin{align}
       \raisebox{-.5\height}{
    \includegraphics[page=1]{figures/figures_1-form_abelian_quivers.pdf}
    } 
\end{align}

Explicit character expansions indicate the following global forms
\begin{alignat}{2}
&\substack{N=3\\k=1} & \quad  
G_t &= 
\begin{cases}
\sorm(3)_y \times \urm(1)_Q \, ,\  \text{w/ $Q$ of $\Z_2$-charge $0$} & q=2,4,6,8 \\
\frac{\surm(2)_y \times \urm(1)_Q }{ \Z_2}\, ,\  \text{w/ $Q$ of $\Z_2$-charge $+1$}  & q = 3,5,7,9
\end{cases} \\
&\substack{N=4\\k=1} & \quad  
G_t &= 
\begin{cases}
\psurm(3)_y \times \urm(1)_Q  \, ,\  \text{w/ $Q$ of $\Z_3$-charge $0$} & q=3,6 \\
\frac{\surm(3)_u \times \urm(1)_Q }{ \Z_3}\, ,\  \text{w/ $Q$ of $\Z_3$-charge $+1$}  & q = 4,7 \\
\frac{\surm(3)_u \times \urm(1)_Q }{ \Z_3}\, ,\  \text{w/ $Q$ of $\Z_3$-charge $+2$}  & q = 2,5
\end{cases} \\
&\substack{N=4\\k=2} & \quad  
G_t &= 
\begin{cases}
\sorm(3)_x \times \urm(1)_Q \times \sorm(3)_u  \, ,\  \text{w/ $Q$ of $\Z_2\times \Z_2$-charge $(0,0)$} & q=2,4 \\
\frac{\surm(2)_x \times \urm(1)_Q  \times \surm(2)_u}{ \Z_2 \times \Z_2}\, ,\  \text{w/ $Q$ of $\Z_2\times \Z_2$-charge $(+1,+1)$}  & q = 3,5
\end{cases} \\
&\substack{N=5\\k=1} & \quad  
G_t &= 
\begin{cases}
\frac{\surm(4)_u \times \urm(1)_Q }{\Z_4} \, ,\  \text{w/ $Q$ of $\Z_4$-charge $(+2)$} & q=2 \\
\frac{\surm(4)_u \times \urm(1)_Q }{\Z_4} \, ,\  \text{w/ $Q$ of $\Z_4$-charge $(+3)$} & q=3 \\
\psurm(4)_u \times \urm(1)_Q \, ,\  \text{w/ $Q$ of $\Z_4$-charge $(0)$} & q=4 \\
\frac{\surm(4)_u \times \urm(1)_Q }{\Z_4} \, ,\  \text{w/ $Q$ of $\Z_4$-charge $(+1)$} & q=5 
\end{cases} \\
&\substack{N=5\\k=2} & \quad  
G_t &= 
\begin{cases}
\frac{\surm(2)_x \times \urm(1)_Q \times \surm(3)_u }{\Z_2 \times\Z_3} \, ,\  \text{w/ $Q$ of $\Z_2\times \Z_3$-charge $(0,+2)$} & q=2 \\
\frac{\surm(2)_x \times \urm(1)_Q  }{\Z_2 } \times \psurm(3)_u \, ,\  \text{w/ $Q$ of $\Z_2\times \Z_3$-charge $(+1,0)$} & q=3 \\
\psurm(2)_x \times \frac{ \urm(1)_Q \times \surm(3)_u }{\Z_3} \, ,\  \text{w/ $Q$ of $\Z_2\times \Z_3$-charge $(0,+1)$} & q=4 \\ 
\frac{\surm(2)_x \times \urm(1)_Q \times \surm(3)_u }{\Z_2 \times\Z_3} \, ,\  \text{w/ $Q$ of $\Z_2\times \Z_3$-charge $(+1,+2)$} & q=5 \\
\psurm(2)_x \times \urm(1)_Q \times \psurm(3)_u  \, ,\  \text{w/ $Q$ of $\Z_2\times \Z_3$-charge $(0,0)$} & q=6 \\
\frac{\surm(2)_x \times \urm(1)_Q \times \surm(3)_y }{\Z_2 \times\Z_3} \, ,\  \text{w/ $Q$ of $\Z_2\times \Z_3$-charge $(+1,+1)$} & q=7 
\end{cases}
\end{alignat}
which then confirms the general formula \eqref{eq:sym_group_abelian_quiver}.

To exemplify, we provide the Hilbert series for $N=5$, $k=2$, $q=2$ here.
The perturbative expansion reads
\begin{align}
\HS= 
1
&+t \left(\phi _{1,1}+\chi _2+1\right)\\
&+t^2 \left(Q \chi _2 \phi _{0,2}+\frac{\chi _2 \phi _{2,0}}{Q}+\chi _2 \phi _{1,1}+\phi _{1,1}+\phi _{2,2}+\chi _2+\chi _4+1\right) \notag\\
&+t^3 \bigg(
Q \left(\chi _2 \phi _{0,2}+\chi _4 \phi _{0,2}+\chi _2 \phi _{1,3}\right)
+\frac{\chi _2 \phi _{2,0}+\chi _4 \phi _{2,0}+\chi _2 \phi _{3,1}}{Q} \notag\\
&\qquad +\chi _2 \phi _{1,1}+\chi _2 \phi _{2,2}+\chi _4 \phi _{1,1}+\phi _{1,1}+\phi _{2,2}+\phi _{3,3}+\chi _2+\chi _4+\chi _6+1\bigg)
\notag\\
&+\ldots 
\notag
\end{align}
with $\phi_{m_1,m_2}=\phi_{m_1,m_2}(u_i)$ and $\chi_{n_1}=\chi_{n_1}(x_1)$ the $\surm(3)$ and $\surm(2)$ characters of irreps $[m_1,m_2]$ and $[n_1]$, respectively.
This follows via the Higgs branch fugacity map
\begin{align}
y_1\to Q^{-\frac{1}{5}} u_1\,, \quad
y_2\to Q^{-\frac{1}{5}} \frac{u_2}{u_1}\,, \quad
y_3\to Q^{-\frac{1}{5}} \frac{1}{u_2}\,, \quad
y_4\to Q^{\frac{3}{10}}  x_1  \,, \quad
y_5\to  Q^{\frac{3}{10}} \frac{ 1}{x_1}
\end{align}
and the Coulomb branch fugacity map
\begin{align}
w_1\to x_1^2\,,\quad 
w_2\to \frac{Q}{x_1^2 u_1^2}\,,\quad 
w_3\to \frac{u_1^2}{u_2}\,,\quad 
w_4\to \frac{u_2^2}{u_1} \,.
\end{align}
Here, $\{u_i\}$ and $x_1$ are the corresponding weight space fugacities. $Q$ denotes the $\urm(1)$ fugacity.
In fact, due to the abelian nature, one can even compute the full highest weight generating function (HWG)
\begin{align}
\mathrm{HWG}_{\text{standard}}&=\frac{1}{1-k_1 k_4 t} \\
\mathrm{HWG}_{\Z_2 \text{ gauging}}&=\frac{1-m_1^2 m_2^2 n_1^4 t^4}{\left(1-t\right) \left(1-m_1 m_2 t\right) \left(1-n_1^2 t\right) \left(1-\frac{m_1^2 n_1^2 t^2}{Q}\right) \left(1-m_2^2 n_1^2 Q t^2\right)}
\end{align}
with $\{k_i\}_{i=1,\ldots,4}$, $\{m_j\}_{j=1,2,3}$, and $n_1$ the $\surm(5)$, $\surm(4)$, and $\surm(2)$ highest weight fugacities, respectively. 

\subsection{\texorpdfstring{$T[\surm(N)]$ theories}{TSUN theories}}
\label{app:TSUN_calc}
We move on to the examples of $T[\surm(N)]$ theories. The employed fugacity maps follow from \eqref{eq:mirror_paras_TSUN} in combination with \eqref{eq:HB_fugacities_TSUN}.

\subsubsection{\texorpdfstring{$T[\surm(3)]$ theories}{TSU3 theories}}

\begin{align}
    \raisebox{-.5\height}{
    \includegraphics[page=1]{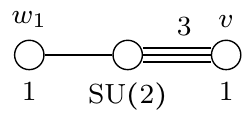}
    } 
    \xleftrightarrow{\quad \text{mirror}\quad }
    \raisebox{-.5\height}{
    \includegraphics[page=2]{figures/figures_1-form_appendix.pdf}
    }  \,.
\end{align}
Use the fugacity map
\begin{align}
\begin{aligned}
 &\Coulomb: \qquad   w_1\to x_1^2 \;, \quad 
    v\to Q^{-1} \\
 &\Higgs: \qquad
 y_1\to Q^{\frac{1}{6}} x_1 \;,\quad 
 y_2\to Q^{\frac{1}{6}} x_1^{-1} \;,\quad
 y_3\to Q^{-\frac{1}{3}}  
\end{aligned}
\end{align}
to an $x_1$ weight space fugacity for $A_1$ and a $\urm(1)$ variable $Q$.
The Coulomb branch Hilbert series of the left quiver (and Higgs branch Hilbert series of the right quiver) reads
\begin{small}
\begin{align}
\label{eq:HS_TSU3_SU2}
    \HS&=
    1 +t \left(\chi _2+1\right)
    +t^2 \left( \textcolor{red}{\left( Q + Q^{-1} \right) \chi_2}
    +2 \chi _2+\chi _4+2\right) 
    \\
    &+t^3 \bigg(
    \left(Q + Q^{-1} \right) \left(2 \chi _2+\chi _4+1\right)
    +4 \chi _2+2 \chi _4+\chi _6+2
    \bigg) \notag 
    + \ldots
\end{align}
\end{small}
here $\chi_{n_1}$ are the $\surm(2)_{x_1}$ characters for irreps with Dynkin labels $[n_1]$. The term in red corresponds to the operator $\Ocal$ in \eqref{eq:HB_GIO_TSUN}.
The symmetry algebra is $\surmL(2)_{x_1}\oplus \urmL(1)_Q$. The $\surm(2)$ centre symmetry $\Z_2$ acts trivial on irreps $[2\cdot n_1]$ for $n_1 \in\N$. Thus \eqref{eq:HS_TSU3_SU2} suggests that the symmetry group is $\sorm(3)_{x_1}\times \urm(1)_Q$.

\subsubsection{\texorpdfstring{$T[\surm(4)]$ theories}{TSU4 theories}}

\paragraph{Gauging a $\Z_3$.}
\begin{align}
    \raisebox{-.5\height}{
    \includegraphics[page=3]{figures/figures_1-form_appendix.pdf}
    } 
    \xleftrightarrow{\quad \text{mirror}\quad }
    \raisebox{-.5\height}{
    \includegraphics[page=4]{figures/figures_1-form_appendix.pdf}
    }  \,.
\end{align}
Use the fugacity map
\begin{align}
\begin{aligned}
&\Coulomb: \qquad 
    w_1\to \frac{x_1^2}{x_2} \;, \quad 
    w_2\to \frac{x_2^2}{x_1} \;, \quad 
    v\to Q^{-1} \\
&\Higgs: \qquad    
y_1 \to Q^{\frac{1}{12}} x_1 \;, \quad 
y_2 \to Q^{\frac{1}{12}} \frac{x_2}{x_1} \;, \quad
y_3 \to Q^{\frac{1}{12}} \frac{1}{x_2}\;, \quad 
y_4 \to Q^{-\frac{1}{4}}
\end{aligned}    
\end{align}
to $x_i$ weight space fugactites for $A_2$ and a $\urm(1)$ variable $Q$.
The Coulomb branch Hilbert series of the left quiver (and Higgs branch Hilbert series of the right quiver) reads
\begin{small}
\begin{align}
\label{eq:HS_TSU4_SU3}
\HS=&1
+t \left(\chi _{1,1}+1\right) 
+t^2 \left(3 \chi _{1,1}+\chi _{2,2}+2\right)
\\
&+t^3 \bigg(
\textcolor{red}{
Q \chi _{3,0}
+Q^{-1}\chi _{0,3}}
+2 \chi _{0,3}+6 \chi _{1,1}+3 \chi _{2,2}+2 \chi _{3,0}+\chi _{3,3}+3\bigg)\notag\\
&+t^4 \bigg(
Q \left(\chi _{1,1}+\chi _{2,2}+2 \chi _{3,0}+\chi _{4,1}\right)
+Q^{-1}\left(2 \chi _{0,3}+\chi _{1,1}+\chi _{1,4}+\chi _{2,2}\right) \notag\\
&\qquad +4 \chi _{0,3}+11 \chi _{1,1}+2 \chi _{1,4}+9 \chi _{2,2}+4 \chi _{3,0}+3 \chi _{3,3}+2 \chi _{4,1}+\chi _{4,4}+4
\bigg)\notag 
+ \ldots 
\end{align}
\end{small}
here $\chi_{k,n}$ are the $\surm(3)_{x_1 x_2}$ characters for irreps with Dynkin labels $[k,n]$. The terms in red corresponds to the operator $\Ocal$ (and its conjugate) in \eqref{eq:HB_GIO_TSUN}.
The symmetry algebra is $\surmL(3)_{x_1,x_2}\oplus \urmL(1)_Q$. The $\surm(3)$ centre symmetry $\Z_3$ acts trivial on irreps $[n_1,n_2]$ with $n_1 -n_2 = 0 \mod 3$. Thus \eqref{eq:HS_TSU4_SU3} suggests that the symmetry group is $\psurm(3)_{x_1,x_2}\times \urm(1)_Q$.
\paragraph{Gauging a $\Z_2$.}
\begin{align}
    \raisebox{-.5\height}{
    \includegraphics[page=5]{figures/figures_1-form_appendix.pdf}
    } 
    \xleftrightarrow{\quad \text{mirror}\quad }
    \raisebox{-.5\height}{
    \includegraphics[page=6]{figures/figures_1-form_appendix.pdf}
    }  \,.
\end{align}
Use the fugacity map
\begin{align}
\begin{aligned}
 &\Coulomb:\qquad    
 w_1\to x_1^2 \;, \quad 
    w_3\to u_1^2 \;, \quad 
    v\to \frac{Q^{-1}}{(u_1)^4} \\
    &\Higgs:  \qquad 
    y_1 \to Q^{\frac{1}{4}} x_1 \;, \quad 
    y_2 \to Q^{\frac{1}{4}} x_1^{-1} \;, \quad 
    y_3 \to Q^{-\frac{1}{4}} u_1 \;,\quad
    y_4 \to Q^{-\frac{1}{4}} u_1^{-1}
\end{aligned}
\end{align}
to an $x_1$ weight space fugactity of $A_1$, $u_1$ the weight space fugacity of another $A_1$,  and a $\urm(1)$ variable $Q$.
The Coulomb branch Hilbert series of the left quiver (and Higgs branch Hilbert series of the right quiver) reads
\begin{small}
    \begin{align}
    \label{eq:HS_TSU4_SU2}
 \HS &=1
 +t\left(\varphi _2+\chi _2+1\right) \\
  &+ t^2\left(
  \left(Q+ Q^{-1} \right)
  \left(1+ \textcolor{red}{ \varphi _2 \chi _2}\right) 
  +2 \varphi _2+\varphi _4+2 \varphi _2 \chi _2+2 \chi _2+\chi _4+3\right)\notag  \\
       &+t^3\bigg(
       \left(Q +Q^{-1}\right)
       \left(3 \chi _2 \varphi _2+\chi _4 \varphi _2+2 \varphi _2+\varphi _4 \chi _2+2 \chi _2+1\right)  \notag \\
       &\qquad +6 \varphi _2+2 \varphi _4+\varphi _6+6 \varphi _2 \chi _2+2 \varphi _4 \chi _2+6 \chi _2+2 \varphi _2 \chi _4+2 \chi _4
       +\chi _6+4
       \bigg) 
  + \ldots \notag
\end{align}
\end{small}%
here $\chi_{n_1}$ are the $\surm(2)_{x_1}$ characters for irreps with Dynkin label $[n_1]$. While $\varphi_{k_1}$ are the $\surm(2)_{u_1}$ characters for irreps $[k_1]$. The term in red corresponds to the operator $\Ocal$ in \eqref{eq:HB_GIO_TSUN}.
The symmetry algebra is $\surmL(2)_{x_1}\oplus \surmL(2)_{u_1}\oplus \urmL(1)_Q$. The $\surm(2)$ centre symmetries $\Z_2$ act trivial on irreps $[n_1]$ with $n_1 = 0 \mod 2$ and $[k_1]$ with $k_1 =0 \mod 2$, respectively. Thus \eqref{eq:HS_TSU4_SU2} suggests that the symmetry group is $\sorm(3)_{x_1}\times \sorm(3)_{u_1}\times  \urm(1)_Q$.
\subsubsection{\texorpdfstring{$T[\surm(5)]$ theories}{TSU5 theories}}

\paragraph{Gauging a $\Z_4$.}
\begin{align}
    \raisebox{-.5\height}{
    \includegraphics[page=7]{figures/figures_1-form_appendix.pdf}
    } 
    \xleftrightarrow{\quad \text{mirror}\quad }
    \raisebox{-.5\height}{
    \includegraphics[page=8]{figures/figures_1-form_appendix.pdf}
    }  \,.
\end{align}
Use the fugacity map
\begin{align}
&\Coulomb: \qquad 
    w_1\to \frac{x_1^2}{x_2} \;, \quad
    w_2\to \frac{x_2^2}{x_1 x_3} \;, \quad
    w_3\to \frac{x_3^2}{x_2} \;, \quad
    v\to  Q^{-1} \\
&\Higgs: \qquad 
 y_1 \to Q^{\frac{1}{20}} x_1 \;, \quad 
  y_2 \to Q^{\frac{1}{20}} \frac{x_2}{x_1} \;, \quad 
  y_3 \to Q^{\frac{1}{20}} \frac{x_3}{x_2} \;, \quad 
  y_4 \to Q^{\frac{1}{20}} \frac{1}{x_3} \;, \quad 
  y_5 \to Q^{-\frac{1}{5}} 
\notag 
\end{align}
to $x_i$ weight space fugacities of $A_3$, and a $\urm(1)$ variable $Q$.
The Coulomb branch Hilbert series of the left quiver (and Higgs branch Hilbert series of the right quiver) reads
\begin{small}
\begin{align}
\label{eq:HS_TSU5_SU4}
    \HS=1
    &+t \left(\chi _{1,0,1}+1\right)
    +t^2 \left(\chi _{0,2,0}+3 \chi _{1,0,1}+\chi _{2,0,2}+2\right) \\
     &+t^3 \left(2 \chi _{0,1,2}+2 \chi _{0,2,0}+7 \chi _{1,0,1}+\chi _{1,2,1}+3 \chi _{2,0,2}+2 \chi _{2,1,0}+\chi _{3,0,3}+3
 \right)
 \notag \\
 &+t^4 \bigg(
 \textcolor{red}{Q^{-1}\chi _{0,0,4} }+ \textcolor{red}{Q \chi _{4,0,0}}+6 \chi _{0,1,2}+6 \chi _{0,2,0}+\chi _{0,4,0}+13 \chi _{1,0,1}+2 \chi _{1,1,3}+4 \chi _{1,2,1} \notag \\
 &\qquad 
 +10 \chi _{2,0,2}+6 \chi _{2,1,0}+\chi _{2,2,2}+3 \chi _{3,0,3}+2 \chi _{3,1,1}+\chi _{4,0,4}+5\bigg) 
 +\ldots
 \notag
\end{align}
\end{small}%
here $\chi_{n_1,n_2,n_3}$ are the $\surm(4)_{x_1, x_2,x_3}$ characters for irreps with Dynkin labels $[n_1,n_2,n_3]$. The terms in red correspond to the operator $\Ocal$ (and its conjugate) in \eqref{eq:HB_GIO_TSUN}.
The symmetry algebra is $\surmL(4)_{x_1,x_2,x_3}\oplus \urmL(1)_Q$. The $\surm(4)$ centre symmetry $\Z_4$ act trivial on irreps $[n_1,n_2,n_3]$ with $n_1 +2n_2 -n_3 = 0 \mod 4$. Thus \eqref{eq:HS_TSU5_SU4} suggests that the symmetry group is $\psurm(4)_{x_1,x_2,x_3}\times  \urm(1)_Q$.
\paragraph{Gauging a $\Z_3$.}
\begin{align}
    \raisebox{-.5\height}{
    \includegraphics[page=9]{figures/figures_1-form_appendix.pdf}
    } 
    \xleftrightarrow{\quad \text{mirror}\quad }
    \raisebox{-.5\height}{
    \includegraphics[page=10]{figures/figures_1-form_appendix.pdf}
    }  \,.
\end{align}
Use the fugacity map
\begin{align}
&\Coulomb: \qquad 
    w_1\to \frac{x_1^2}{x_2} \;, \quad
    w_2\to \frac{x_2^2}{x_1} \;, \quad 
    w_4\to u_1^2 \;, \quad 
    v\to \frac{Q^{-1}}{( u_1)^5} \\
&\Higgs: \qquad
y_1 \to Q^{\frac{2}{15}} x_1 \;,\quad 
y_2 \to Q^{\frac{2}{15}} \frac{x_2}{x_1} \;,\quad y_3 \to Q^{\frac{2}{15}} \frac{1}{x_2} \;,\quad
y_4 \to Q^{-\frac{1}{5}} u_1 \;,\quad 
y_5 \to Q^{-\frac{1}{5}} \frac{1}{u_1}
\notag
\end{align}
to $x_i$ weight space fugacities of $A_2$, $u_1$ the weight space fugacity of $A_1$, and a $\urm(1)$ variable $Q$.
The Hilbert series are
\begin{small}
\begin{align}
\label{eq:HS_TSU5_SU3}
    \HS&=  1 
    +t \left(\chi _{1,1}+\phi _2+1\right) 
        +t^2 \left(3 \chi _{1,1}+\chi _{2,2}+2 \phi _2 \chi _{1,1}+2 \phi _2+\phi _4+3\right)  \\
    &+t^3 \bigg(
    Q^{-1} \left(\phi _1 \chi _{1,1}
    +\textcolor{red}{\phi _3 \chi _{3,0}}\right)
    +Q\left(
    \textcolor{red}{\phi _3 \chi _{0,3}}
    +\phi _1 \chi _{1,1}\right) \notag \\
    &\qquad +2 \chi _{0,3}+8 \chi _{1,1}+3 \chi _{2,2}+2 \chi _{3,0}+\chi _{3,3} +\phi _2 \chi _{0,3}+8 \phi _2 \chi _{1,1}+2 \phi _4 \chi _{1,1}\notag \\ 
    &\quad +2 \phi _2 \chi _{2,2}+\phi _2 \chi _{3,0}+6 \phi _2+2 \phi _4+\phi _6+5\bigg) 
+ \ldots \notag
\end{align}
\end{small}%
here $\chi_{n_1,n_2}$ are the $\surm(3)_{x_1, x_2}$ characters for irreps with Dynkin labels $[n_1,n_2]$. The $\phi_{k_1}$ are the $\surm(2)_{u_1}$ characters for irreps with Dynkin label $[k_1]$. The terms in red correspond to the operator $\Ocal$ (and its conjugate) in \eqref{eq:HB_GIO_TSUN}.
The symmetry algebra is $\surmL(3)_{x_1,x_2}\oplus \surmL(2)_{u_1}\oplus \urmL(1)_Q$. The $\surm(3)$ centre symmetry $\Z_3$ act trivial on irreps $[n_1,n_2]$ with $n_1  -n_2 = 0 \mod 3$; while the $\surm(2)$ centre $\Z_2$ acts trivial on irreps $[k_1]$ with $k_1 =0 \mod 2$. Thus \eqref{eq:HS_TSU5_SU3} suggests that the symmetry group is $\psurm(3)_{x_1,x_2}\times  \left(\surm(2)_{u_1} \times \urm(1)_Q \right)\slash \Z_2$. $\Z_2 \subset \urm(1)_Q$ such that $Q$ has charge $1$.
\paragraph{Gauging a $\Z_2$.}
\begin{align}
    \raisebox{-.5\height}{
    \includegraphics[page=11]{figures/figures_1-form_appendix.pdf}
    } 
    \xleftrightarrow{\quad \text{mirror}\quad }
    \raisebox{-.5\height}{
    \includegraphics[page=12]{figures/figures_1-form_appendix.pdf}
    }  \,.
\end{align}
Use the fugacity map
\begin{align}
&\Coulomb: \qquad
    w_1\to x_1^2 \;, \quad
    w_3\to \frac{u_1^2}{u_2} \;, \quad 
    w_4\to \frac{u_2^2}{u_1} \;, \quad 
    v\to \frac{Q^{-1}}{(u_2)^5} \\
&\Higgs: \qquad
y_1 \to Q^{\frac{3}{10}} x_1 \;,\quad 
y_2 \to Q^{\frac{3}{10}} \frac{1}{x_1} \;,\quad y_3 \to Q^{-\frac{1}{5}} u_1 \;,\quad
y_4 \to Q^{-\frac{1}{5}} \frac{u_2}{u_1} \;,\quad y_5 \to Q^{-\frac{1}{5}} \frac{1}{u_2}
\notag
\end{align}
to an $x_1$ weight space fugacity of $A_1$, $u_i$ the weight space fugacities of $A_2$, and a $\urm(1)$ variable $Q$.
The Coulomb branch Hilbert series of the left quiver (and Higgs branch Hilbert series of the right quiver) reads
\begin{small}
\begin{align}
\label{eq:HS_TSU5_SU2}
    \HS&=
       1+\left(\chi _2+\phi _{1,1}+1\right) t\\
    &+ t^2 \bigg(
     Q \left( 
     \textcolor{red}{\chi _2 \phi _{0,2}}
     +\phi _{1,0}\right)
    +Q^{-1} \left(\phi _{0,1}+
    \textcolor{red}{
    \chi _2 \phi _{2,0}}
    \right)
    +2 \chi _2+\chi _4+2 \chi _2 \phi _{1,1}+3 \phi _{1,1}+\phi _{2,2}+3\bigg) \notag \\
    &+t^3 \bigg(
    Q \left(3 \chi _2 \phi _{0,2}+\chi _4 \phi _{0,2}+2 \phi _{0,2}+2 \chi _2 \phi _{1,0}+2 \phi _{1,0}+\chi _2 \phi _{1,3}+\chi _2 \phi _{2,1}+\phi _{2,1}\right)  \notag \\
    &\qquad +Q^{-1}\left(2 \chi _2 \phi _{0,1}+2 \phi _{0,1}+\chi _2 \phi _{1,2}+\phi _{1,2}+3 \chi _2 \phi _{2,0}+\chi _4 \phi _{2,0}+2 \phi _{2,0}+\chi _2 \phi _{3,1}\right) \notag \\
    &\qquad +6 \chi _2+2 \chi _4+\chi _6+\chi _2 \phi _{0,3}+2 \phi _{0,3}+8 \chi _2 \phi _{1,1}+2 \chi _4 \phi _{1,1}+8 \phi _{1,1}+2 \chi _2 \phi _{2,2} \notag\\
    &\qquad \qquad +3 \phi _{2,2}+\chi _2 \phi _{3,0}+2 \phi _{3,0}+\phi _{3,3}+5
    \bigg) 
   + \ldots \notag
\end{align}
\end{small}%
here $\chi_{n_1}$ are the $\surm(2)_{x_1}$ characters for irreps with Dynkin labels $[n_1]$. The $\phi_{k_1,k_2}$ are the $\surm(3)_{u_1,u_2}$ characters for irreps with Dynkin label $[k_1,k_2]$. The terms in red correspond to the operator $\Ocal$ (and its conjugate) in \eqref{eq:HB_GIO_TSUN}.
The symmetry algebra is $\surmL(2)_{x_1}\oplus \surmL(3)_{u_1,u_2}\oplus \urmL(1)_Q$. The $\surm(3)$ centre symmetry $\Z_3$ act trivial on irreps $[k_1,k_2]$ with $k_1  -k_2 = 0 \mod 3$; while the $\surm(2)$ centre $\Z_2$ acts trivial on irreps $[n_1]$ with $n_1 =0 \mod 2$. Thus \eqref{eq:HS_TSU5_SU2} suggests that the symmetry group is $\psurm(2)_{x_1}\times  \left(\surm(3)_{u_1,u_2} \times \urm(1)_Q \right)\slash \Z_3$. $\Z_3 \subset \urm(1)_Q$ such that $Q$ has charge $2$.

\subsubsection{\texorpdfstring{$T[\surm(6)]$ theories}{TSU6 theories}}

\paragraph{Gauging a $\Z_5$.}
\begin{align}
    \raisebox{-.5\height}{
    \includegraphics[page=13]{figures/figures_1-form_appendix.pdf}
    } 
    \xleftrightarrow{\; \text{mirror}\; }
    \raisebox{-.5\height}{
    \includegraphics[page=14]{figures/figures_1-form_appendix.pdf}
    }  \,.
\end{align}
Use the fugacity map
\begin{align}
    &\Coulomb: \qquad
    w_1\to \frac{x_1^2}{x_2} \;, \quad
    w_2\to \frac{x_2^2}{x_1 x_3} \;, \quad 
    w_3\to \frac{x_3^2}{x_2 x_4} \;, \quad 
    w_4\to \frac{x_4^2}{x_3} \;, \quad
    v\to Q^{-1} \\
    &\Higgs: \qquad
y_1 \to Q^{\frac{1}{30}} x_1 \;,\quad 
y_2 \to Q^{\frac{1}{30}} \frac{x_2}{x_1} \;,\quad
y_3 \to Q^{\frac{1}{30}} \frac{x_3}{x_2} \;,\quad
y_4 \to Q^{\frac{1}{30}} \frac{x_4}{x_3} \;,\quad
y_5 \to Q^{\frac{1}{30}} \frac{1}{x_4} \;, \notag \\
&\qquad \qquad y_6 \to Q^{-\frac{1}{6}}  
\notag
\end{align}
to $\{x_i\}$ the weight space fugacities of $A_4$, and a $\urm(1)$ variable $Q$. The perturbative Coulomb/Higgs branch Hilbert series reads
\begin{small}
\begin{align}
\label{eq:HS_TSU6_SU5}
    \HS=1
    &+t \left(\chi _{1,0,0,1}+1\right) 
    +t^2 \left(
    \chi _{0,1,1,0}
    +3 \chi _{1,0,0,1}
    +\chi _{2,0,0,2}+2\right) 
    \\
&+t^3 \left(
2 \chi _{0,1,0,2}
+3 \chi _{0,1,1,0}
+7 \chi _{1,0,0,1}
+\chi _{1,1,1,1}
+3 \chi _{2,0,0,2}
+2 \chi _{2,0,1,0}
+\chi _{3,0,0,3}+3\right) \notag \\
    &+t^4 \bigg(
    2 \chi _{0,0,2,1}
    +6 \chi _{0,1,0,2}
    +8 \chi _{0,1,1,0}
    +\chi _{0,2,2,0}
    +14 \chi _{1,0,0,1}
    +2 \chi _{1,1,0,3}
    +5 \chi _{1,1,1,1} \notag \\
    &\qquad 
    +2 \chi _{1,2,0,0}
    +10 \chi _{2,0,0,2}
    +6 \chi _{2,0,1,0}
    +\chi _{2,1,1,2}
    +3 \chi _{3,0,0,3}
    +2 \chi _{3,0,1,1}
    +\chi _{4,0,0,4}+5
    \bigg)
    \notag \\
    &+ t^5 
    \bigg(
    \textcolor{red}{
Q^{-1} \chi _{0,0,0,5}
+Q \chi _{5,0,0,0}}
+\chi _{0,0,1,3}
+6 \chi _{0,0,2,1}
+17 \chi _{0,1,0,2}
+17 \chi _{0,1,1,0}
+2 \chi _{0,2,1,2} \notag \\
 &\qquad
+3 \chi _{0,2,2,0}
+2 \chi _{0,3,0,1}
+25 \chi _{1,0,0,1}
+2 \chi _{1,0,2,2}
+2 \chi _{1,0,3,0}
+9 \chi _{1,1,0,3}
+18 \chi _{1,1,1,1} \notag \\
     &\qquad
+6 \chi _{1,2,0,0}
+\chi _{1,2,2,1}
+23 \chi _{2,0,0,2}
+17 \chi _{2,0,1,0}
+2 \chi _{2,1,0,4}
+5 \chi _{2,1,1,2}
+2 \chi _{2,1,2,0} \notag \\
     &\qquad
+2 \chi _{2,2,0,1}
+10 \chi _{3,0,0,3}
+9 \chi _{3,0,1,1}
+\chi _{3,1,0,0}
+\chi _{3,1,1,3}
+3 \chi _{4,0,0,4}
+2 \chi _{4,0,1,2} \notag \\
     &\qquad
+\chi _{5,0,0,5}+7
    \bigg)
    +\ldots \notag 
\end{align}
\end{small}%
here $\chi_{n_1,n_2,n_3,n_4}$ are the $\surm(5)_{x_i}$ characters for irreps with Dynkin labels $[n_1,n_2,n_3,n_4]$. The terms in red correspond to the operator $\Ocal$ (and its conjugate) in \eqref{eq:HB_GIO_TSUN}.
The algebra is $\surmL(5)_{x_i}\oplus \urm(1)_Q$. The $\surm(4)$ centre symmetry $\Z_5$ acts with charge $1$ in the fundamental $[1,0,0,0]$. The appearing characters in \eqref{eq:HS_TSU6_SU5} are all neutral under the centre, which suggests the global form $\psurm(5)_{x_i}\times \urm(1)_Q$.

\paragraph{Gauging a $\Z_4$.}
\begin{align}
    \raisebox{-.5\height}{
    \includegraphics[page=15]{figures/figures_1-form_appendix.pdf}
    } 
    \xleftrightarrow{\; \text{mirror}\; }
    \raisebox{-.5\height}{
    \includegraphics[page=16]{figures/figures_1-form_appendix.pdf}
    }  \,.
\end{align}
Use the fugacity map
\begin{align}
 &\Coulomb: \qquad
    w_1\to \frac{x_1^2}{x_2} \;, \quad
    w_2\to \frac{x_2^2}{x_1 x_3} \;, \quad 
    w_3\to \frac{x_3^2}{x_2} \;, \quad
    w_5\to u_1^2 \;, \quad 
    v\to \frac{Q^{-1}}{(u_1)^6} \\
       &\Higgs: \qquad
y_1 \to Q^{\frac{1}{12}} x_1 \;,\quad 
y_2 \to Q^{\frac{1}{12}} \frac{x_2}{x_1} \;,\quad
y_3 \to Q^{\frac{1}{12}} \frac{x_3}{x_2} \;,\quad
y_4 \to Q^{\frac{1}{12}} \frac{1}{x_3} \;, \notag \\
&\qquad \qquad
y_5 \to Q^{-\frac{1}{6}} u_1  \;, \quad 
 y_6 \to Q^{-\frac{1}{6}}\frac{1}{u_1}  
\notag
\end{align}
to $\{x_i\}$ the weight space fugacities of $A_3$, $u_1$ the weight space fugacity of $A_1$, and a $\urm(1)$ variable $Q$. The Hilbert series reads
\begin{small}
\begin{align}
\HS =1
&+t \left(\phi _{1,0,1}+\chi _2+1\right)\\
&+t^2 \left(
2 \chi _2 \phi _{1,0,1}
+\phi _{0,2,0}
+3 \phi _{1,0,1}
+\phi _{2,0,2}
+2 \chi _2+\chi _4+3\right) \notag \\
&+t^3 \bigg(
\chi _2 \phi _{0,1,2}
+2 \chi _2 \phi _{0,2,0}
+8 \chi _2 \phi _{1,0,1}
+2 \chi _2 \phi _{2,0,2}
+\chi _2 \phi _{2,1,0}
+2 \chi _4 \phi _{1,0,1}
+2 \phi _{0,1,2} \notag \\
&\qquad 
+2 \phi _{0,2,0}
+9 \phi _{1,0,1}
+\phi _{1,2,1}
+3 \phi _{2,0,2}
+2 \phi _{2,1,0}
+\phi _{3,0,3}
+6 \chi _2+2 \chi _4+\chi _6+5
\bigg) \notag \\
&+t^4 \bigg(
Q^{-1} \left(\textcolor{red}{\chi _4 \phi _{0,0,4}}
+\chi _2 \phi _{0,1,2}+\phi _{0,2,0}\right)
+Q\left(\chi _2 \phi _{2,1,0}+\textcolor{red}{\chi _4 \phi _{4,0,0}}+\phi _{0,2,0}\right)
+8 \chi _2 \phi _{0,1,2} \notag \\ 
&\qquad
+7 \chi _2 \phi _{0,2,0}
+26 \chi _2 \phi _{1,0,1}
+\chi _2 \phi _{1,1,3}
+3 \chi _2 \phi _{1,2,1}
+10 \chi _2 \phi _{2,0,2}
+8 \chi _2 \phi _{2,1,0} \notag \\
&\qquad 
+2 \chi _2 \phi _{3,0,3}
+\chi _2 \phi _{3,1,1}
+\chi _4 \phi _{0,1,2}
+2 \chi _4 \phi _{0,2,0}
+9 \chi _4 \phi _{1,0,1}
+2 \chi _6 \phi _{1,0,1}
+3 \chi _4 \phi _{2,0,2}
\notag \\
&\qquad 
+\chi _4 \phi _{2,1,0}
+7 \phi _{0,1,2}
+9 \phi _{0,2,0}
+\phi _{0,4,0}
+21 \phi _{1,0,1}
+2 \phi _{1,1,3}
+4 \phi _{1,2,1}
+12 \phi _{2,0,2}
\notag \\
&\qquad 
+7 \phi _{2,1,0}+\phi _{2,2,2}+3 \phi _{3,0,3}+2 \phi _{3,1,1}+\phi _{4,0,4}+12 \chi _2+7 \chi _4+2 \chi _6+\chi _8+11
\bigg)
+\ldots
\notag 
\end{align}
\end{small}%
The symmetry algebra is $\surmL(4)_{x_1,x_2,x_3}\times \surmL(2)_{u_1} \times \urm(1)_Q$. The terms in red correspond to the operator $\Ocal$ (and its conjugate) in \eqref{eq:HB_GIO_TSUN}.
The appearing characters suggest that the global form is $\psurm(4)_{x_1,x_2,x_3,x_4} \times \sorm(3)_{u_1} \times\urm(1)_Q$, i.e.\ the centre symmetries act trivially.
\paragraph{Gauging a $\Z_3$.}
\begin{align}
    \raisebox{-.5\height}{
    \includegraphics[page=17]{figures/figures_1-form_appendix.pdf}
    } 
    \xleftrightarrow{\; \text{mirror}\;}
    \raisebox{-.5\height}{
    \includegraphics[page=18]{figures/figures_1-form_appendix.pdf}
    }  \,.
\end{align}
Use the fugacity map
\begin{align}
&\Coulomb: \qquad
    w_1\to \frac{x_1^2}{x_2} \;, \quad
    w_2\to \frac{x_2^2}{x_1} \;, \quad
    w_4\to \frac{u_1^2}{u_2} \;, \quad 
    w_5\to \frac{u_2^2}{u_1} \;, \quad 
    v\to \frac{Q^{-1}}{(u_2)^5} \\
&\Higgs: \qquad
y_1 \to Q^{\frac{1}{6}} x_1 \;,\quad 
y_2 \to Q^{\frac{1}{6}} \frac{x_2}{x_1} \;,\quad
y_3 \to Q^{\frac{1}{6}} \frac{1}{x_2}  \notag \\
&\qquad \qquad
y_4 \to Q^{-\frac{1}{6}} u_1  \;, \quad
y_5 \to Q^{-\frac{1}{6}} \frac{u_2}{u_1} \;,
 y_6 \to Q^{-\frac{1}{6}}\frac{1}{u_2}  
\notag
\end{align}
to $x_i$ weight space fugacities of $A_2$, $u_i$ the weight space fugacities of another $A_2$, and a $\urm(1)$ variable $Q$. The perturbative Coulomb/Higgs branch Hilbert series is evaluated to
\begin{small}
\begin{align}
    \HS&=  1 +t \left(\chi _{1,1}+\phi _{1,1}+1\right) \\
    &+t^2 \left(3 \chi _{1,1}+\chi _{2,2}+2 \chi _{1,1} \phi _{1,1}+3 \phi _{1,1}+\phi _{2,2}+3\right) \notag \\
   &+t^3 \bigg(Q^{-1} \left( 
   \textcolor{red}{\chi _{0,3} \phi _{3,0}}
   +\chi _{1,1} \phi _{1,1}+1\right)
   +Q\left( \chi _{1,1} \phi _{1,1}
   + \textcolor{red}{\chi _{3,0} \phi _{0,3}}+1 \right)   \notag \\
   &+2 \chi _{0,3}+8 \chi _{1,1}+3 \chi _{2,2}+2 \chi _{3,0}+\chi _{3,3}+\chi _{1,1} \phi _{0,3}+\chi _{0,3} \phi _{1,1}+10 \chi _{1,1} \phi _{1,1}+2 \chi _{1,1} \phi _{2,2} \notag\\ 
   &+\chi _{1,1} \phi _{3,0}+2 \chi _{2,2} \phi _{1,1}+\chi _{3,0} \phi _{1,1}+2 \phi _{0,3}+8 \phi _{1,1}+3 \phi _{2,2}+2 \phi _{3,0}+\phi _{3,3}+6\bigg)
   + \ldots
   \notag
\end{align}
\end{small}%
The symmetry algebra is $\surmL(3)_{x_1,x_2}\oplus \surmL(3)_{u_1,u_2}\oplus \urmL(1)$. The terms in red correspond to the operator $\Ocal$ (and its conjugate) in \eqref{eq:HB_GIO_TSUN}.
The appearing characters indicate that all irreps are trivial under the centre symmetries, such that the global form is $\psurm(3)_{x_1,x_2}\times \psurm(3)_{u_1,u_2} \times \urm(1)_Q$.
\paragraph{Gauging a $\Z_2$.}
\begin{align}
    \raisebox{-.5\height}{
    \includegraphics[page=19]{figures/figures_1-form_appendix.pdf}
    } 
    \xleftrightarrow{\; \text{mirror}\; }
    \raisebox{-.5\height}{
    \includegraphics[page=20]{figures/figures_1-form_appendix.pdf}
    }  \,.
\end{align}
Use the fugacity map
\begin{align}
 &\Coulomb: \qquad
    w_1\to x_1^2 \;, \quad
    w_3\to \frac{u_1^2}{u_2} \;, \quad 
    w_4\to \frac{u_2^2}{u_2 u_3} \;, \quad 
    w_5\to \frac{u_3^2}{u_2} \;, \quad 
    v\to \frac{Q^{-1}}{(u_3)^5}\\
    &\Higgs: \qquad
y_1 \to Q^{\frac{1}{3}} x_1 \;,\quad 
y_2 \to Q^{\frac{1}{3}} \frac{1}{x_1} \;,  \notag \\
&\qquad \qquad
y_3 \to Q^{-\frac{1}{6}} u_1  \;, \quad
y_4 \to Q^{-\frac{1}{6}} \frac{u_2}{u_1} \;, \quad 
y_5 \to Q^{-\frac{1}{6}} \frac{u_3}{u_2} \;, \quad 
 y_6 \to Q^{-\frac{1}{6}}\frac{1}{u_3}  
\notag
\end{align}
to an $x_1$ weight space fugacity of $A_1$, $u_i$ the weight space fugacities of $A_3$, and a $\urm(1)$ variable $Q$. The Coulomb branch (or Higgs branch) Hilbert series reads
\begin{small}
\begin{align}
    \HS=
    1
    +t \left(\phi _{1,0,1}+\chi _2+1\right)
    &+t^2 \bigg(
    Q \left( \textcolor{red}{\chi _2 \phi _{0,0,2}}+\phi _{0,1,0} \right) 
    +Q^{-1} \left( \textcolor{red}{\chi _2 \phi _{2,0,0}}+\phi _{0,1,0}\right)  \\
    &\qquad +2 \chi _2 \phi _{1,0,1}+\phi _{0,2,0}+3 \phi _{1,0,1}+\phi _{2,0,2}+2 \chi _2+\chi _4+3\bigg)
    + \ldots \notag
\end{align}
\end{small}%
The terms in red correspond to the operator $\Ocal$ (and its conjugate) in \eqref{eq:HB_GIO_TSUN}. The 
global symmetry is $\psurm(2)_{x_1} \times \left( \surm(4)_{u_1,u_2,u_3} \times \urm(1)_Q \right) \slash \Z_4$ and $Q$ has $\Z_4$ charge $2\bmod 4$.

\subsection{\texorpdfstring{Some $T[\surm(N)]$ examples with higher charges}{Some TSUN examples with higher charges}}

\label{app:TSUN_higher_charges}
Consider the quiver theories in \eqref{eq:TSU5_SU3_charge_2} and \eqref{eq:TSU5_SU2_charge_3}. Redefine fugacities as 
\begin{alignat}{2}
&\eqref{eq:TSU5_SU3_charge_2}:\qquad  &
w_1 &\to \frac{y_1^2}{y_2} \, , \quad 
w_2 \to \frac{y_2^2}{y_1}\, , \quad 
w_4\to x_1^2\, , \quad 
v\to \frac{Q}{x_1^{10}}
\\
&\eqref{eq:TSU5_SU2_charge_3}: \qquad  &
w_1&\to x_1^2 \, , \quad 
w_3 \to \frac{y_1^2}{y_2}\, , \quad 
w_4 \to \frac{y_2^2}{y_1}\, , \quad 
v \to \frac{Q^{-1}}{y_2^{15}}\, , \quad 
\end{alignat}
such that $x_1$ is an $A_1$ fugacity and $\{y_{1,2}\}$ are $A_2$ fugacities.
The perturbative monopole formula for the left-hand-side quivers reads
\begin{small}
\begin{align}
\label{eq:HS_TSU5_with_charges}
    \HS&=
       1 +t\left(\chi _2+\phi _{1,1}+1\right)  \\
       &+t^2\left(2 \phi _{1,1} \chi _2+2 \chi _2+\chi _4+3 \phi _{1,1}+\phi _{2,2}+3\right) \notag  \\
    &+t^3\bigg(\phi _{0,3} \chi _2+8 \phi _{1,1} \chi _2+2 \phi _{2,2} \chi _2+\phi _{3,0} \chi _2+6 \chi _2+2 \chi _4+\chi _6+2 \phi _{0,3}+2 \chi _4 \phi _{1,1} \notag \\
    &\qquad +8 \phi _{1,1}+3 \phi _{2,2}+2 \phi _{3,0}+\phi _{3,3}+5\bigg) \notag  \\
    &+ t^4\bigg(7 \phi _{0,3} \chi _2+24 \phi _{1,1} \chi _2+\phi _{1,4} \chi _2+10 \phi _{2,2} \chi _2+7 \phi _{3,0} \chi _2+2 \phi _{3,3} \chi _2+\phi _{4,1} \chi _2 \notag \\
    &\qquad +12 \chi _2+7 \chi _4+2 \chi _6+\chi _8+\chi _4 \phi _{0,3}+5 \phi _{0,3}+9 \chi _4 \phi _{1,1}+2 \chi _6 \phi _{1,1}+19 \phi _{1,1}\notag \\
    &\qquad+2 \phi _{1,4}+3 \chi _4 \phi _{2,2}+11 \phi _{2,2}+\chi _4 \phi _{3,0}+5 \phi _{3,0}+3 \phi _{3,3}+2 \phi _{4,1}+\phi _{4,4}+10\bigg) \notag \\
      &+t^5\bigg(22 \phi _{0,3} \chi _2+60 \phi _{1,1} \chi _2+9 \phi _{1,4} \chi _2+38 \phi _{2,2} \chi _2+\phi _{2,5} \chi _2+22 \phi _{3,0} \chi _2+10 \phi _{3,3} \chi _2 \notag \\
    &\qquad+9 \phi _{4,1} \chi _2+2 \phi _{4,4} \chi _2+\phi _{5,2} \chi _2+25 \chi _2+15 \chi _4+7 \chi _6+2 \chi _8+\chi _{10}+9 \chi _4 \phi _{0,3} \notag \\
    &\qquad+\chi _6 \phi _{0,3}+16 \phi _{0,3}+30 \chi _4 \phi _{1,1}+9 \chi _6 \phi _{1,1}+2 \chi _8 \phi _{1,1}+40 \phi _{1,1}+2 \chi _4 \phi _{1,4}+8 \phi _{1,4} \notag \\
    &\qquad+15 \chi _4 \phi _{2,2}+3 \chi _6 \phi _{2,2}+28 \phi _{2,2}+2 \phi _{2,5}+9 \chi _4 \phi _{3,0}+\chi _6 \phi _{3,0}+16 \phi _{3,0}+3 \chi _4 \phi _{3,3}\notag \\
    &\qquad+11 \phi _{3,3}+2 \chi _4 \phi _{4,1}+8 \phi _{4,1}+3 \phi _{4,4}+2 \phi _{5,2}+\phi _{5,5}+15\bigg) \notag \\
    &+t^6\bigg(
     Q \left(\chi _6 \phi _{0,6}+\chi _4 \phi _{1,4}+\chi _2 \phi _{2,2}+\phi _{3,0}\right)
      +\frac{\phi _{0,3}+\chi _2 \phi _{2,2}+\chi _4 \phi _{4,1}+\chi _6 \phi _{6,0}}{Q} \notag \\
    &\qquad +62 \phi _{0,3} \chi _2+2 \phi _{0,6} \chi _2+132 \phi _{1,1} \chi _2+37 \phi _{1,4} \chi _2+107 \phi _{2,2} \chi _2+9 \phi _{2,5} \chi _2+62 \phi _{3,0} \chi _2\notag \\
    &\qquad
    +41 \phi _{3,3} \chi _2+\phi _{3,6} \chi _2+37 \phi _{4,1} \chi _2+10 \phi _{4,4} \chi _2+9 \phi _{5,2} \chi _2+2 \phi _{5,5} \chi _2+2 \phi _{6,0} \chi _2+\phi _{6,3} \chi _2\notag \\
    &\qquad
    +44 \chi _2+33 \chi _4+16 \chi _6+7 \chi _8+2 \chi _{10}+\chi _{12}+31 \chi _4 \phi _{0,3}+9 \chi _6 \phi _{0,3}+\chi _8 \phi _{0,3}+36 \phi _{0,3}\notag \\
    &\qquad
    +\chi _4 \phi _{0,6}+3 \phi _{0,6}+81 \chi _4 \phi _{1,1}+31 \chi _6 \phi _{1,1}+9 \chi _8 \phi _{1,1}+2 \chi _{10} \phi _{1,1}+77 \phi _{1,1}+16 \chi _4 \phi _{1,4}\notag \\
    &\qquad
    +2 \chi _6 \phi _{1,4}+25 \phi _{1,4}+59 \chi _4 \phi _{2,2}+16 \chi _6 \phi _{2,2}+3 \chi _8 \phi _{2,2}+70 \phi _{2,2}+2 \chi _4 \phi _{2,5}+8 \phi _{2,5}\notag \\
    &\qquad
    +31 \chi _4 \phi _{3,0}+9 \chi _6 \phi _{3,0}+\chi _8 \phi _{3,0}+36 \phi _{3,0}
    +17 \chi _4 \phi _{3,3}+4 \chi _6 \phi _{3,3}+32 \phi _{3,3}+2 \phi _{3,6}\notag \\
    &\qquad
    +16 \chi _4 \phi _{4,1}+2 \chi _6 \phi _{4,1}+25 \phi _{4,1}+3 \chi _4 \phi _{4,4}+11 \phi _{4,4}+2 \chi _4 \phi _{5,2}+8 \phi _{5,2}+3 \phi _{5,5}\notag \\
    &\qquad
    +\chi _4 \phi _{6,0}+3 \phi _{6,0}
   +2 \phi _{6,3}+\phi _{6,6}+28\bigg) 
    + \ldots \notag 
\end{align}
\end{small}%
wherein $\phi_{m_1,m_2}$ are the characters of $A_2$ irreps $[m_1,m_2]$ and $\chi_{n_1}$ are the $A_1$ characters for $[n_1]$ irreps.
The global form of the Coulomb branch isometry group is $\psurm(2)_{x_1}\times \psurm(3)_{y_{1,2}}\times\urm(1)_Q$.

\subsection{\texorpdfstring{Some $T_{\rho}^{\sigma}[\surm(N)]$ examples}{Some T rho sigma examples}}

\label{app:T-rho-sigma_calc}

\paragraph{1st example.}
Consider the example 
\begin{align}
      \raisebox{-.5\height}{
    \includegraphics[page=14]{figures/figures_1-form_T_rho_sigma.pdf}
    } 
\end{align}
and use the fugacity map
\begin{align}
    v_1 \to x_1^2 
    \;,\quad 
    v_2 \to \frac{Q_1}{x_1}
    \;,\quad 
    v_3 \to Q_2
    \;,\quad 
    v_3 \to Q_3\,.
\end{align}
The Coulomb branch Hilbert series reads
\begin{small}
\begin{align}
\label{eq:HS_U1-U2-SU2-U1}
    \HS&=
    1+t \left(\chi _2+3\right) 
    +t^2 \bigg(
    \left( Q_1 +  Q_1^{-1}\right) \chi_1
    + \left(Q_2+Q_2^{-1} \right)+3 \chi _2+\chi _4+8\bigg)  \\
    &+t^3 \bigg(
    \left(Q_1+ Q_1^{-1} \right) \left(4 \chi _1+\chi _3\right)
    + \left(Q_2+ Q_2^{-1} \right)\left(\chi _2+3\right) \notag \\
    &\qquad \qquad +Q_2 Q_3 Q_1^2
    +\frac{1}{Q_1^2 Q_2 Q_3}
    +8 \chi _2+3 \chi _4+\chi _6+17 \bigg) \notag \\
    &+t^4 \bigg(
    Q_1^2 \left(Q_2 Q_3 \left(\chi _2+4\right)+Q_3+\chi _2\right)
    +\frac{\frac{\chi _2+4}{Q_2 Q_3}+\frac{1}{Q_3}+\chi _2}{Q_1^2} \notag \\
    &\qquad +Q_1 \left(\frac{\chi _1}{Q_2}+Q_2 \left(Q_3 \chi _1+\chi _1\right)+12 \chi _1+4 \chi _3+\chi _5\right)
    +\frac{\frac{\frac{\chi _1}{Q_3}+\chi _1}{Q_2}+Q_2 \chi _1+12 \chi _1+4 \chi _3+\chi _5}{Q_1} \notag \\
    &\quad +Q_2 \left(3 \chi _2+\chi _4+9\right)+\frac{3 \chi _2+\chi _4+9}{Q_2}+Q_2^2+\frac{1}{Q_2^2}+18 \chi _2+8 \chi _4+3 \chi _6+\chi _8+34
    \bigg) \notag 
    +\ldots
\end{align}
\end{small}%
and the global symmetry is given by $\left( \surm(2)_x \times \urm(1)_{Q_1}\right)\slash \Z_2 \times \urm(1)_{Q_2} \times \urm(1)_{Q_3}$ where the $\Z_2$ centre symmetry acts with charge $+1$ on $Q_1$ and trivial on all other $Q_i$.

\paragraph{2nd example.}
Next, modify the example slightly and consider
\begin{align}
      \raisebox{-.5\height}{
    \includegraphics[page=15]{figures/figures_1-form_T_rho_sigma.pdf}
    } 
\end{align}
together with the fugacity map
\begin{align}
    v_1 \to x_1^2 
    \;,\quad 
    v_2 \to Q_1
    \;,\quad 
    v_3 \to Q_2
    \;,\quad 
    v_3 \to Q_3\,.
\end{align}
The Coulomb branch Hilbert series reads
\begin{small}
\begin{align}
\label{eq:HS_U1-SU2-U2-U1}
    \HS&=
    1+t \left(\chi _2+3\right) 
     + t^{3/2}\left(Q_1+\frac{1}{Q_1}\right)  
    +t^2 \left(Q_2+\frac{1}{Q_2}+3 \chi _2+\chi _4+8\right)  \\
  &+t^{5/2} \left(\frac{\frac{1}{Q_2}+\chi _2+4}{Q_1}+Q_1 \left(Q_2+\chi _2+4\right)\right) \notag \\
      &+t^3 \left(Q_2 \left(\chi _2+3\right)+\frac{\chi _2+3}{Q_2}+Q_1^2+\frac{1}{Q_1^2}+8 \chi _2+3 \chi _4+\chi _6+17\right)\notag  \\
   &+t^{7/2} \left(\frac{\frac{\chi _2+4}{Q_2}+Q_2+4 \chi _2+\chi _4+11}{Q_1}+Q_1 \left(Q_2 \left(\chi _2+4\right)+\frac{1}{Q_2}+4 \chi _2+\chi _4+11\right)\right) \notag \\
   &+t^4 \bigg(Q_1^2 \left(Q_2 \left(Q_3 \chi _2+1\right)+\chi _2+4\right)+\frac{\frac{\frac{\chi _2}{Q_3}+1}{Q_2}+\chi _2+4}{Q_1^2} 
   +Q_2^2+\frac{1}{Q_2^2} \notag \\
   &\qquad +Q_2 \left(3 \chi _2+\chi _4+9\right)
   +\frac{3 \chi _2+\chi _4+9}{Q_2}
   +18 \chi _2+8 \chi _4+3 \chi _6+\chi _8+34\bigg) \notag
   +\ldots
\end{align}
\end{small}%
and the global symmetry is given by $ \psurm(2)_x \times \prod_{i=1}^3 \urm(1)_{Q_i}$ where the $\Z_2$ centre symmetry acts trivial on all $Q_i$.

\paragraph{3rd example.}
The Hilbert series for the example in \eqref{eq:T-sigma-rho-ex3} with notation \eqref{eq:T-sigma-rho-ex3_labels}
\begin{align}
\label{eq:HS_T-sigma_rho_ex3}
\HS =
1
&+t \left(\chi _{1,1}+1\right) 
+t^{\frac{3}{2}} \left(\frac{\chi _{0,1}}{Q}+Q \chi _{1,0}\right)  
+t^2 \left(3 \chi _{1,1}+\chi _{2,2}+3\right)  \\
&+t^{\frac{5}{2}} \left(\frac{3 \chi _{0,1}+\chi _{1,2}+\chi _{2,0}}{Q}+Q \left(\chi _{0,2}+3 \chi _{1,0}+\chi _{2,1}\right)\right)
+ \ldots \notag
\end{align}
with $\chi_{n_1,n_2}$ the $A_2$ characters for irreps $[n_1,n_2]$. The $\Z_3$ centre charge of the $\urm(1)_Q$ is determined to be $-1 \bmod 3$, such that the symmetry group is $(\surm(3)\times \urm(1)) \slash \Z_3 \cong \urm(3)$.

After gauging a discrete $\Z_2$ $0$-form symmetry, the labelling becomes
\begin{align}
  \raisebox{-.5\height}{
    \includegraphics[page=24]{figures/figures_1-form_T_rho_sigma.pdf}
    }
\qquad \longleftrightarrow\qquad
    \raisebox{-.5\height}{
    \includegraphics[page=25]{figures/figures_1-form_T_rho_sigma.pdf}
    }
    \label{eq:T-sigma-rho-ex3_gauged_label}
\end{align}
with fugacity map
\begin{align}
w_1 = \frac{Q_1}{u_1} 
\; , \qquad 
w_2 = x_1^2
\; , \qquad 
v = Q_0
\end{align}
with $x_1$ an $A_1$ weight space fugacity and $Q_{0,1}$ two $\urm(1)$ fugacities. The Hilbert series becomes
\begin{align}
\label{eq:HS_T-sigma_rho_ex3_gauged}
\HS =1
&+t \left(\chi _2+2\right) 
+t^{3/2} \left(Q_1+\frac{1}{Q_1}\right)  \chi _1 \\
&+t^2 \left(\left(Q_0 Q_1^2+\frac{1}{Q_0 Q_1^2}+4\right) \chi _2+\chi _4+7\right) \notag \\
&+t^{5/2} \left(\left(Q_0 Q_1+5 Q_1+\frac{1}{Q_0 Q_1}+\frac{5}{Q_1}\right) \chi _1+\left(Q_1+\frac{1}{Q_1}\right) \chi _3\right) 
+ \ldots \notag 
\end{align}
where $\chi_{n_1}$ denotes $A_1$ characters for irreps $[n_1]$. It is apparent that the $\Z_2$ centre charges of $(Q_0,Q_1)$ are $(0,-1\bmod 2) $.

\subsection{\texorpdfstring{$T[\sorm(2N)]$ theories}{TSO2N theories}}
\label{app:TSO2N}
In this appendix, computational details for the $T[\sorm(2N)]$ theories are provided. 
For orthosymplectic quivers, the topological symmetries visible in the UV Lagrangian are severely limited. For an $\sorm(k)$ gauge group, there is only a $\Z_2$ valued topological 0-form symmetry. For $\sorm(2)$, there exists a whole $\urm(1)$ topological 0-form symmetry. Thus, to confirm mirror symmetry after such a $\Z_2$ is gauged, one needs to identify the $\Z_2$ in the original mirror pair. Therefore, it is sufficient to provide the $\Z_2$-refined Hilbert series of the original mirror pair to demonstrate agreement after gauging.
The Hilbert series after gauging the discrete $\Z_2$ symmetry is simply obtained by averaging over $\Z_2$.

\subsubsection{\texorpdfstring{$T[\sorm(6)]$ theories}{TSO6 theories}}
For $T[\sorm(6)]$, one can gauge a $\mathbb{Z}_2^t \subset \psorm(6)_t$, which corresponds to gauging the $\mathbb{Z}_2^f$ factor inside the flavour symmetry of the mirror theory, which is identified by a ``2+1'' splitting of the fundamental flavours.
Before gauging, the discrete fugacities are attributed as follows:
\begin{align}
      \raisebox{-.5\height}{
    \includegraphics[page=1]{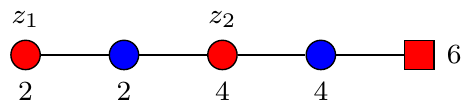}
    } 
  \quad  \longleftrightarrow \quad 
      \raisebox{-.5\height}{
    \includegraphics[page=2]{figures/figures_1-form_TSO2N_app.pdf}
    } 
\end{align}
 The Coulomb branch Hilbert series of the left-hand side (which equals the Higgs branch Hilbert series of the right theory) reads 
 \begin{align}
     \HS=1&+(7+8z_2)t+(63+56z_2)t^2+(328+336z_2)t^3+(1476 + 1448 z_2)t^4\\ \notag
     &+(5390 + 5424 z_2)t^5+(17500 + 17416 z_2)t^6  + \ldots \notag
 \end{align}
 with the $\Z_2$ fugacity $z_2=a$.

\subsubsection{\texorpdfstring{$T[\sorm(8)]$ theories}{TSO8 theories}}
For $T[\sorm(8)]$, one can gauge $\mathbb{Z}_2 \subset \psorm(8)_t$, which corresponds to gauge a $\mathbb{Z}_2$ factor inside the flavour symmetry for the mirror theory. Now one can choose to gauge this $\Z_2^t$ for the $\sorm(4)$ or $\sorm(6)$ gauge node.

\paragraph{$\Z_2^t$ of $\sorm(4)$.}
 Gauging the $\Z_2^t$ of the $\sorm(4)$ gauge node leads to a ``2+2'' splitting of the fundamental flavours. Before gauging, the discrete fugacities are attributed as follows:
\begin{align}
      &\raisebox{-.5\height}{
    \includegraphics[page=3]{figures/figures_1-form_TSO2N_app.pdf}
    }  \\
  &\quad  \longleftrightarrow \quad 
      \raisebox{-.5\height}{
    \includegraphics[page=4]{figures/figures_1-form_TSO2N_app.pdf}
    } \notag 
\end{align}
 The Coulomb branch Hilbert series of the left-hand side (which equals the Higgs branch Hilbert series of the right theory) reads 
  \begin{align}
     \HS&=1+(12+16z_2)t+(213+192z_2)t^2+(1984+2048z_2)t^3+\ldots \notag
 \end{align}
 with the $\Z_2$ fugacity $z_2=a$.

\paragraph{$\Z_2^t$ of $\sorm(6)$.}
 Gauging  the $\Z_2^t$ of the $\sorm(6)$ gauge node leads to ``3+1'' splitting of the fundamental flavours in the mirror theory. Before gauging, the discrete fugacities are attributed as follows:
\begin{align}
      &\raisebox{-.5\height}{
    \includegraphics[page=5]{figures/figures_1-form_TSO2N_app.pdf}
    } \\
  &\quad  \longleftrightarrow \quad 
      \raisebox{-.5\height}{
    \includegraphics[page=6]{figures/figures_1-form_TSO2N_app.pdf}
    } \notag 
\end{align}
The Coulomb branch Hilbert series of the left-hand side (and the Higgs branch Hilbert series of the right-hand theory) reads
   \begin{align}
     \HS&=1+(12z_3+16)t+(213+192z_3)t^2+(1984z_3+2048)t^3+\ldots\notag
 \end{align}
 with the $\Z_2$ fugacity $z_3=a$.

\subsection{\texorpdfstring{$\sprm(k)$ SQCD and orthosymplectic mirrors}{Sp(k) SQCD  and orthosymplectic mirrors}}
\label{app:Sp_SQCD}
The logic is as in Appendix \ref{app:TSO2N}, to evaluate the Hilbert series of the theories after $\Z_2$ gauging, one evaluates the $\Z_2$-refined Hilbert series of $\sprm(k)$ SQCD and its mirror theory. Once agreement is found, the theories after gauging have agreeing the Hilbert series by construction. 

\subsubsection{\texorpdfstring{$\sprm(2)$ SQCD, 5 flavours}{Sp(k) SQCD, 5 flavours}}
Consider $\sprm(2)$ with 5 fundamental flavours.

\paragraph{$\Z_2^t$ of $\sorm(2)$.}
Gauging the $\Z_2^t$ of one of the  $\sorm(2)$ gauge nodes corresponds to  a ``1+4''-splitting of the fundamental flavours in the mirror theory. Before gauging, the discrete fugacities are assigned as follows:
\begin{align}
      \raisebox{-.5\height}{
    \includegraphics[page=1]{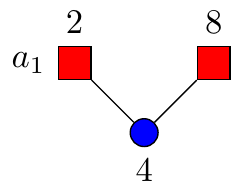}
    } 
   \quad  \longleftrightarrow \quad 
      \raisebox{-.5\height}{
    \includegraphics[page=2]{figures/figures_1-form_Sp_SQCD.pdf}
    } 
\end{align}
The Coulomb branch Hilbert series of the right theory (which agrees with the Higgs branch Hilbert series of the left theory) reads
\begin{align}
    \HS_{\text{1+4}}=1+(16a_1+29)t+(448a_1+532)t^2+\ldots
\end{align}
with the $\Z_2$-fugacity $a_1$.

\paragraph{$\Z_2^t$ inside $\sorm(4)$.}
Gauging the $\Z_2^t$ of one of the  $\sorm(4)$ gauge nodes corresponds to  a ``2+3''-splitting of the fundamental flavours in the mirror theory. Before gauging, the discrete fugacities are assigned as follows:
\begin{align}
      \raisebox{-.5\height}{
    \includegraphics[page=3]{figures/figures_1-form_Sp_SQCD.pdf}
    } 
   \quad  \longleftrightarrow \quad 
      \raisebox{-.5\height}{
    \includegraphics[page=4]{figures/figures_1-form_Sp_SQCD.pdf}
    } 
\end{align}
The Coulomb branch Hilbert series of the right theory (which agrees with the Higgs branch Hilbert series of the left theory) reads
\begin{align}
    \HS_{\text{2+3}}=1+(24a_2+21)t+(480a_2+500)t^2+\ldots
\end{align}
with $\Z_2$-fugacity $a_2$.

\subsubsection{\texorpdfstring{$\sprm(2)$ SQCD, 6 flavours}{Sp(k) SQCD, 6 flavours}}

\paragraph{$\Z_2^t$ of $\sorm(2)$.}
Gauging the $\Z_2^t$ of one of the  $\sorm(2)$ gauge nodes corresponds to  a ``1+5''-splitting of the fundamental flavours in the mirror theory. Before gauging, the discrete fugacities are assigned as follows:
\begin{align}
      \raisebox{-.5\height}{
    \includegraphics[page=7]{figures/figures_1-form_Sp_SQCD.pdf}
    } 
   \quad  \longleftrightarrow \quad 
      \raisebox{-.5\height}{
    \includegraphics[page=8]{figures/figures_1-form_Sp_SQCD.pdf}
    } 
\end{align}
The Coulomb branch Hilbert series of the right theory (which agrees with the Higgs branch Hilbert series of the left theory) reads
\begin{align}
    \text{HS}_{\text{1+5}}=1+(20a_1+46)t+(900a_1+1233)t^2+\ldots
\end{align}
with the $\Z_2$-fugacity $a_1$.

\paragraph{$\Z_2^t$ of $\sorm(4)$.}
Gauging the $\Z_2^t$ of one of the  $\sorm(4)$ gauge nodes corresponds to  a ``2+4''-splitting of the fundamental flavours in the mirror theory. Before gauging, the discrete fugacities are assigned as follows:
\begin{align}
      \raisebox{-.5\height}{
    \includegraphics[page=9]{figures/figures_1-form_Sp_SQCD.pdf}
    } 
   \quad  \longleftrightarrow \quad 
      \raisebox{-.5\height}{
    \includegraphics[page=10]{figures/figures_1-form_Sp_SQCD.pdf}
    } 
\end{align}
The Coulomb branch Hilbert series of the right theory (which agrees with the Higgs branch Hilbert series of the left theory) reads
\begin{align}
    \text{HS}_{\text{2+4}}=1+(32a_2+34)t+(1056a_2+1077)t^2+\ldots
\end{align}
with the $\Z_2$-fugacity $a_2$.

\paragraph{$\Z_2^t$ of $\sorm(5)$.}
Gauging the $\Z_2^t$ of the  $\sorm(5)$ gauge node corresponds to a ``3+3''-splitting of the fundamental flavours in the mirror theory. Before gauging, the discrete fugacities are assigned as follows:
\begin{align}
      \raisebox{-.5\height}{
    \includegraphics[page=11]{figures/figures_1-form_Sp_SQCD.pdf}
    } 
   \quad  \longleftrightarrow \quad 
      \raisebox{-.5\height}{
    \includegraphics[page=12]{figures/figures_1-form_Sp_SQCD.pdf}
    } 
\end{align}
and the Coulomb branch Hilbert series of the right theory (which agrees with the Higgs branch Hilbert series of the left theory) reads
\begin{align}
    \text{HS}_{\text{3+3}}=1+(36a_3+30)t+(1044a_3+1089)t^2+\ldots
\end{align}
with the $\Z_2$-fugacity $a_3$.
\subsection{\texorpdfstring{$\sprm(k)$ SQCD and $D$-type mirrors}{Sp(k) SQCD and D-type mirrors}}
\label{app:Sp_SQCD_D-type_mirror}
In this appendix, computational evidence for the general results of Section \ref{sec:D-type} is provided. The relevant mirror map is discussed in Appendix \ref{app:Sp_SQCD}.
\subsubsection{\texorpdfstring{$D_7$ Dynkin quiver}{D7 Dynkin quiver}}
Consider $\sprm(2)$ SQCD with $N=7$ flavours and its $D_7$ Dynkin mirror quiver, see \eqref{eq:Sp_SQCD_unitary_mirror}.
\paragraph{2nd node.} The mirror pairs is defined by 
\begin{align}
    \raisebox{-.5\height}{
    \includegraphics[page=7]{figures/figures_1-form_Sp_SQCD_uni.pdf}
    } 
    \quad \longleftrightarrow \quad
     \raisebox{-.5\height}{
    \includegraphics[page=8]{figures/figures_1-form_Sp_SQCD_uni.pdf}
    } 
    \;.
\end{align}
and the fugacity map is
\begin{align}
w_1 = u_1^2  \; , \qquad 
(w_3,w_4,w_5,w_6,w_7)_i= \prod_{j=1}^5   x_j^{C^{D_5}_{ij}}  
\; , \qquad
v= \frac{u_2^2}{x_2}
\end{align}
with the weight space fugacities $x_i$, $u_a$ of $\sormL(10)$ and $\sormL(4)$, respectively. (See the end of Section \ref{sec:D-type} for the global symmetry enhancement.) The flavour fugacities follow from \eqref{eq:mirror_map_D-type_after_gauging}.
The Higgs/Coulomb branch Hilbert series then evaluates to
\begin{align}
\label{eq:HS_D7_node2}
\HS = 1
&+t \left(\chi _{0,1,0,0,0}+\phi _{0,2}+\phi _{2,0}\right)  \\
&+
t^2 \bigg(
\chi _{0,0,0,1,1}
+\chi _{0,2,0,0,0}
+\chi _{2,0,0,0,0}
+2 \chi _{0,1,0,0,0} \phi _{0,2}
+2 \chi _{0,1,0,0,0} \phi _{2,0} \notag \\
&\qquad +\chi _{2,0,0,0,0} \phi _{2,2}+\phi _{0,4}+\phi _{2,2}+\phi _{4,0}+2
\bigg) + \ldots
\notag
\end{align}
with $\chi$, $\phi$ denoting $\sormL(10)$, $\sormL(4)$ characters, respectively. The global form of the isometry group is $\psorm(4)\times \psorm(10)$.
\paragraph{3rd node.} The mirror pair is 
\begin{align}
    \raisebox{-.5\height}{
    \includegraphics[page=9]{figures/figures_1-form_Sp_SQCD_uni.pdf}
    } 
    \quad \longleftrightarrow \quad
     \raisebox{-.5\height}{
    \includegraphics[page=10]{figures/figures_1-form_Sp_SQCD_uni.pdf}
    } 
    \;.
\end{align}
and the fugacity map is
\begin{align}
(w_1,w_2)_a = \prod_{b=1}^2   u_b^{C^{A_2}_{ab}}    \; , \qquad 
(w_4,w_5,w_6,w_7)_i= \prod_{j=1}^4   x_j^{C^{D_4}_{ij}}  
\; , \qquad
v= \frac{Q}{x_1}
\end{align}
with the weight space fugacities $x_i$, $u_a$ of $\sormL(8)$ and $\surmL(3)$, respectively. The flavour fugacities follow from \eqref{eq:mirror_map_D-type_after_gauging}. 
The Higgs/Coulomb branch Hilbert series then evaluates to
\begin{align}
\HS =1 
&+t \left(\chi _{0,1,0,0}+\phi _{1,1}+1\right) \\
&+
t^2 \bigg(
Q \chi _{1,0,0,0}
+\frac{\chi _{1,0,0,0}}{Q}
+Q \chi _{1,0,0,0} \phi _{1,1}
+\frac{\chi _{1,0,0,0} \phi _{1,1}}{Q} \notag \\
&\qquad
+\chi _{0,0,0,2}
+\chi _{0,0,2,0}
+2 \chi _{0,1,0,0}
+\chi _{0,2,0,0}
+\chi _{2,0,0,0}\notag \\
&\qquad
+2 \chi _{0,1,0,0} \phi _{1,1}
+\chi _{2,0,0,0} \phi _{1,1}
+3 \phi _{1,1}
+\phi _{2,2}+3
\bigg) + \ldots \notag 
\end{align}
with $\chi$, $\phi$ denoting $\sormL(8)$, $\surmL(3)$ characters, respectively. The global form is given by $\psurm(3) \times \frac{\urm(1)_Q \times \spin(8) }{ \Z_2 \times \Z_2}$ with $Q$ having $\Z_2 \times \Z_2$ centre charges $(0,1 \bmod 2)$. Thus, the isometry group is $\psurm(3) \times \frac{\urm(1)_Q \times \sorm(8) }{\Z_2}$.
\paragraph{4th node.} The mirror pair is
\begin{align}
    \raisebox{-.5\height}{
    \includegraphics[page=11]{figures/figures_1-form_Sp_SQCD_uni.pdf}
    } 
    \quad \longleftrightarrow \quad
     \raisebox{-.5\height}{
    \includegraphics[page=12]{figures/figures_1-form_Sp_SQCD_uni.pdf}
    } 
    \;.
\end{align}
and the fugacity map is
\begin{align}
(w_1,w_2,w_3)_a = \prod_{b=1}^3   u_b^{C^{A_3}_{ab}}    \; , \qquad 
(w_5,w_6,w_7)_i= \prod_{j=1}^3   x_j^{C^{D_3}_{ij}}  
\; , \qquad
v= Q
\end{align}
with the weight space fugacities $x_i$, $u_a$ of $\sormL(6)$ and $\surmL(4)$, respectively.  The flavour fugacities follow from \eqref{eq:mirror_map_D-type_after_gauging}.
The Higgs/Coulomb branch Hilbert series then evaluates to
\begin{align}
\HS =1
&+t \left(\chi _{0,1,1}+\phi _{1,0,1}+1\right)
\\
&+t^2 \bigg(
Q \phi _{0,2,0}
+\frac{\phi _{0,2,0}}{Q}
+3 \chi _{0,1,1}
+\chi _{0,2,2}
+\chi _{2,0,0}
+2 \chi _{0,1,1} \phi _{1,0,1} \notag \\
&\qquad 
+\chi _{2,0,0} \phi _{1,0,1}
+2 \phi _{0,2,0}
+3 \phi _{1,0,1}
+\phi _{2,0,2}
+Q+\frac{1}{Q}+3
\bigg)
+\ldots \notag 
\end{align}
with $\chi$, $\phi$ denoting $\sormL(6)$, $\surmL(4)$ characters, respectively. The global form is $\psurm(4) \times \urm(1)_Q \times \psorm(6)$.
\paragraph{5th node.} The mirror pair is
\begin{align}
    \raisebox{-.5\height}{
    \includegraphics[page=13]{figures/figures_1-form_Sp_SQCD_uni.pdf}
    } 
    \quad \longleftrightarrow \quad
     \raisebox{-.5\height}{
    \includegraphics[page=14]{figures/figures_1-form_Sp_SQCD_uni.pdf}
    } 
    \;.
\end{align}
and the fugacity map is
\begin{align}
(w_1,w_2,w_3,w_4)_a = \prod_{b=1}^4   u_b^{C^{A_4}_{ab}}    \; , \qquad 
(w_6,w_7)_i= \prod_{j=1}^2   x_j^{C^{D_2}_{ij}}  
\; , \qquad
v= \frac{Q}{u_4}
\end{align}
with the weight space fugacities $x_i$, $u_a$ of $\sormL(4)$ and $\surmL(5)$, respectively.  The flavour fugacities follow from \eqref{eq:mirror_map_D-type_after_gauging}.
The Higgs/Coulomb branch Hilbert series then evaluates to
\begin{align}
\HS =1 &+t \left(\chi _{0,2}+\chi _{2,0}+\phi _{1,0,0,1}+1\right)
\\
&+ t^2 \bigg(
\frac{\phi _{0,0,0,1}}{Q}
+Q \phi _{1,0,0,0}
+Q \phi _{0,0,2,0}
+\frac{\phi _{0,2,0,0}}{Q}
+2 \chi _{0,2}
+\chi _{0,4}\notag \\
&\qquad 
+2 \chi _{2,0} 
+\chi _{2,2}
+\chi _{4,0}
+2 \chi _{0,2} \phi _{1,0,0,1}
+2 \chi _{2,0} \phi _{1,0,0,1} \notag \\
&\qquad 
+\chi _{2,2} \phi _{1,0,0,1} 
+2 \phi _{0,1,1,0}
+3 \phi _{1,0,0,1}
+\phi _{2,0,0,2}
+4\bigg)
+\ldots
\notag 
\end{align}
with $\chi$, $\phi$ denoting $\sormL(4)$, $\surmL(5)$ characters, respectively. The global form is $\frac{\surm(5)\times \urm(1)_Q}{\Z_5} \times \psorm(4)$ with $Q$ having $\Z_5$ centre charge $4 \bmod 5$; i.e.\ the isometry group is $\urm(5) \times \psorm(4)$.
\paragraph{6th node.} The mirror pair reads
\begin{align}
    \raisebox{-.5\height}{
    \includegraphics[page=15]{figures/figures_1-form_Sp_SQCD_uni.pdf}
    } 
    \quad \longleftrightarrow \quad
     \raisebox{-.5\height}{
    \includegraphics[page=16]{figures/figures_1-form_Sp_SQCD_uni.pdf}
    } 
    \;.
\end{align}
and the fugacity map is
\begin{align}
(w_1,w_2,w_3,w_4,w_5,w_7)_a = \prod_{b=1}^6   u_b^{C^{A_6}_{ab}}    \; , \qquad 
v= \frac{Q}{u_4}
\end{align}
with the weight space fugacities $u_a$ of  $\surmL(7)$, respectively.  The flavour fugacities follow from \eqref{eq:mirror_map_D-type_after_gauging}.
The Higgs/Coulomb branch Hilbert series then evaluates to
\begin{align}
\HS = 1
&+t \left(\phi _{1,0,0,0,0,1}+1\right) \\
&+t^2 \bigg(
Q \phi _{0,0,0,0,2,0}
+\frac{\phi _{0,0,0,1,0,0}}{Q}
+Q \phi _{0,0,1,0,0,0}
+\frac{\phi _{0,2,0,0,0,0}}{Q} \notag \\
&\qquad 
+2 \phi _{0,1,0,0,1,0}+2 \phi _{1,0,0,0,0,1}+\phi _{2,0,0,0,0,2}+2\bigg)
+\ldots \notag
\end{align}
with $\phi$ denoting $\surmL(7)$ characters. The global symmetry is $\frac{\surm(7)\times \urm(1)_Q}{\Z_7}$ with $Q$ having $\Z_7$ centre charges $4\bmod 7$.

\subsubsection{\texorpdfstring{$D_8$ Dynkin quiver}{D8 Dynkin quiver}}
Consider $\sprm(2)$ SQCD with $N=8$ flavours and its $D_8$ Dynkin mirror quiver, see \eqref{eq:Sp_SQCD_unitary_mirror}.
\paragraph{2nd node.} The mirror pair is defined by
\begin{align}
    \raisebox{-.5\height}{
    \includegraphics[page=17]{figures/figures_1-form_Sp_SQCD_uni.pdf}
    } 
    \quad \longleftrightarrow \quad
     \raisebox{-.5\height}{
    \includegraphics[page=18]{figures/figures_1-form_Sp_SQCD_uni.pdf}
    } 
    \;.
\end{align}
and the fugacity map is
\begin{align}
w_1 = u_1^2   \; , \qquad 
(w_3,w_4,w_5,w_6,w_7,w_8)_i = \prod_{j=1}^6   x_j^{C^{D_6}_{ij}}    \; , \qquad 
v= \frac{u_2^2}{x_2}
\end{align}
with the weight space fugacities $u_a$ and $x_j$ of $\sormL(4)$ and  $\sormL(12)$, respectively.  The flavour fugacities follow from \eqref{eq:mirror_map_D-type_after_gauging}.
The Higgs/Coulomb branch Hilbert series then evaluates to
\begin{align}
\label{eq:HS_D8_node2}
\HS = 1
&+t \left(\chi _{0,1,0,0,0,0}+\phi _{0,2}+\phi _{2,0}\right)  \\
&+t^2 \bigg(
\chi _{0,0,0,1,0,0}
+\chi _{0,2,0,0,0,0}
+\chi _{2,0,0,0,0,0}
+2 \chi _{0,1,0,0,0,0} \phi _{0,2} \notag \\
&\qquad 
+2 \chi _{0,1,0,0,0,0} \phi _{2,0}+\chi _{2,0,0,0,0,0} \phi _{2,2}+\phi _{0,4}+\phi _{2,2}+\phi _{4,0}+2\bigg)
+\ldots
\notag
\end{align}
and $\phi$ are $\sormL(4)$ characters and $\chi$ are $\sormL(12)$ characters. The global form is read off to be $\psorm(4) \times \psorm(12)$.

\paragraph{3rd node.} The mirror pair is
\begin{align}
    \raisebox{-.5\height}{
    \includegraphics[page=19]{figures/figures_1-form_Sp_SQCD_uni.pdf}
    } 
    \quad \longleftrightarrow \quad
     \raisebox{-.5\height}{
    \includegraphics[page=20]{figures/figures_1-form_Sp_SQCD_uni.pdf}
    } 
    \;.
\end{align}
and the fugacity map is
\begin{align}
(w_1,w_2)_a = \prod_{b=1}^2   u_b^{C^{A_2}_{ab}}  \; , \qquad 
(w_4,w_5,w_6,w_7,w_8)_i = \prod_{j=1}^5   x_j^{C^{D_5}_{ij}}    \; , \qquad 
v= \frac{Q}{x_1}
\end{align}
with the weight space fugacities $u_a$ and $x_j$ of $\surmL(3)$ and  $\sormL(10)$, respectively. The flavour fugacities follow from \eqref{eq:mirror_map_D-type_after_gauging}.  
The Higgs/Coulomb branch Hilbert series then evaluates to
\begin{align}
\HS &= 1+t \left(\chi _{0,1,0,0,0}+\phi _{1,1}+1\right) \\
&+ t^2 \bigg(
Q \chi _{1,0,0,0,0}
+\frac{\chi _{1,0,0,0,0}}{Q}
+Q \chi _{1,0,0,0,0} \phi _{1,1}
+\frac{\chi _{1,0,0,0,0} \phi _{1,1}}{Q} \notag \\
&\qquad 
+\chi _{0,0,0,1,1}+2 \chi _{0,1,0,0,0}+\chi _{0,2,0,0,0}+\chi _{2,0,0,0,0}+2 \chi _{0,1,0,0,0} \phi _{1,1} \notag \\
&\qquad +\chi _{2,0,0,0,0} \phi _{1,1}+3 \phi _{1,1}+\phi _{2,2}+3\bigg) + \ldots
\notag
\end{align}
and $\phi$ denotes $\surmL(3)$ characters, while $\chi$ are $\sormL(10)$ characters. The global symmetry is $\psurm(3)\times \frac{\urm(1)_Q\times\spin(10) }{\Z_4}$ where the $Q$ can be assigned $\Z_4$ charge $2 \bmod 4$.

\paragraph{4th node.} The mirror pair reads
\begin{align}
    \raisebox{-.5\height}{
    \includegraphics[page=21]{figures/figures_1-form_Sp_SQCD_uni.pdf}
    } 
    \quad \longleftrightarrow \quad
     \raisebox{-.5\height}{
    \includegraphics[page=22]{figures/figures_1-form_Sp_SQCD_uni.pdf}
    } 
    \;.
\end{align}
and the fugacity map is
\begin{align}
(w_1,w_2,w_3)_a = \prod_{b=1}^3   u_b^{C^{A_3}_{ab}}  \; , \qquad 
(w_5,w_6,w_7,w_8)_i = \prod_{j=1}^4   x_j^{C^{D_4}_{ij}}    \; , \qquad 
v= Q
\end{align}
with the weight space fugacities $u_a$ and $x_j$ of $\surmL(4)$ and  $\sormL(8)$, respectively.  The flavour fugacities follow from \eqref{eq:mirror_map_D-type_after_gauging}.
The Higgs/Coulomb branch Hilbert series then evaluates to
\begin{align}
\HS =1
&+t \left(\chi _{0,1,0,0}+\phi _{1,0,1}+1\right)
\\
&+t^2 \bigg(
Q \phi _{0,2,0}
+\frac{\phi _{0,2,0}}{Q}
+\chi _{0,0,0,2}
+\chi _{0,0,2,0}
+2 \chi _{0,1,0,0}
+\chi _{0,2,0,0} 
+\chi _{2,0,0,0}\notag \\
&\qquad
+2 \chi _{0,1,0,0} \phi _{1,0,1}
+\chi _{2,0,0,0} \phi _{1,0,1}
+2 \phi _{0,2,0}
+3 \phi _{1,0,1}
+\phi _{2,0,2}
+Q+\frac{1}{Q}+3
\bigg)
 + \ldots
\notag
\end{align}
with $\phi$ denoting $\surmL(4)$ characters and $\chi$ are $\sormL(8)$ characters. The global symmetry is $\psurm(4)\times \urm(1)_Q \times \psorm(8)$.
\paragraph{5th node.} The mirror pair is
\begin{align}
    \raisebox{-.5\height}{
    \includegraphics[page=23]{figures/figures_1-form_Sp_SQCD_uni.pdf}
    } 
    \quad \longleftrightarrow \quad
     \raisebox{-.5\height}{
    \includegraphics[page=24]{figures/figures_1-form_Sp_SQCD_uni.pdf}
    } 
    \;.
\end{align}
and the fugacity map is
\begin{align}
(w_1,w_2,w_3,w_4)_a = \prod_{b=1}^4   u_b^{C^{A_4}_{ab}}  \; , \qquad 
(w_6,w_7,w_8)_i = \prod_{j=1}^3   x_j^{C^{D_3}_{ij}}    \; , \qquad 
v= \frac{Q}{u_4}
\end{align}
with the weight space fugacities $u_a$ and $x_j$ of $\surmL(5)$ and  $\sormL(6)$, respectively.  The flavour fugacities follow from \eqref{eq:mirror_map_D-type_after_gauging}.
The Higgs/Coulomb branch Hilbert series then evaluates to
\begin{align}
\HS = 1&+t \left(
\chi _{0,1,1}
+\phi _{1,0,0,1}+1\right)\\
&+t^2 \bigg(
\frac{\phi _{0,0,0,1}}{Q}
+Q \phi _{1,0,0,0}
+Q \phi _{0,0,2,0}
+\frac{\phi _{0,2,0,0}}{Q}
+3 \chi _{0,1,1}
+\chi _{0,2,2}
+\chi _{2,0,0} \notag \\
&\qquad 
+2 \chi _{0,1,1} \phi _{1,0,0,1}
+\chi _{2,0,0} \phi _{1,0,0,1}
+2 \phi _{0,1,1,0}
+3 \phi _{1,0,0,1}
+\phi _{2,0,0,2}+3
\bigg)
+ \ldots
\notag
\end{align}
and $\phi$, $\chi$ denote $\surmL(5)$, $\sormL(6)$ characters, respectively. The isometry group is $\frac{\surm(5) \times \urm(1)_Q}{\Z_5} \times \psorm(6)$ where the $\Z_5$ charge of $Q$ is $4 \bmod 5$. The global form is then $\urm(5) \times \psorm(6)$.
\paragraph{6th node.} The mirror pair is 
\begin{align}
    \raisebox{-.5\height}{
    \includegraphics[page=25]{figures/figures_1-form_Sp_SQCD_uni.pdf}
    } 
    \quad \longleftrightarrow \quad
     \raisebox{-.5\height}{
    \includegraphics[page=26]{figures/figures_1-form_Sp_SQCD_uni.pdf}
    } 
    \;.
\end{align}
and the fugacity map is
\begin{align}
(w_1,w_2,w_3,w_4,w_5)_a = \prod_{b=1}^5   u_b^{C^{A_5}_{ab}}  \; , \qquad 
(w_7,w_8)_i = \prod_{j=1}^2   x_j^{C^{D_2}_{ij}}    \; , \qquad 
v= \frac{Q}{u_4}
\end{align}
with the weight space fugacities $u_a$ and $x_j$ of $\surmL(6)$ and  $\sormL(2)$, respectively. The flavour fugacities follow from \eqref{eq:mirror_map_D-type_after_gauging}. 
The Higgs/Coulomb branch Hilbert series then evaluates to
\begin{align}
\HS =1&+t \left(\chi _{0,2}+\chi _{2,0}+\phi _{1,0,0,0,1}+1\right)  \\
&+t^2 \bigg(
\frac{\phi _{0,0,0,1,0}}{Q}
+Q \phi _{0,1,0,0,0}
+Q \phi _{0,0,0,2,0}
+\frac{\phi _{0,2,0,0,0}}{Q}
+2 \chi _{0,2}
+\chi _{0,4}
+2 \chi _{2,0} \notag \\
&\qquad 
+\chi _{2,2}
+\chi _{4,0}
+2 \chi _{0,2} \phi _{1,0,0,0,1}
+2 \chi _{2,0} \phi _{1,0,0,0,1}
+\chi _{2,2} \phi _{1,0,0,0,1}
+2 \phi _{0,1,0,1,0}
\notag \\
&\qquad 
+3 \phi _{1,0,0,0,1}
+\phi _{2,0,0,0,2}
+4\bigg) +\ldots
\notag 
\end{align}
here $\phi$ denote $\surmL(6)$ characters and $\chi$ denote $\sormL(4)$ characters.  The global form is $\frac{\surm(6) \times \urm(1)_Q}{\Z_6} \times \psorm(4)$ where the $\Z_6$ charge of $Q$ is $4 \bmod 6$.
\paragraph{7th node.} The mirror pair is given by
\begin{align}
    \raisebox{-.5\height}{
    \includegraphics[page=27]{figures/figures_1-form_Sp_SQCD_uni.pdf}
    } 
    \quad \longleftrightarrow \quad
     \raisebox{-.5\height}{
    \includegraphics[page=28]{figures/figures_1-form_Sp_SQCD_uni.pdf}
    } 
    \;.
\end{align}
and the fugacity map is
\begin{align}
(w_1,w_2,w_3,w_4,w_5,w_6,w_8)_a = \prod_{b=1}^7   u_b^{C^{A_7}_{ab}}  \; , 
v= \frac{Q}{u_4}
\end{align}
with the weight space fugacities $u_a$ of $\surmL(8)$. The flavour fugacities follow from \eqref{eq:mirror_map_D-type_after_gauging}.  
The Higgs/Coulomb branch Hilbert series then evaluates to
\begin{align}
\HS = 
1
&+t \left(\phi _{1,0,0,0,0,0,1}+1\right) \\
& +t^2 \bigg(
Q \phi _{0,0,0,0,0,2,0}
+Q \phi _{0,0,0,1,0,0,0}
+\frac{\phi _{0,0,0,1,0,0,0}}{Q}
+\frac{\phi _{0,2,0,0,0,0,0}}{Q}  \notag \\
&\qquad +2 \phi _{0,1,0,0,0,1,0}+2 \phi _{1,0,0,0,0,0,1}+\phi _{2,0,0,0,0,0,2}+2\bigg) 
+ \ldots \notag
\end{align}
where $\phi$ denotes $\surmL(8)$ characters. The global form is $\frac{\surm(8)\times \urm(1)_Q}{\Z_8}$ and $Q$ has $\Z_8$ charge $4 \bmod 8$.
\subsection{\texorpdfstring{$\orm(2k)$ SQCD and $C$-type mirrors}{O(2k) SQCD and C-type mirrors}}
This appendix contains explicit Hilbert series for non-simply laced Dynkin quivers and their $\orm(2k)$ mirror SQCD theories. 
For concreteness,  $\orm(2)$ SQCD with 4 fundamental flavours is considered. The mirror is a $C_4$ balanced Dynkin quiver.

\paragraph{Example 1.}
 Gauging the topological symmetry of the gauge node at the long edge leads to the mirror pair
\begin{align}
 \raisebox{-.5\height}{
    \includegraphics[page=12]{figures/figures_1-form_non-simply_laced.pdf}
    } 
    \qquad \longleftrightarrow \qquad 
     \raisebox{-.5\height}{
    \includegraphics[page=13]{figures/figures_1-form_non-simply_laced.pdf}
    } 
    \label{eq:C-type_ex1_appendix}
\end{align}
and the magnetic fluxes for the right-hand side quiver in \eqref{eq:C-type_ex1_appendix} take values in 
\begin{align}
(m_1,\bm{m}_2,\bm{m}_3,l,h) \in \Z \times \Z^2 \times \Z^2 \times \Z \times \Z \,.
\end{align}
The fugacity map is given by
\begin{subequations}
\begin{alignat}{4}
w_1 &= \frac{x_1^2}{x_2} \,, \quad 
&
w_2 &= \frac{x_2^2}{x_1 x_3}  \,, \quad 
&
w_3 &= \frac{x_3^2}{x_2}  \,, \quad 
&
v &= \frac{Q^{-1}}{x_2^2} \\
y_1 &= Q^{\frac{1}{4}} x_1 \,, \quad 
&
y_2 &= Q^{\frac{1}{4}} \frac{x_2}{x_1} \,, \quad 
&
y_3 &= Q^{\frac{1}{4}} \frac{x_3}{x_2}  \,, \quad 
&
y_4 &= Q^{\frac{1}{4}} \frac{1}{x_3}
\end{alignat}
\end{subequations}
with the $A_3$ weight space fugacities $x_i$.
One evaluates the Hilbert series to read
\begin{align}
\label{eq:HS_C4_quiver_SU_4th}
\HS =
1&+t \left(\phi _{1,0,1}+1\right) \\
&+t^2 \left(\frac{\phi _{0,0,4}+\phi _{0,2,0}}{Q}+Q \left(\phi _{0,2,0}+\phi _{4,0,0}\right)+\phi _{0,2,0}+2 \phi _{1,0,1}+2 \phi _{2,0,2}+2\right) \notag \\
&+t^3 \bigg(
\frac{2 \phi _{0,0,4}+\phi _{0,1,2}+\phi _{0,2,0}+\phi _{1,0,5}+\phi _{1,1,3}+\phi _{1,2,1}}{Q} \notag \\
&\qquad +Q \left(\phi _{0,2,0}+\phi _{1,2,1}+\phi _{2,1,0}+\phi _{3,1,1}+2 \phi _{4,0,0}+\phi _{5,0,1}\right) 
\notag \\ 
&\qquad +\phi _{0,1,2}+\phi _{0,2,0}+3 \phi _{1,0,1}+\phi _{1,1,3}+\phi _{1,2,1}+4 \phi _{2,0,2}+\phi _{2,1,0}+2 \phi _{3,0,3}+\phi _{3,1,1}+2
\bigg) \notag  \\
&+\ldots \notag 
\end{align}
where $\phi_{n_1,n_2,n_3}$ denotes $A_3$ characters for irreps $[n_1,n_2,n_3]$. The $\urm(1)_Q$ is trivial under the $\Z_4$ centre of $\surm(4)$, such that the global symmetry group becomes $\psurm(4)\times \urm(1)_Q$.

\paragraph{Example 2.}
Gauging the topological symmetry for the node closest to the non-simply laced edge on the short side leads to
\begin{align}
 \raisebox{-.5\height}{
    \includegraphics[page=14]{figures/figures_1-form_non-simply_laced.pdf}
    } 
    \qquad \longleftrightarrow \qquad 
     \raisebox{-.5\height}{
    \includegraphics[page=15]{figures/figures_1-form_non-simply_laced.pdf}
    } 
    \label{eq:C-type_ex2_appendix}
\end{align}
and the magnetic fluxes for the right-hand side quiver in \eqref{eq:C-type_ex2_appendix} are defined via
\begin{align}
(m_1,\bm{m}_2,l,\bm{m}_4,h) \in 
\bigcup_{i=0}^1 \left( 
\Z \times \Z^2 \times \Z \times \left(\Z + \frac{i}{2} \right)^2 \times \left(\Z + \frac{i}{2} \right) \right) \,.
\end{align}
The fugacity map is given by
\begin{subequations}
\begin{alignat}{4}
w_1 &= \frac{x_1^2}{x_2} \;, \qquad &
w_2 &= \frac{x_2^2}{x_1 x_3}  \;, \qquad &
v&= \frac{1}{x_2^2}  \;, \qquad &
w_4 &= u_1^2 \\
y_1 &=  u_1 \;, \qquad &
y_2 &= x_1 \;, \qquad &
y_3 &= \frac{x_2}{x_1} \;, \qquad &
y_4 &= \frac{x_3}{x_2}
\end{alignat}
\end{subequations}
where $x_i$ are $C_3$ weight space fugacities and $u_1$ is a $C_1$ weight space fugacity.
the Hilbert series can be evaluated to read
\begin{align}
\label{eq:HS_C4_quiver_SU_3rd}
\HS=  1
&+t \left(\chi _{2,0,0}+\phi _2\right)
+t^2 \left(\chi _{0,1,0}+\chi _{0,2,0}+\chi _{4,0,0}+2 \phi _2 \chi _{2,0,0}+\phi _4+1\right)\\
&+t^3 
\bigg(\chi _{2,0,0}+\chi _{2,1,0}+\chi _{2,2,0}+\chi _{6,0,0}+\phi _2 \chi _{0,1,0}+\phi _2 \chi _{0,2,0}+\phi _2 \chi _{2,0,0}+\phi _2 \chi _{2,1,0} \notag \\
&\qquad +2 \phi _2 \chi _{4,0,0}+2 \phi _4 \chi _{2,0,0}+\phi _2+\phi _6\bigg)
+\ldots 
\notag
\end{align}
with $\chi_{n_1,n_2,n_3}$ the $C_3$ characters for irreps $[n_1,n_2,n_3]_C$ and $\phi_{k_1}$ the $C_1$ characters for irreps $[k_1]_C$. As all appearing irreps are invariant under the $\Z_2$ centre symmetries for $C_3$ and $C_1$, the global symmetry group is $\psprm(3)\times \psprm(1)$.

\paragraph{Example 3.}
Gauging the topological symmetry of the other $\urm(2)$ gauge node on the short side leads to
\begin{align}
 \raisebox{-.5\height}{
    \includegraphics[page=16]{figures/figures_1-form_non-simply_laced.pdf}
    } 
    \qquad \longleftrightarrow \qquad 
     \raisebox{-.5\height}{
    \includegraphics[page=17]{figures/figures_1-form_non-simply_laced.pdf}
    } 
    \label{eq:C-type_ex3_appendix}
\end{align}
and the magnetic fluxes for the right-hand side quiver in \eqref{eq:C-type_ex3_appendix} take values in
\begin{align}
(m_1,l,\bm{m}_3,\bm{m}_4,h) \in
\bigcup_{i=0}^1 \left( 
\Z \times \Z \times \Z^2 \times \left(\Z + \frac{i}{2} \right)^2 \times \left(\Z + \frac{i}{2} \right) \right)
\end{align}
The relevant fugacity map is given by
\begin{subequations}
\begin{alignat}{4}
w_1 &= \frac{x_1^2}{x_2} \;, \qquad &
v&= \frac{1}{x_2^2} \;, \qquad &
w_3 &= \frac{u_1^2}{u_2} \;, \qquad &
w_4 &= \frac{u_2^2}{u_1^2} \\
y_1 &= u_1 \;, \qquad &
y_2 &= \frac{u_2}{u_1} \;, \qquad &
y_3 &= x_1 \;, \qquad &
y_4 &=  \frac{x_2}{x_1}
\end{alignat}
\end{subequations}
with $x_i$ and $u_i$ two sets of $C_2$ weight space fugacities.
The Hilbert series reads
\begin{align}
\label{eq:HS_C4_quiver_SU_2nd}
\HS =1
&+t \left(\chi _{2,0}+\phi _{2,0}\right) \\
&+t^2 \left(\chi _{0,1}+\chi _{0,2}+\chi _{4,0}+\chi _{0,1} \phi _{0,1}+2 \chi _{2,0} \phi _{2,0}+\phi _{0,1}+\phi _{0,2}+\phi _{4,0}+1\right) \notag \\
&+t^3 \bigg(
\chi _{2,0}+\chi _{2,1}+\chi _{2,2}+\chi _{6,0}+\chi _{0,1} \phi _{2,0}+\chi _{0,2} \phi _{2,0}+\chi _{2,0} \phi _{2,0}+\chi _{2,1} \phi _{2,0}
\notag \\ 
&\qquad
+2 \chi _{4,0} \phi _{2,0}+\chi _{0,1} \phi _{2,1}+\chi _{2,0} \phi _{0,1}+\chi _{2,0} \phi _{0,2}  +\chi _{2,0} \phi _{2,1}+2 \chi _{2,0} \phi _{4,0}
\notag \\ 
&\qquad
+\chi _{2,1} \phi _{0,1}+\phi _{2,0}+\phi _{2,1}+\phi _{2,2}+\phi _{6,0}
\bigg) \notag  +\ldots 
\end{align}
where $\chi_{n_1,n_2}$ and $\phi_{k_1,k_2}$ denote the $C_2$ characters of the irreps labelled by $[n_1,n_2]$ and $[k_1,k_2]$, respectively. All appearing irreps are invariant under the $\Z_2$ centre symmetries; thus, the global symmetry group is $\psprm(2)\times \psprm(2)$.

\bibliographystyle{JHEP}
\bibliography{bibli.bib}

\end{document}